%% file: paper.tex
\documentclass[usenatbib]{mn2e}
\pdfoutput=1
\usepackage[usenames,dvipsnames]{color}
\usepackage{graphicx}
\usepackage{natbib}
\usepackage{longtable}
\usepackage{placeins}
\renewcommand{\sc}{\scshape}

\def\Ni{587} 
\def\Nib{357} 
\def\Nic{278} 
\def\Nicc{124} 
\def\Nii{292} 
\def\Niii{122} 
\def\Niiib{112} 
\def\Niv{3,675} 
\def\Nivb{993} 
\def\Nv{1,533} 
\def\Nvi{4,029} 
\def\Nvii{114,711} 
\def\Nviii{12,582} 
\def\Nix{6,983} 
\def\Nxi{5,776} 
\def\Nxii{5,642} 
\def\Nxiii{17} 
\def\Nxiv{31} 
\def\Nxivb{25} 
\def\Nxv{503} 
\def\Nxvb{402} 
\def\Nxviii{42} 
\def\Nxix{394} 
\def\Nxx{49} 
\def\Nx{493} 
\def\Nxa{412} 
\def\Nxb{81} 
\def\Nxc{345} 
\def\Nxd{218} 
\def\Nxe{127} 
\def\Nxvi{$\simeq$ 900} 
\def\Nxvii{453} 

\title{The \textit{XMM} Cluster Survey: X-ray analysis methodology} 
\author[E. J. Lloyd-Davies et al.] 
{E.~J.~Lloyd-Davies, $^1$\thanks{E.Lloyd-Davies@sussex.ac.uk}
A.~Kathy Romer,$^1$
Nicola Mehrtens,$^1$
\newauthor
Mark Hosmer,$^1$
Michael Davidson,$^2$
Kivanc Sabirli,$^3$
\newauthor 
Robert G. Mann,$^2$
Matt Hilton,$^{4,5}$ 
Andrew~R.~Liddle,$^1$ 
\newauthor 
Pedro~T. P.~Viana,$^{6,7}$
Heather C.~Campbell,$^{1,8}$
Chris~A.~Collins,$^9$
\newauthor 
E.~Naomi Dubois,$^1$
Peter Freeman,$^{10}$
Craig~D.~Harrison,$^{11}$ 
\newauthor 
Ben Hoyle,$^{8,12,13,14}$
Scott~T.~Kay,$^{15}$ 
Emma Kuwertz,$^1$ 
\newauthor
Christopher~J.~Miller,$^{11}$ 
Robert~ C. Nichol,$^8$ 
Martin~Sahl{\'e}n,$^{16}$ 
\newauthor 
S.~A.~ Stanford,$^{17,18}$ 
John~P.~Stott$^9$ 
\\
$^1$~Astronomy Centre, University of Sussex, Falmer, Brighton, BN1 9QH, UK\\
$^2$~SUPA, Institute for Astronomy, University of Edinburgh, Royal
Observatory, Edinburgh, EH9 3HJ, UK\\ 
$^3$~Department of Physics, Carnegie Mellon University, 5000 Forbes
Avenue, Pittsburgh, PA 15213, USA\\ 
$^{4}$~School of Mathematical Sciences, University of KwaZulu-Natal, Private Bag X54001,Durban 4000, South Africa\\
$^{5}$~University of Nottingham, School of Physics \& Astronomy,
Nottingham, NG7 2RD, UK\\ 
$^{6}$~Centro de Astrof\'{\i}sica da Universidade do Porto, Rua das
Estrelas, 4150-762, Porto, Portugal\\ 
$^{7}$~Departamento de F\'{\i}sica e Astronomia, Faculdade de
Ci\^{e}ncias, Universidade do Porto, 4169-007 Porto, Portugal\\
$^8$~Institute of Cosmology and Gravitation, Dennis Sciama Building,
Burnaby Road, Portsmouth, PO1 3FX, UK\\ 
$^{9}$~Astrophysics Research Institute, Liverpool John Moores
University, Twelve Quays House, Egerton Wharf, Birkenhead, CH41 1LD,
UK\\ 
$^{10}$~Department of Statistics, Carnegie Mellon University, 5000
Forbes Avenue, Pittsburgh, PA 15213, USA\\ 
$^{11}$~Astronomy Department, University of Michigan, Ann Arbor, MI
48109, USA\\ 
$^{12}$~Institute of Sciences of the Cosmos (ICCUB-IEEC), Physics
Department, University of Barcelona, Barcelona 08024, Spain.\\
$^{13}$~CSIC, Consejo Superior de Investigaciones CientiÞcas, Serrano 117, Madrid, 28006, Spain.\\
$^{14}$~Helsinki Institute of Physics, P.O. Box 64, FIN-00014 University of Helsinki, Finland.\\
$^{15}$~Jodrell Bank Centre for Astrophysics, School of Physics and Astronomy, The University of Manchester, Manchester M13 9PL \\
$^{16}$~The Oskar Klein Centre for Cosmoparticle Physics, Department
of Physics, Stockholm University, SE-106 91 Stockholm, Sweden\\
$^{17}$~Physics Department, University of California, Davis, CA 95616,
USA\\ 
$^{18}$~Institute of Geophysics and Planetary Physics, Lawrence
Livermore National Laboratory, Livermore, CA 
94551, USA\\
}

\date{Accepted 2011 ??.  
      Received 2010 ??; 
      in original form 2010 October 4}

\pagerange{\pageref{firstpage}--\pageref{lastpage}}
\pubyear{2010}

\begin{document}

\maketitle

\label{firstpage}

\begin{abstract}

The \textit{XMM} Cluster Survey (XCS) is a serendipitous search for
galaxy clusters using all publicly available data in the
\textit{XMM-Newton} Science Archive. Its main aims are to measure
cosmological parameters and trace the evolution of X-ray scaling
relations. In this paper we describe the data processing methodology
applied to the \Nxi\, \textit{XMM} observations used to construct the
current XCS source catalogue. A total of \Niv\, $>4$-$\sigma$ cluster
candidates with $>$50 background-subtracted X-ray counts are extracted
from a total non-overlapping area suitable for cluster searching of
410 deg$^2$. Of these, \Nivb\, candidates are detected with $>$300 
background-subtracted X-ray photon counts, and we demonstrate that 
robust temperature measurements can be obtained down to this count 
limit. We describe in detail the automated
pipelines used to perform the spectral and surface brightness fitting
for these candidates, as well as to estimate redshifts from the X-ray
data alone. A total of \Ni\, (\Niii) X-ray temperatures to a typical
accuracy of $<$40 ($<$10) per cent have been measured to
date. 
We also present the methodology adopted for determining the 
selection function of the survey, and show that the extended 
source detection algorithm is robust to a range of cluster 
morphologies by inserting mock clusters derived from 
hydrodynamical simulations into real \textit{XMM}images.
These tests show that the simple isothermal
$\beta$-profiles is sufficient to capture the
essential details of the cluster population detected in the archival
\textit{XMM} observations. The redshift follow-up of the XCS cluster
sample is presented in a companion paper, together with a first data
release of \Nxv\, optically-confirmed clusters.
\end{abstract}


\section{Introduction}

Clusters of galaxies are massive objects ($10^{13.5-15}M_{\odot}$)
composed of galaxies, hot ionised gas and dark matter. The
gravitational potential is dominated by dark matter, with the mass
ratio of the three components being roughly 3:10:87 respectively, although with a
strong mass dependence in the ratio of gas to stars
\citep{gonzalez07a}. Clusters provide us with the opportunity to
obtain information about the underlying cosmological model and
important insights into the processes that govern structure formation
(see \citealt{voit05a,allen11a} for reviews).

While detailed studies of individual clusters are extremely important,
especially for obtaining insight into the small-scale processes that
influence the evolution of their baryonic components, a full
understanding of the complex nature of cluster formation and evolution
requires the study of the galaxy cluster population as a whole. This
is best achieved, in practice, by undertaking cluster surveys.  The
first large cluster surveys were carried out via eye-ball searches for
galaxy over-densities on optical photographic plates
\citep{abell58a,zwicky68a}, but, nowadays, cluster finding uses
sophisticated automated techniques.

In this paper we describe automated cluster finding at X-ray
wavelengths; the hot ionised gas (or intracluster medium/ICM) emits
soft X-ray radiation in proportion to the square of the electron
density.  However, this is not the only way new clusters are being
discovered.  For example, the effect of cluster sized gravitational
potentials can be seen in the optical/infra-red, via strong or weak
gravitational lensing \citep[e.g.][]{wittman03a}. Increasing numbers
of clusters are also being discovered at millimetre wavelengths
\citep[e.g.][]{staniszewski09a, vanderlinde10a,
menanteau10a,marriage10a,williamson11a,planckcollab11a,Foley11} using the
Sunyaev-Zel'dovich (SZ) effect \citep{sunyaev72a}: the inverse Compton
scattering of photons from the cosmic microwave background (CMB) by
the hot ICM. At longer wavelengths still, one can discover clusters
out to high redshift using radio telescopes, via the unusual signature
of head-tail galaxies \citep{blanton03a}. Due to the advent of large
format CCD detectors, cluster finding using galaxy over-densities is
also currently undergoing a renaissance \citep[e.g.][]{gladders00a,
miller05a, koester07a, wilson09a}.
                                                                                                 
Cluster surveys have already revolutionised our understanding of the
physics of the ICM \citep[e.g.][]{ponman99a, arnaud10a} and delivered
cosmological constraints independent of, and competitive with, those
derived from observations of the CMB \citep[e.g.][]{larson10a,
dunkley10a} and Type 1a supernovae \citep[e.g.][]{kessler09a}. When
combined with these other cosmological probes, clusters are playing an
important role in the quest to understand the nature of dark energy
\citep[e.g.][see \citealt{sahlen09a} for a review of earlier cluster
cosmology studies dating back to \citealt{frenk90a} and
\citealt{oukbir92a}] {vikhlinin09b, mantz10a,
rozo10a,sehgal11a}. Clusters are also being used to test general
relativity on large scales \citep[e.g.][]{rapetti10a}, constrain the
properties of neutrinos \citep[e.g.][]{mantz10b}, and search for
evidence of non-Gaussian primordial density fluctuations
\citep[e.g.][]{hoyle10a}. Future cluster surveys will be wider, more
sensitive and better calibrated than ever before, and so are sure to
deliver significantly improved constraints compared to these existing
works \citep[e.g.][]{predehl06a, majumdar04a, cunha09a, wu10a}.

In this paper we present the \emph{XMM} Cluster Survey (XCS), a search
for serendipitous galaxy clusters in archival \emph{XMM-Newton}
observations.  The original XCS concept and motivation is described in
\citet{romer01a}. The main goals of the survey are ({\it i}) to
measure cosmological parameters, ({\it ii}) to measure the evolution
of the X-ray luminosity--temperature scaling relation ($L_{\rm
X}-T_{\rm X}$ relation, hereafter), ({\it iii}) to study galaxy
properties in clusters to high redshift, and ({\it iv}) to provide the
community with a high quality, homogeneously selected X-ray cluster
sample. The XCS follows a rich tradition of X-ray cluster surveys
dating back almost 30 years using earlier satellites: \citet[][{\it
HEAO I} ]{piccinotti82a}, \citet[][{\it Einstein} ]{gioia90a}, and
several derived from the \emph{ROSAT} All Sky Survey
\citep[RASS;][]{ebeling98a, bohringer00a, ebeling00a, ebeling01a,
ebeling02a, cruddace02a, gioia03a, bohringer04a, henry06a}, and the
\emph{ROSAT} pointed observations archive \citep{rosati98a, romer00a,
perlman02a, mullis03a, burke03a, burenin07a, horner08a}.

The XCS is not the only project currently exploiting the
\emph{XMM--Newton} (\emph{XMM} hereafter) archive for new detections
of clusters. Other projects include: XDCP \citep{mullis05a,
fassbender08a, santos09a, schwope10a, fassbender10a, Suhada11}; XMM-LSS
\citep{pierre06a, bremer06a, pacaud07a, Adami11}; SEXCLAS
\citep{kolokotronis06a}; COSMOS \citep{finoguenov07a}; XMM-BSC
\citep{suhada10a}; SXDS \citep{finoguenov10a}; and one being carried
out by members of the \emph{XMM} Survey Science Center
\citep{schwope04a, lamer08a}. This intense international interest
stems from the fact that \emph{XMM} has several features advantageous 
to cluster searching: in essence it combines sensitivity, and a large field of
view, with spectral imaging capabilities. 

The \emph{XMM} image quality does not match that of \emph{Chandra},
but it is still good enough to allow one to differentiate between
point-like and extended sources over the whole field of view: given
that clusters dominate the extended X-ray source population, this then
allows us to identify cluster candidates efficiently, despite the fact
that clusters only comprise $\simeq$10\% of the total X-ray source
population. Moreover, the spectral capabilities of \emph{XMM} allow
the measurement of the temperature of the hot ICM directly from the
discovery data. These $T_{\rm X}$ measurements allow us to then
estimate cluster masses, something of vital importance to cosmological
studies.  Finally, the mission has been in operation for over 10
years, and has built up a large archive of observations distributed
across the sky. By now there are several hundred square degrees
available that are suitable for a serendipitous cluster survey,
already exceeding that of the largest deep \emph{ROSAT} survey
\citep{burenin07a}. Serendipitous cluster
surveys have also been conducted using the \emph{Chandra} archive
\citep[e.g.][]{barkhouse06a}, although the available area for cluster
searching is significantly smaller in comparison to the \emph{XMM}
archive.

As predicted in \citet{romer01a}, and now demonstrated below, XCS will
deliver the largest number of cluster temperature measurements to
date. Importantly, these clusters will form a homogeneous sample (both
in terms of selection and analysis) and have a well-understood
selection function. In a companion paper (\citealt{mehrtens10a}, M11 hereafter), 
we present our first data release (XCS-DR1) and this includes \Nxvb\, $T_{\rm X}$ 
measurements. By comparison, the largest previous compilations of $T_{\rm X}$ values from
homogeneous samples contain less than 100 clusters each,
e.g. \citet[][63 clusters]{reiprich02a}, 
\citet[][25 clusters]{henry04a}, \citet[][29 clusters]{pacaud07a}, 
\citet[][31 clusters]{pratt09a}, \citet[][85
clusters]{vikhlinin09a}, and \citet[][96 clusters]{mantz10a}.  Larger compilations of clusters with
heterogeneous selection do exist, and some have significantly better
per cluster $T_{\rm X}$ precision than XCS, but even so the largest
published collection is still only 115 strong (\citealt{maughan07a}; a
larger sample, of 273 low-redshift clusters, was put together by
\citealt{horner01a}, but was not made public).

XCS highlights to date include the detection and subsequent
multi-wavelength follow-up of a $z=1.46$ cluster \citep[XMMXCS
J2215.9-1738;][]{stanford06a, hilton07a, hilton09a, hilton10a}, which
for several years held the record for the highest redshift
spectroscopically confirmed cluster \citep[recent discoveries of higher redshift X-ray clusters include][]{tanaka10a, papovich10a, henry10, gobat11}. XCS clusters have
also been used in compilation studies of galaxy evolution in high
redshift clusters \citep{collins09a, stott10a}. Conservative forecasts
of the performance of XCS for cosmological parameter estimation and
cluster scaling relations can be found in \citet{sahlen09a}: we expect
to measure (at 1-$\sigma$ and from clusters alone, i.e. not in
combination with CMB or supernovae observations) $\Omega_{\rm m}$ to
$\pm0.03$ (and $\Omega_{\Lambda}$ to the same accuracy assuming
flatness), and $\sigma_8$ to $\pm 0.05$, whilst also constraining the
normalisation and slope of the $L_{\rm X}-T_{\rm X}$ relation to $\pm
6$ and $\pm 13$ per cent, respectively.

\begin{figure*}
\includegraphics[width=9.5cm]{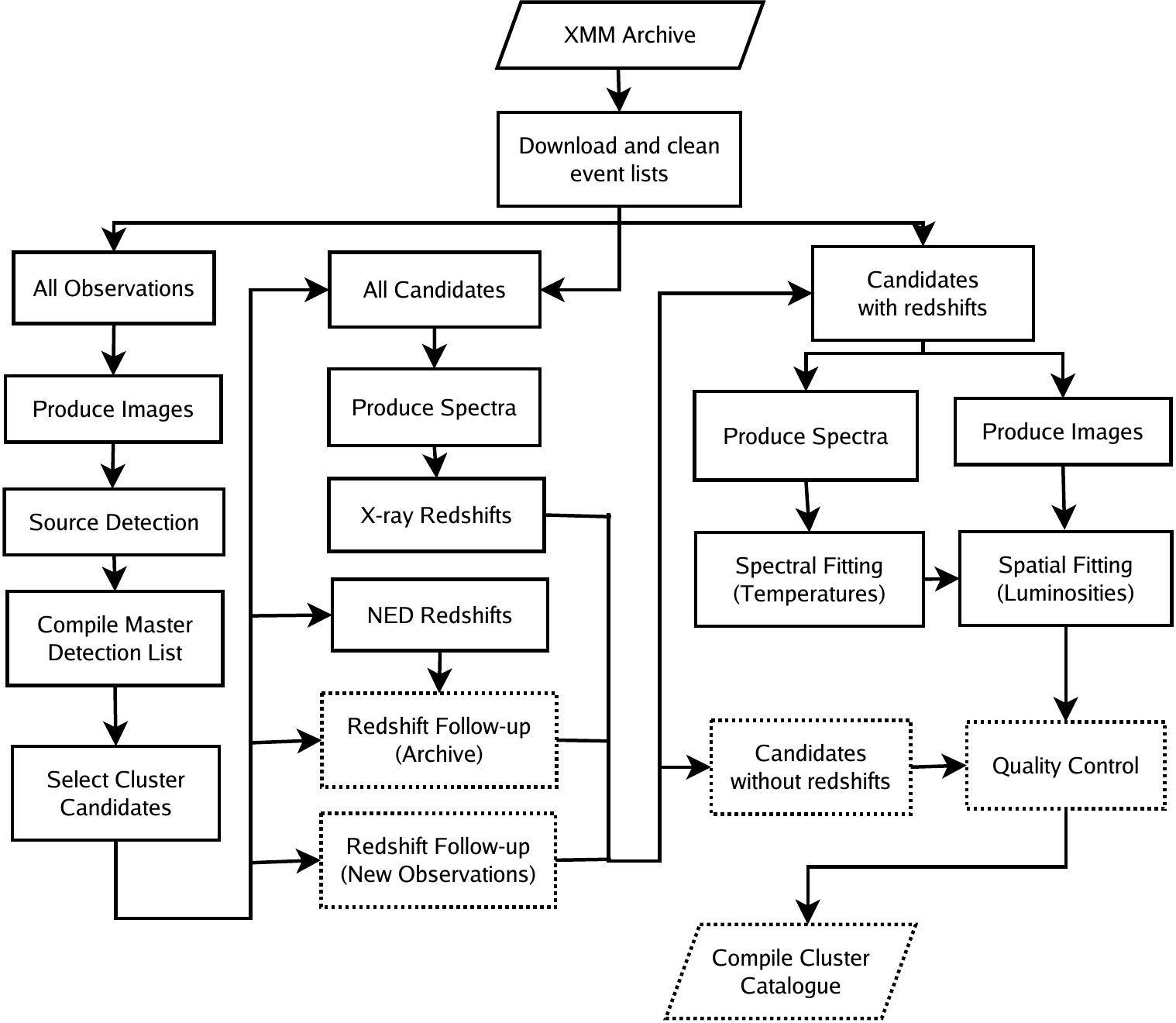}
\caption{Flowchart showing an overview of the XCS analysis methodology. This 
illustrates the sequence by which data from the \emph{XMM} archive is used
to create a catalogue of galaxy clusters.
\label{fig:overview_flow}}
\end{figure*}

In this paper, we present an overview of the XCS data analysis
strategy, from acquiring the data to producing a catalogue. A
schematic of our approach is shown in Fig.~\ref{fig:overview_flow},
although note that components indicated with dashed outlines are
discussed in M11.  The paper is broken up into 3 main sections.  
Section~\ref{sec:xmmdata} describes data acquisition, reduction 
and image generation. Section~\ref{sec:xapa} describes source 
detection, the compilation of candidate lists, and simulations 
of the survey selection function.  Section~\ref{sec:postproc} 
describes how we use \emph{XMM} data to measure X-ray redshifts,
temperatures and luminosities for the candidates.

\section{XMM Data Reduction}
\label{sec:xmmdata}

The \emph{XMM} archive contains thousands of public observations
suitable for conducting the XMM Cluster Survey. Such a large volume of
data means we have to carry out most of the XCS in a fully automated
manner -- the only parts that are not automated are the mask making
(Section~\ref{sec:masks}), optical follow-up, and quality control
(M11). While this automation presents a number of challenges, in terms
of handling the variety and complexity of the archival data, it also
has a number of benefits: not only has the entire data set been
treated in a consistent and systematic way, but we are also able to
run realistic simulations of our selection function.

In this section we describe how the raw \emph{XMM} archive is
manipulated into science-grade image files. First the data are
downloaded from the remote storage facility at the European Space
Astronomy Centre (ESAC) near Madrid to the University of Sussex
(Section~\ref{sec:downloads}). Then the data are calibrated and cleaned of
periods of high background contamination
(Section~\ref{sec:reduction}). Next, images are produced
(Section~\ref{sec:images}) and flux conversion factors calculated
(Section~\ref{sec:ecf_desc}). We begin this section with an overview of
some of the salient features of the \emph{XMM} mission.

\subsection{The \emph{XMM--Newton} Mission}

The \emph{XMM} mission \citep{jansen97a} consists of three co-aligned
Wolter Type I \citep{wolter52a,wolter52b} X-ray telescopes mounted on
the same spacecraft. The mission was undertaken by the European Space
Agency (ESA) and the spacecraft was launched on 10th December,
1999. The mission configuration, with three separate telescopes
simultaneously illuminating three cameras, means that most exposures
generate data with potential for serendipitous cluster finding: by
comparison \emph{Chandra} \citep{weisskopf99a}, has a single
telescope that illuminates only one of several instruments at any
given time, and not all those instruments are suitable for cluster
finding.

The European Photon Imaging Camera \citep[EPIC:][]{villa96a} consists
of three separate cameras, each in the focal plane of a separate X-ray
telescope. Each camera consists of an array of charge-coupled devices
\citep[CCDs:][]{boyle70a} in different configurations. Two cameras,
the EPIC-mos1 and 2, consist of arrays of 7 metal oxide semi-conductor
CCDs illuminated by 44\% of the light from their respective telescopes
(the rest is redirected to the Reflection Grating Spectrometers).  The
EPIC-pn camera consists of 12 back-illuminated CCDs. These CCDs are
not only more sensitive than those in the EPIC-mos cameras, but the
EPIC-pn receives all the light from its respective telescope. Thus,
the EPIC-pn camera has more than twice the sensitivity of the EPIC-mos
cameras.

One disappointing aspect of both \emph{XMM} and \emph{Chandra} has
been the unexpectedly high background in their CCD cameras.  Both
these missions are in similar, highly-elliptical orbits, and it was
only after their launch that it was realised that these orbits
intersect a population of low-energy protons trapped in the Earth's
magnetosphere.  The lower energy protons can be funnelled by the
grazing incidence mirrors onto the detectors and this has resulted in
a significantly higher background than was expected before
launch. Consequently, certain aspects of XCS have proved to be more
challenging than was anticipated in our pre-launch predictions
\citep{romer01a}. In addition to the enhanced background, there have
been a number of incidents of damage to the EPIC cameras while in
orbit, but in only one case has this resulted in a significant loss of
detector area \citep{abbey06a}.
 
\subsubsection{\emph{XMM--Newton} Point-Spread Function}\label{psf_desc}

A crucial issue for the detection of extended sources by XCS is the
treatment of the \emph{XMM} Point-Spread Function (PSF). The PSF is a
strong function of off-axis angle and photon energy (where off-axis
angle is the angle between the source location and the centre of
the field of view). As the off-axis angle increases, the PSF shape
morphs from being circularly symmetric to ellipsoidal and finally
bow-tie shaped.  There have been a number of attempts to characterise
the \emph{XMM} PSF including: simulations based on measurements of the
shape of the mirrors \citep{gondoin96a}; measurements taken on the
ground by passing X-ray beams from synchrotron sources through
\emph{XMM} mirror modules \citep{stockman97a,gondoin98a}; and fitting
1-dimensional profiles to observations of bright X-ray sources
\citep{gondoin00a,ghizzardi01a,ghizzardi02a,read04a}.  Unfortunately,
thus far, this has not resulted in a complete and reliable
characterisation of the \emph{XMM} PSF. Currently four PSF models are
available: the Low, Medium, High and Extended Accuracy Models
\citep{altieri04a}. Of these, only the Medium Accuracy Model (MAM) is
2-dimensional, but as it is based on simulations that relied on
pre-launch measurements of the mirrors, it suffers from a number of
deficiencies. The Extended Accuracy Model (EAM) is a 1-dimensional
model based on in-orbit measurements of real sources, and is
considered the most accurate but obviously does not encapsulate the
complex 2-dimensional structure observed in the PSF at large off-axis
angles. Currently in XCS, we use the EAM when measuring source extents
for both real sources and simulated ones used to create the survey
selection function (Section~\ref{sec:extent}), and when carrying out
spatial fits to cluster surface brightness profiles
(Section~\ref{sec:spatial_fits}), and we use the MAM when creating simulated
sources for the selection function (Section~\ref{sec:sf_descr}). In the 
future we hope to include the new 2-d model under development
by the \emph{XMM} Science Operations Centre \citep{read10a}. This
improved model will more acturately encode the off-axis, azimuthal and
energy dependencies of the PSF.
  
\subsection{Data Acquisition}
\label{sec:downloads}

\begin{figure}
\includegraphics[width=8.5cm]{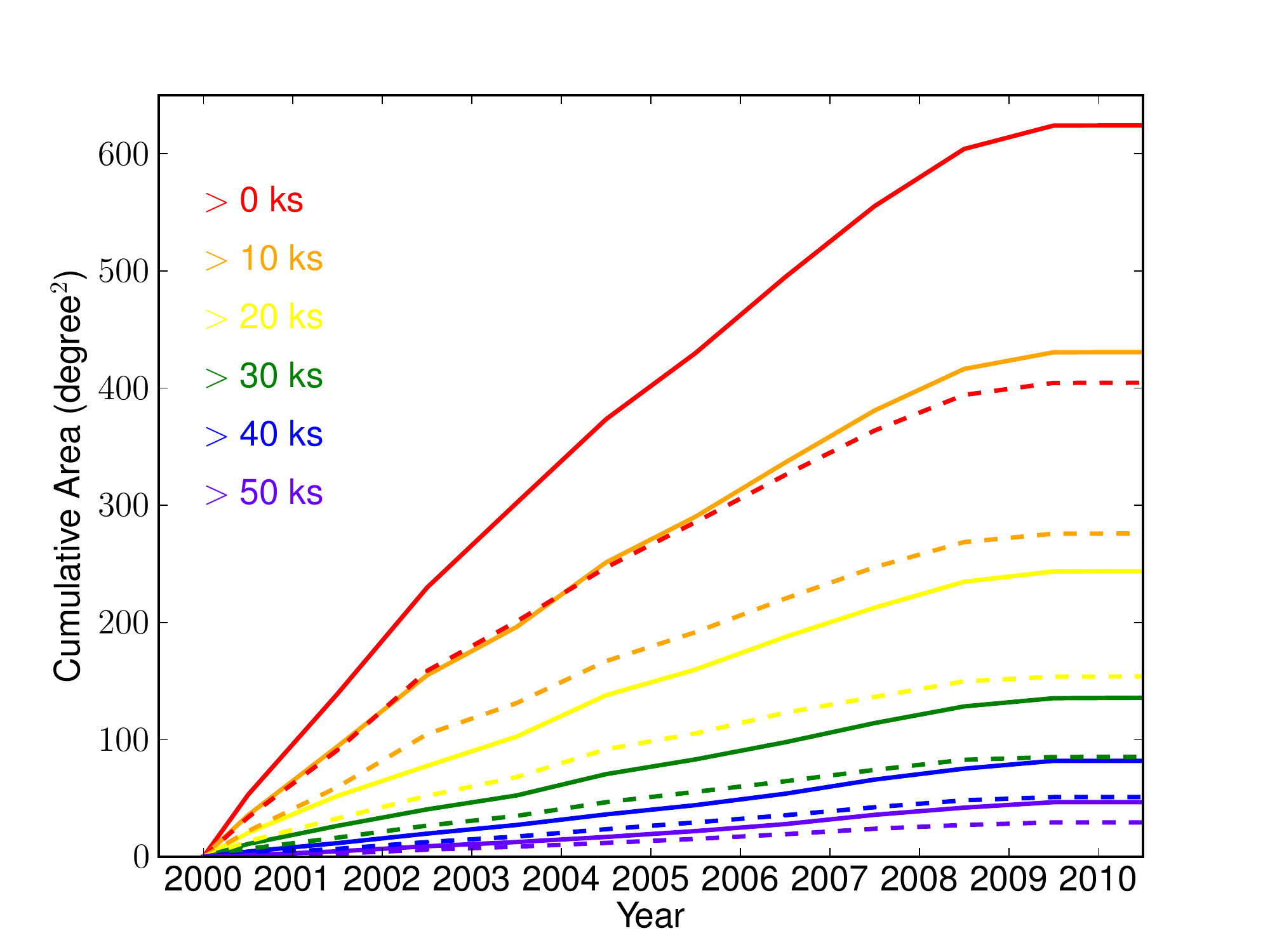}
\caption{Cumulative sky area covered by public data in the \emph{XMM}
archive as a function of exposure time, for the whole sky
(solid) and excluding the Galactic plane and Magellanic Clouds
(dashed) and for a variety of different exposure time cuts, at the
time of the most recent XCS download in July 2010. The flattening of
the curves mid-way through 2009 reflects the fact that proprietary
observations only become public a year after they are completed, so
that most data taken after that time was still proprietary at the time
of the download. \label{fig:area-time}}
\end{figure}

\begin{figure*}
\includegraphics[width=14.0cm]{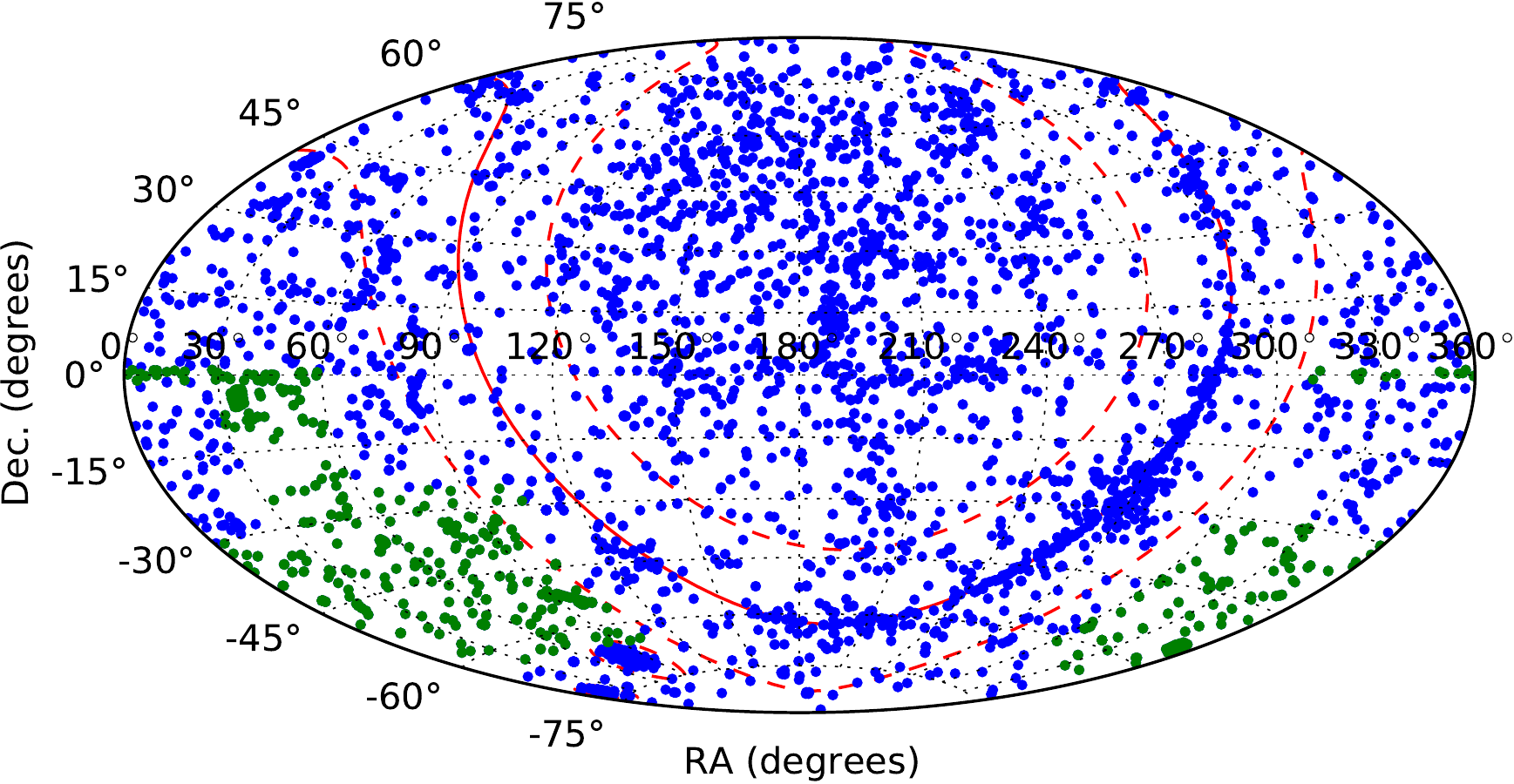}
\caption{The distribution on the sky of the \Nxi\, ObsIDs in the
 \emph{XMM} archive as of 21st July 2010. Locations in green [blue]
 are inside [outside] the proposed footprint of the Dark Energy Survey
 (darkenergysurvey.org). The Galactic plane and locations of the
 Magellanic Clouds are highlighted by the red dashed line (we do not
 carry out cluster searches within those regions).
\label{fig:skydistribution}}
\end{figure*}

In Fig.~\ref{fig:area-time} we illustrate how the non-overlapping area
in the public \emph{XMM} archive has grown over the past ten years,
both in terms of total area and in terms of area suitable for the
discovery of clusters, i.e. outside the Galaxy ($|b|>20^{\circ}$) and
Magellanic Clouds ($>6^{\circ}$ [3$^{\circ}$] of the Large [Small]
Magellanic Clouds). We note that these calculations take into account
other, smaller, regions deemed by XCS to be unsuitable for
serendipitous source detection (see Section~\ref{sec:masks}).  By now there
are over 600 deg$^2$ of the sky covered by \emph{XMM}, but of that,
only $\simeq$50deg$^2$, 280 deg$^2$ and 410 deg$^2$, at $>40$ ks,
$>10$ ks and $>0$ ks depths respectively (exposure times are those
after flare cleaning, Section~\ref{sec:flares}), are in regions suitable
for cluster searching. This area is distributed across the sky
(Fig.~\ref{fig:skydistribution}) rather than as a contiguous region.

As shown in Fig.~\ref{fig:area-time}, new data enters the archive
almost every day, but due to practical constraints we have only
processed the data in a small number of large batches, corresponding
to all the public EPIC data available at that particular time.  The
downloads take advantage of the Archive InterOperability System
\citep[AIO:][]{arviset04a}; this protocol allows the \emph{XMM}
Science Archive \citep[XSA:][]{clavel98a,arviset02a} to be searched in
an automated fashion.  At the time of writing, the most recent
download was completed on 21st July 2010, corresponding to \Nxi\,
separate \emph{XMM} observations. Their locations are shown in
Fig.~\ref{fig:skydistribution}. Each of these observations (including
those broken down into multiple exposures) has a unique identification number, 
or ObsID.  In the following, we use the term ObsID to refer to the set of 
Observation Data Files (ODF) that contains all the observation-specific data. We
note that, even with appropriate compression etc., the XCS archive, of
raw and processed data products, amounts to on the order of 4
terabytes.

\subsection{Data Reduction}
\label{sec:reduction}

The data reduction was carried out in a fairly standard manner
\citep[see for instance section 3 of][]{read03a}. Only events with
patterns (characterisations of how many CCD pixels are involved in an
event) 0-4 were used for the EPIC-pn and 0-12 for the EPIC-mos. A
schematic of the data reduction procedure is shown in
Fig.~\ref{reduction_flow}.

\subsubsection{Calibration}

The reduction and analysis of \emph{XMM} data requires calibration
information detailing how the telescopes and instruments behave,
e.g. the effective area of the \emph{XMM} telescopes and the detection
efficiency of the instruments (both being functions of photon energy
and detector position), plus the instrumental uncertainty associated
with measuring photon energies. The most up-to-date version of the
\emph{XMM} Current Calibration Files (CCF), as of 21st July 2010, were
used for the analysis presented herein.

\subsubsection{Software Versions}

Several different software packages are deployed for XCS analysis:
version 10.0.0 of the Science Analysis Software
\citep[SAS:][]{gabriel04a}; version 6.9 of HEASOFT
\citep{blackburn95a}; version 4.2 of CIAO
\citep{doe01a,deponteevans08a}; and version 12.6.0i of XSPEC
\citep{arnaud96a}. In order for these packages to be used in the
automated batch manner needed for XCS, several different wrapper
programmes were written in scripting languages. For the work described
in Sections~\ref{sec:xmmdata} and \ref{sec:postproc}, version 2.6.4 of
Python (docs.python.org) was used to write these wrapper programmes,
whereas version 7.1 of IDL (www.ittvis.com) was used for the work
presented in Section~\ref{sec:xapa}.

\subsubsection{Flare Cleaning}\label{sec:flares}

One important aspect of our pipeline reduction was the treatment of
background flares. It is well documented
\citep{lumb02a,read03a,pradas05a} that \emph{XMM} observations often
suffer from periods of enhanced particle background, caused mostly by
variations in solar activity in conjunction with the position of the
spacecraft in its orbit.  To increase the signal-to-noise of the data,
we have designed an automated procedure to remove periods of high
background. This was achieved by creating a lightcurve, divided into
50-second bins. The bin size was chosen to balance a reasonable time
resolution with minimising shot noise. The lightcurve was first
generated, and cleaned, using the high-energy events (12-15 keV for
the EPIC-pn and 10-12 keV for the EPIC-mos cameras), because these
events are more likely to be from the particle background than from
astronomical sources. The cleaning process is then repeated, using a
soft-energy lightcurve (0.2-1.0 keV), to account for periods of
elevated background coming from soft protons.

The cleaning process for each energy band involved an iterative
3-$\sigma$ clipping procedure that selected which 50\,s bins to
exclude. The mean and standard deviation of the lightcurve were
calculated and bins more than $\pm$3-$\sigma$ from the mean were
removed. The 3-$\sigma$ limits were then re-calculated and the process
repeated up to 50 times or until a stable state is reached, whereby
the bins that are being excluded are not changing (note that
previously excluded bins can be re-instated in subsequent iterations
if the 3-$\sigma$ limits become larger). The maximum of 50 iterations
was set to avoid cases where the stable solution oscillates between
two or more similar states.

\begin{figure*}
\includegraphics[width=8cm]{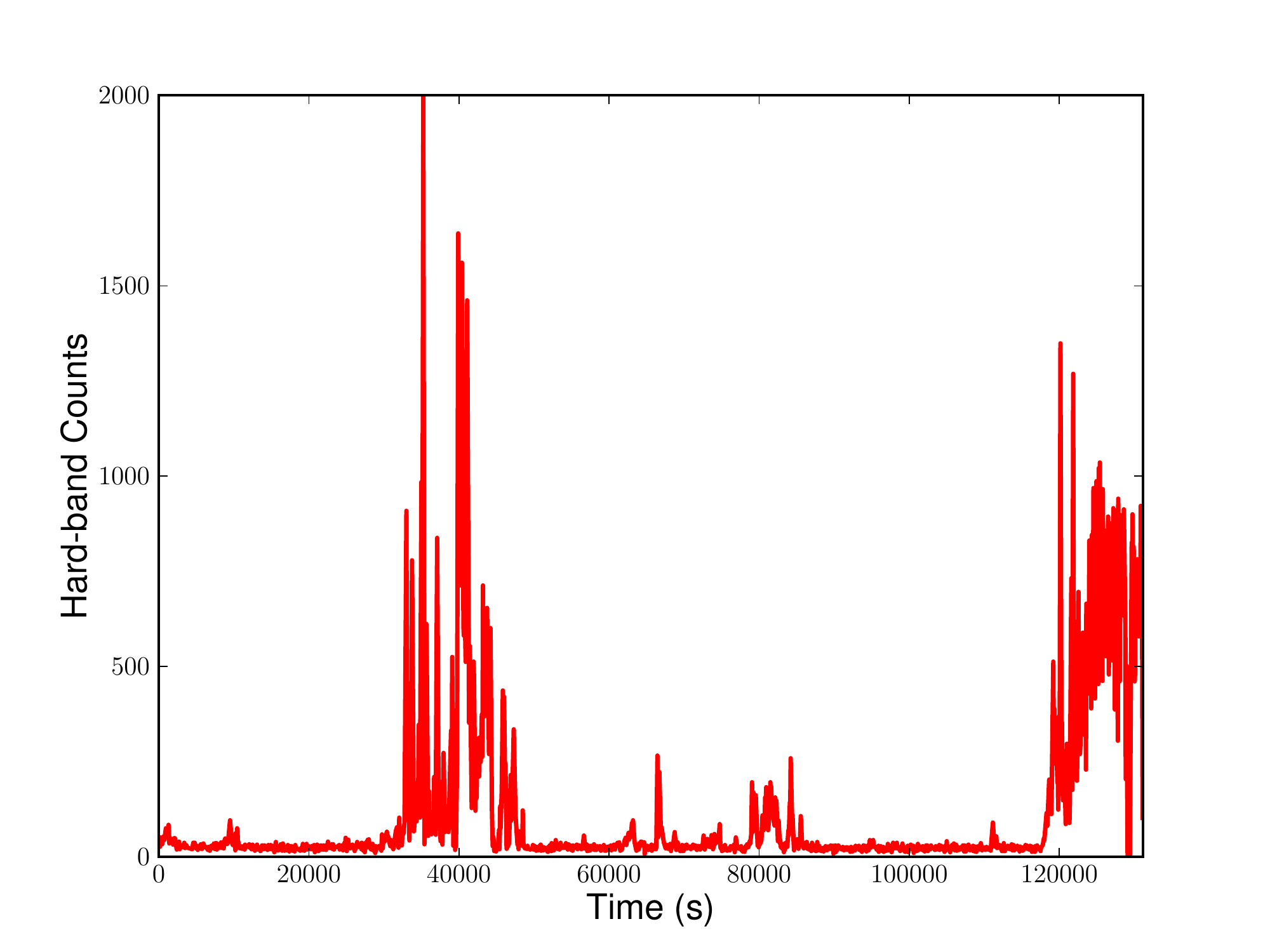}
\includegraphics[width=8cm]{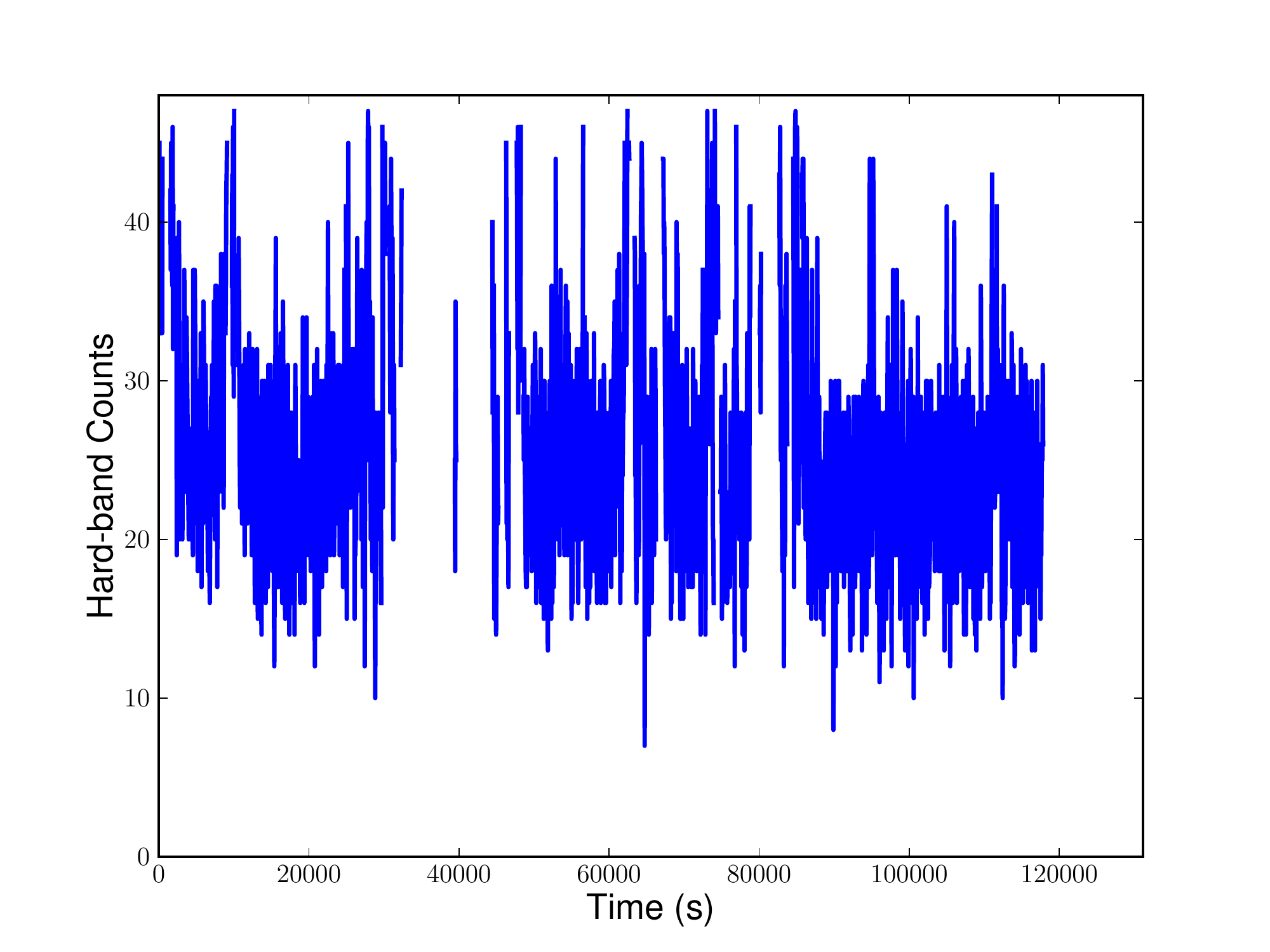}
\caption{EPIC-pn example hard-band lightcurve with 50\,s
bins. \emph{Left panel:} Raw events before cleaning. \emph{Right panel:}
Cleaned events with periods of high background removed.
\label{cleaning_example}}
\end{figure*}

\begin{figure*}
\includegraphics[width=8.5cm]{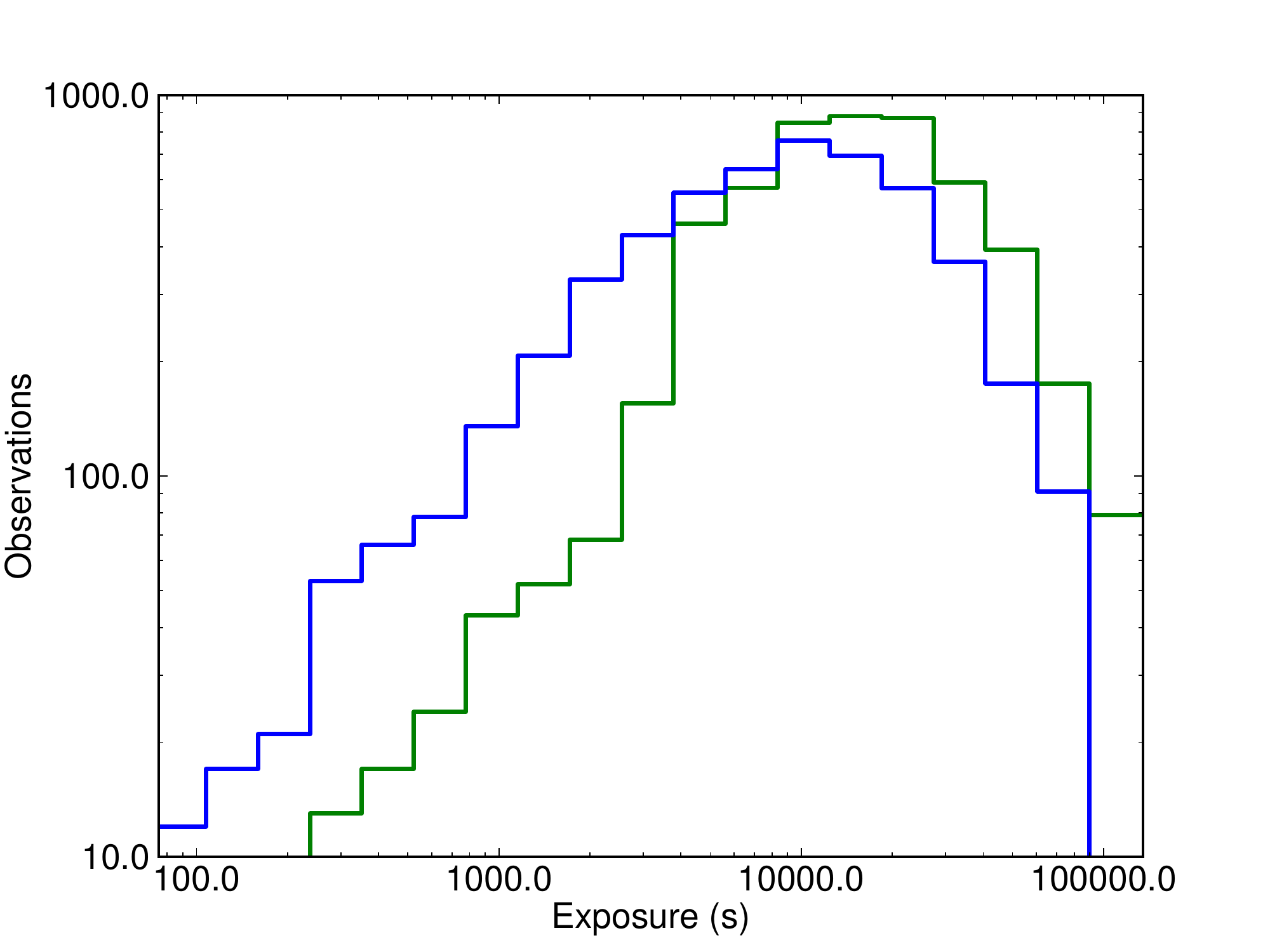}
\includegraphics[width=8.5cm]{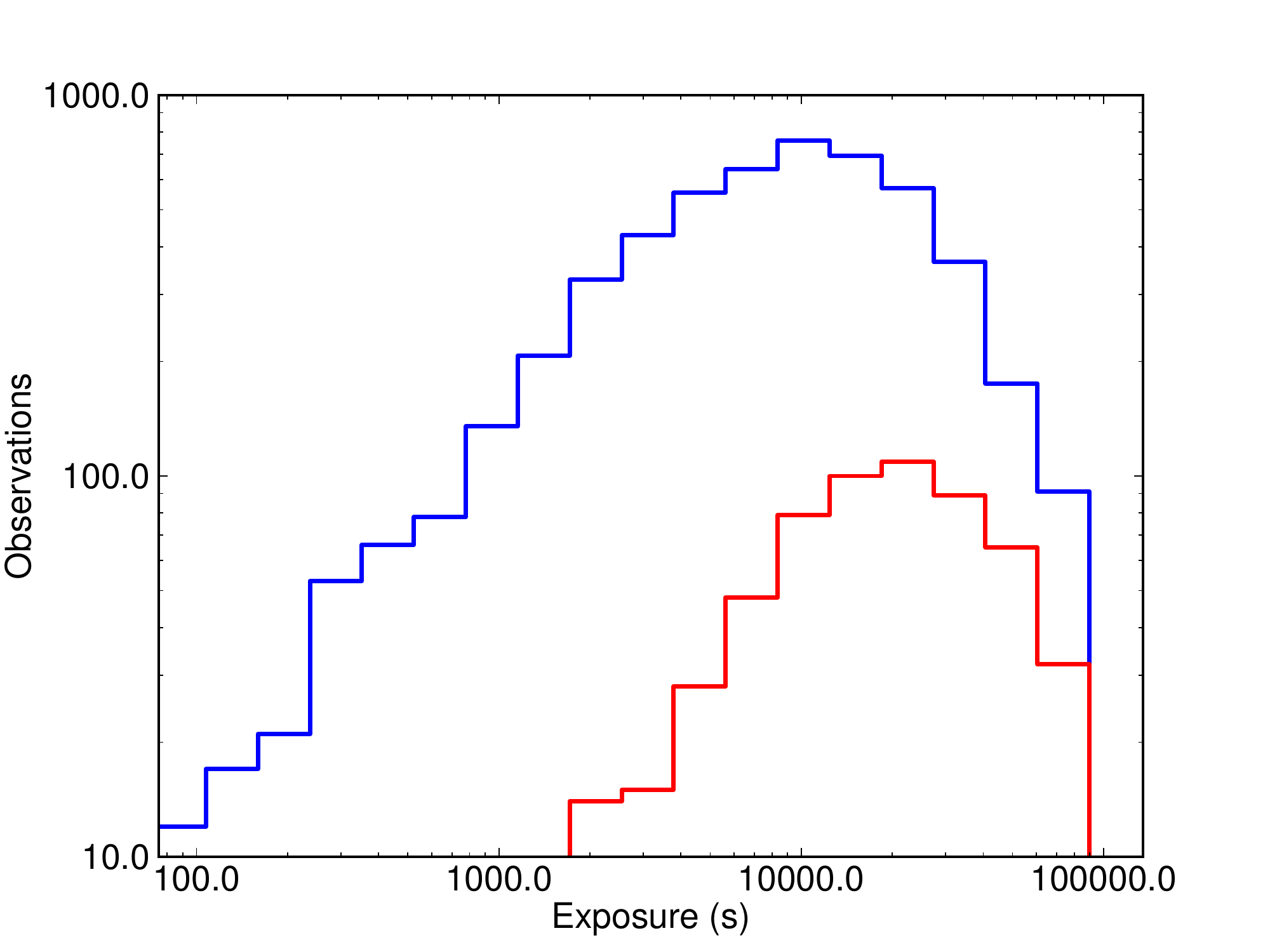}
\caption{The distribution of ObsID exposure times. {\it Left panel}:
  The number of ObsIDs before (green) and after (blue) the 
  process of flare cleaning. {\it
  Right panel}: The number of ObsIDs in which extended XCS sources with 300 or
  more counts were detected (red), compared with all ObsIDs (after
  flare cleaning).
\label{cleaning_hist}}
\end{figure*}

We note that before the first 3-$\sigma$ clipping took place, an
initial maximum-rate threshold is used to `clip' the
lightcurve. This threshold is the greater of either 50 counts per bin
for the EPIC-pn (and half this for the EPIC-mos cameras) or 125
per cent of the highest value in the lowest 5 per cent of the bins. This
initial filtering was found to improve the flare cleaning results when
flares accounted for a large fraction of the total exposure time. A
flowchart illustrating the flare cleaning steps is shown in
Fig.~\ref{cleaning_flow}. Fig.~\ref{cleaning_example} shows an
example hard-band lightcurve before and after cleaning. 

The combination of the excluded bins for the hard and soft-background
lightcurves is then used to define the good time intervals (GTI) used
to filter the raw event files. Fig.~\ref{cleaning_hist} shows the
distribution of ObsID exposure times before and after the
process of flare cleaning. The filtered event files are used several
times during XCS analysis. They are used to produce the images
(Section~\ref{sec:images}), used for the initial XCS source detection
(Section~\ref{sec:wavedetect}), and then again to determine spectroscopic
(Section~\ref{sec:Txmethod}, \ref{sec:zxmethod}) and spatial parameters
(Section~\ref{sec:spatialmethod}) for the cluster candidates.

\subsection{Image Production}
\label{sec:images}

Starting with the cleaned event lists described above
(Section~\ref{sec:flares}), the individual camera exposures were spatially
binned, with a pixel size of 4.35 arcsec, to generate images. This
pixel size was chosen because it is smaller than the PSF, at all
detector locations and photon energies.  Images were produced in two
bands, soft (0.5-2.0 keV) and hard (2-10 keV). Exposure maps were also
created for each image. The exposure maps encode the impact of
vignetting on the image sensitivity and also record the locations of
chip gaps, bad rows, etc.

The EPIC cameras do not have shutters, so events received while 
an observation is reading out, the so called out-of-time events, will be
assigned incorrect positions and energies. For XCS, only
EPIC-pn images were corrected for out of time events, because the 
EPIC-mos cameras have a much lower readout rate and negligible 
out-of-time events. The EPIC-pn corrections were done in
the standard way, i.e. the event file was recreated assuming
all the events are out-of-time and assigning them new positions along
the CCD column at random. These are then used to create
out-of-time images that can be subtracted off the true images (with
the appropriate correction for the fraction of out-of-time events).

\begin{figure*}
\includegraphics[width=8cm]{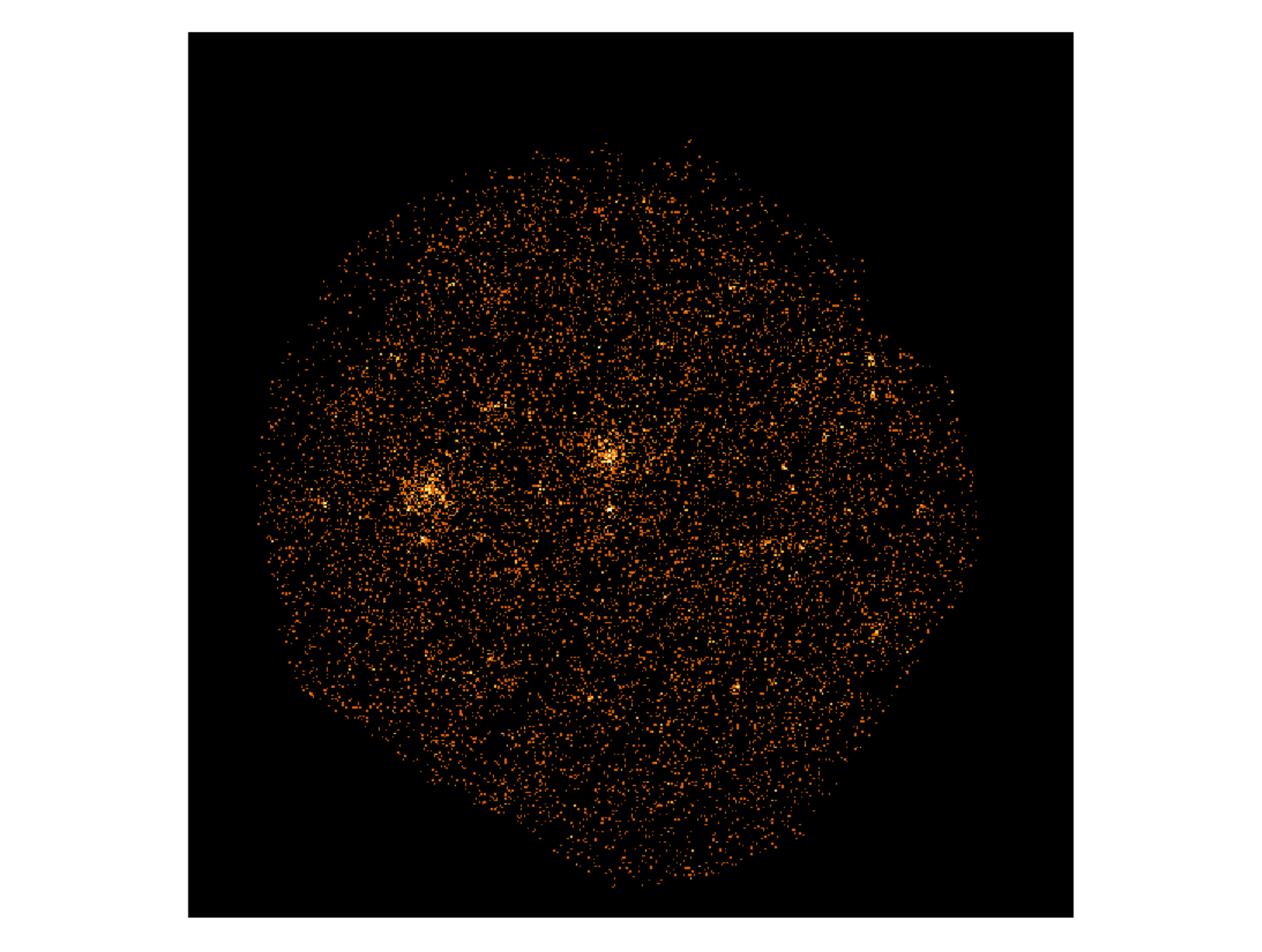}
\includegraphics[width=8cm]{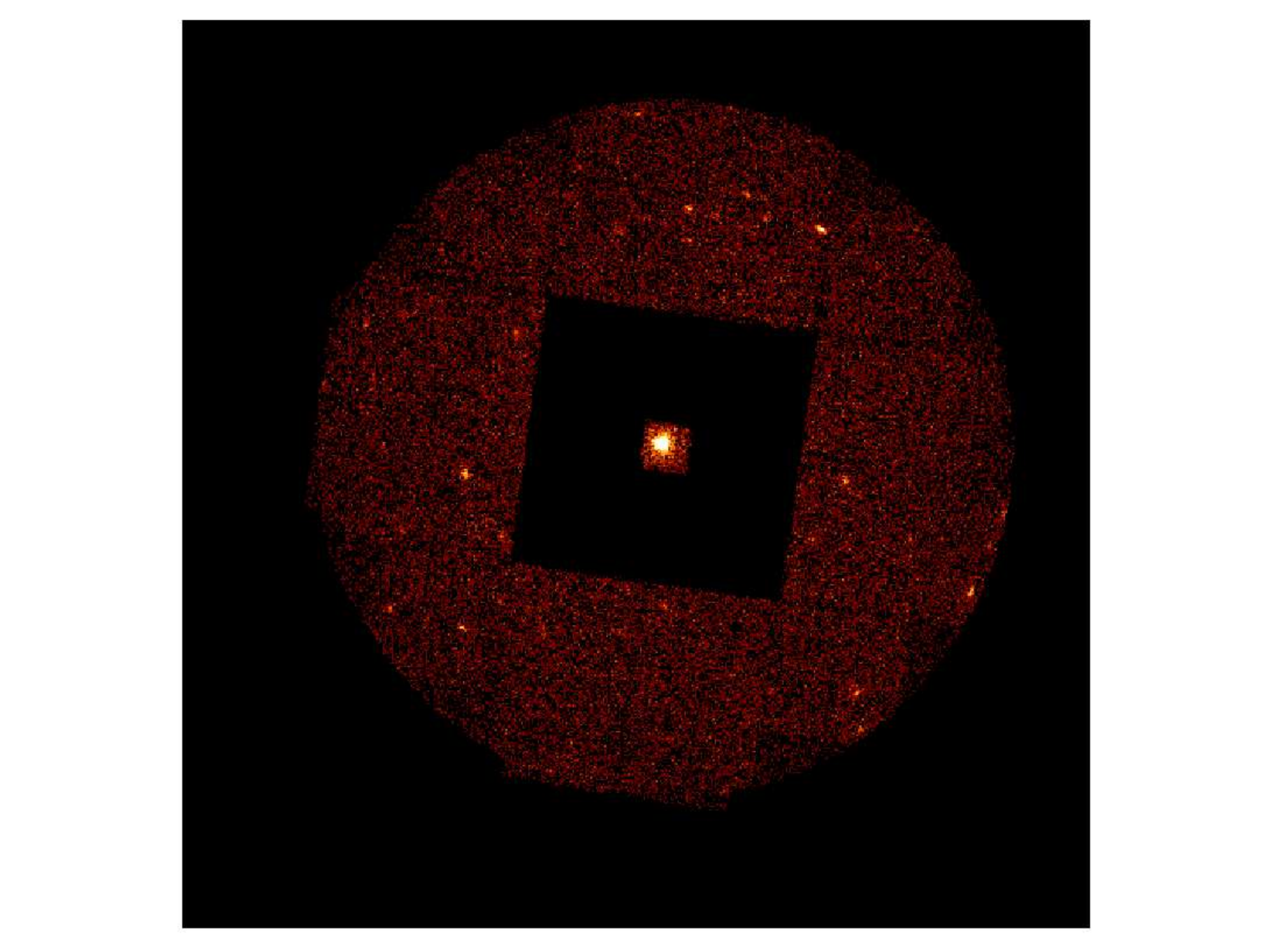}\\
\includegraphics[width=8cm]{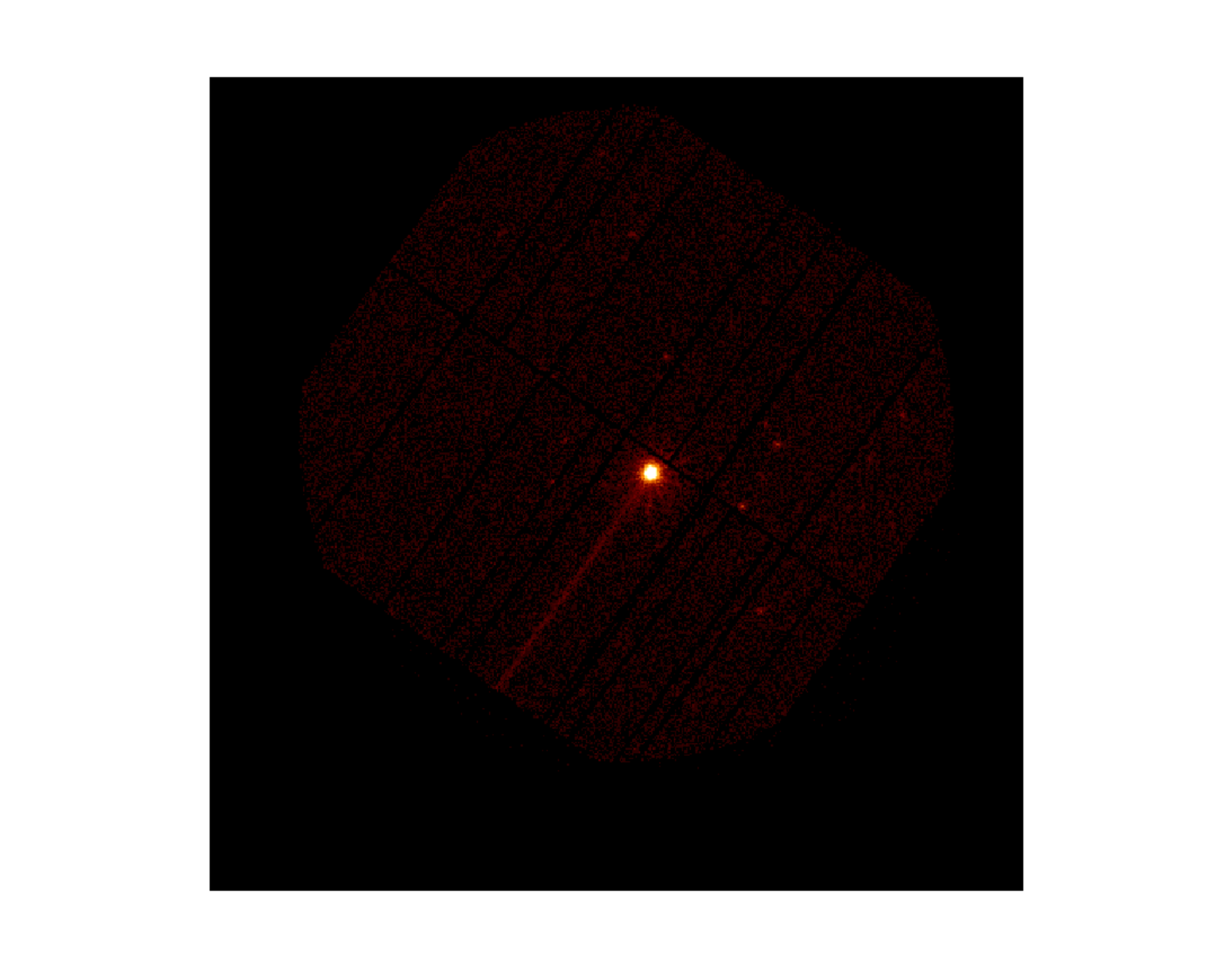}
\includegraphics[width=8cm]{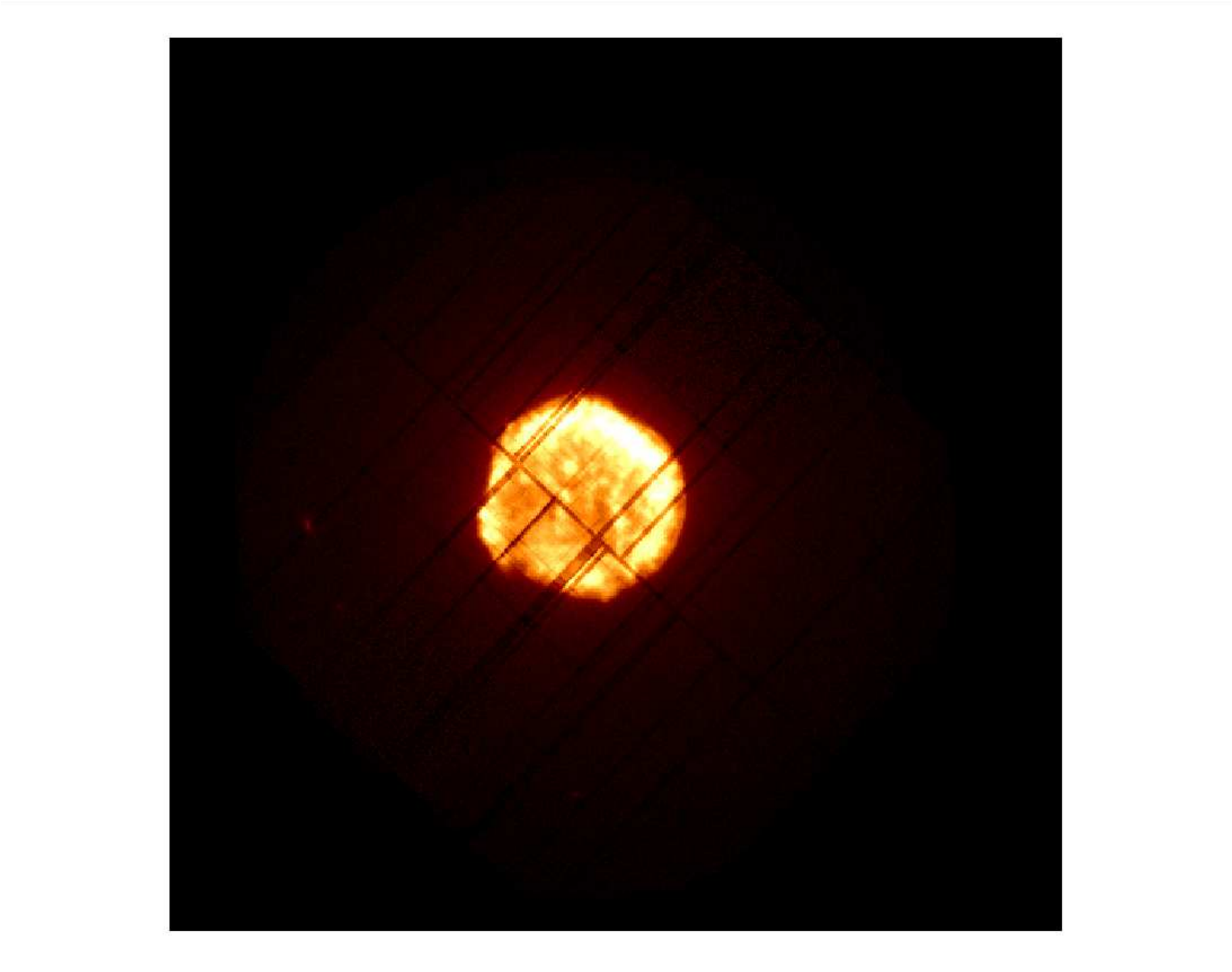}
\caption{Examples of reduced and merged \emph{XMM} images with a variety of different target types.
\label{fig:OBSids}}
\end{figure*}

The images and exposure maps for the individual cameras were merged to
create a single image and exposure map per ObsID. For this, the pixel values in the
EPIC-mos maps were scaled to that of the EPIC-pn camera using the
previously calculated ECFs (Section~\ref{sec:ecf_desc}). Examples of XCS
generated exposure maps and images can be seen in
Fig.~\ref{fig:indiv_exps} and \ref{fig:OBSids}. A total of \Nxii\,
image files have been generated from the \Nxi\, \emph{XMM} ObsIDs
that make up the current XCS dataset (a small number of ObsIDs
in the archive are not suitable for automated image generation for a
variety of technical reasons such as telemetry and calibration issues,
etc.).

\begin{figure*}
\includegraphics[width=8cm]{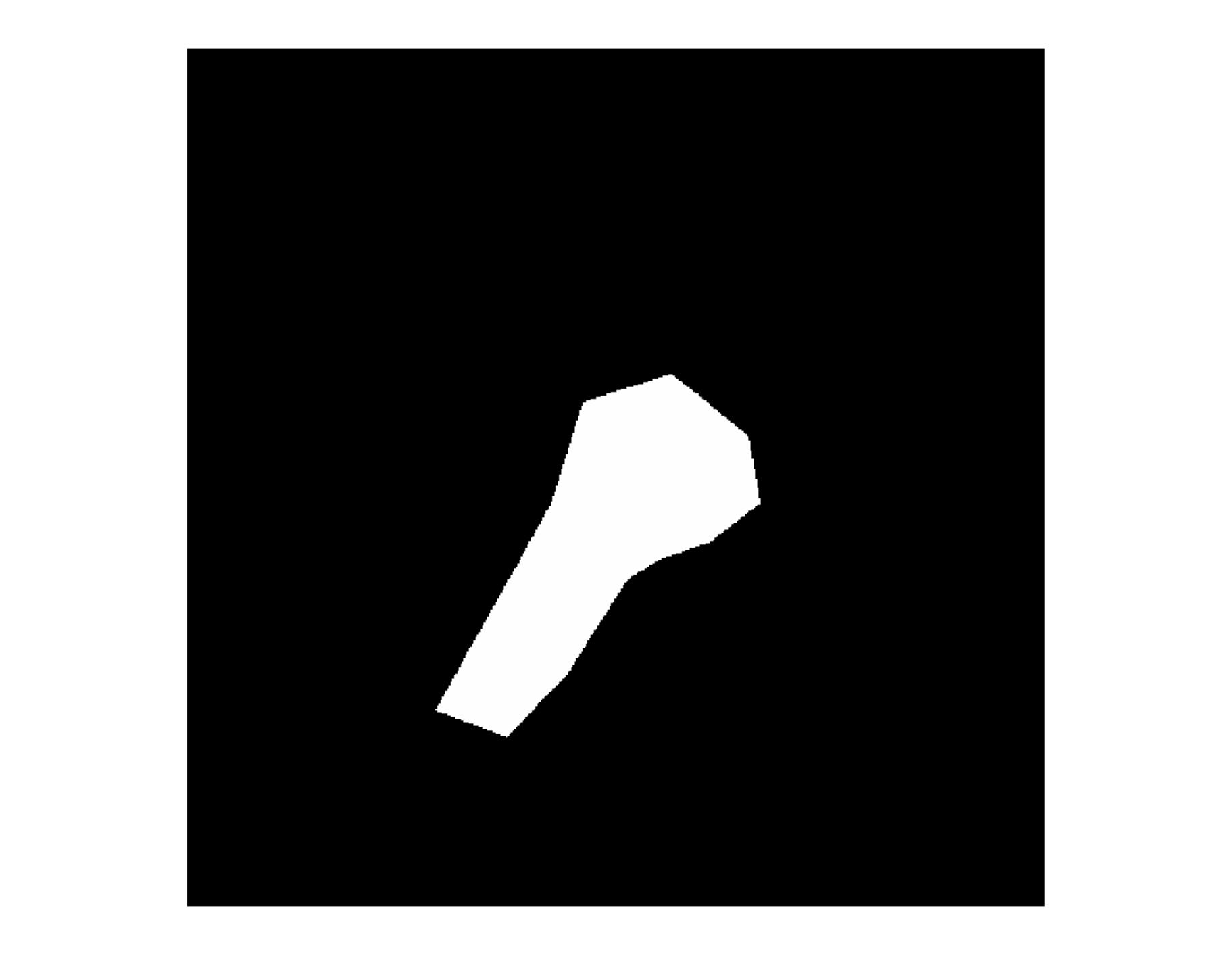}
\includegraphics[width=8cm]{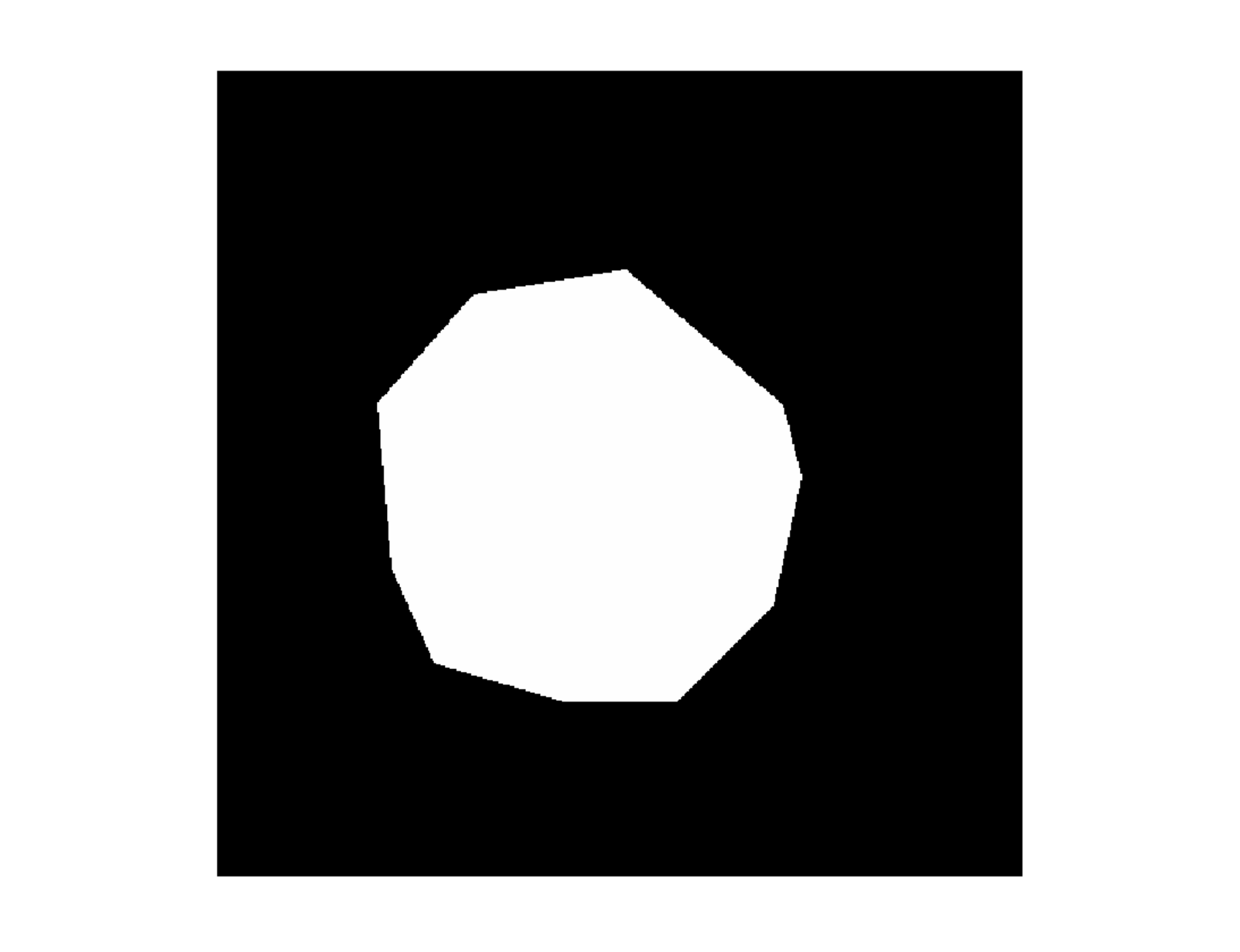}
\caption{Examples of masks created for the lower two images in Fig.~\ref{fig:OBSids}.
\label{fig:maskexamples}}
\end{figure*}

\subsubsection{Image Masking} \label{sec:masks}

The production of images is an automated process, however they do need
to be checked by eye before passing them to the source detection
pipeline (Section~\ref{sec:wavedetect}). This is because we download all
public data, regardless of the intended (by the PI) target. As a
result, the XCS image archive includes ObsIDs with very extended
targets (such as low-redshift clusters or Galactic supernova remnants)
and ObsIDs with very bright targets (such as luminous AGN). The very
extended targets will enhance the background level over the majority
of the \emph{XMM} field of view, and thus reduce our ability to make
serendipitous detections of sources. The very bright sources will
generate artefacts in the images, such as radial spikes and
out-of-time bleed trails; those artefacts could then be falsely
identified as additional sources. The eye-balling process identifies
ObsIDs that should be completely excluded from the other stages of the
XCS pipelines. It also allows us to mask out regions of ObsIDs that
are only partially afflicted by bright/extended targets. Approximately
one-third of ObsIDs require some degree of masking, with the
median area lost being around 4 per cent (though this can be as high as
80 per cent in extreme cases). The mask files are of the same
dimensions as the image files and are used during the source detection
and also when creating backgrounds for the spectral and spatial
fitting. We show some examples of XCS images that require full or
partial masking in Fig.~\ref{fig:maskexamples}.

\subsection{Energy Conversion Factors}
\label{sec:ecf_desc}

In order to be able to convert image source counts into energy fluxes,
energy conversion factors (ECFs) need to be calculated. These are
necessarily model dependent and are affected not only by the source
and instrument properties but also by the HI column, $n_{\rm H}$
hereafter, along the line of sight.  In our survey, the source
properties are not known in advance, so a generic model has to be
assumed. Since the vast majority of the sources detected by XCS are
point sources, and point sources are likely to have power-law spectra,
the model used to calculate the conversion is an absorbed power law
with a canonical AGN index of 1.7 \citep{mushotzky93a}. The
photoelectric absorption is set to the appropriate $n_{\rm H}$ value
for the field (Section~\ref{sec:nh_desc}). The ECFs were calculated, using
the XSPEC spectral fitting package and the on-axis spectral responses,
for each camera exposure related to a particular ObsID. For the specified model,
the ratio of the resulting flux and count-rate is stored as the ECF
for that exposure. ECFs are not exposure time dependent, but due to
variations in $n_{\rm H}$, the choice of optical blocking filter and
the effective area of the instrument, ECFs in XCS still vary from
exposure to exposure and from ObsID to ObsID. They generally range
from 4.4 to 6.6 for the EPIC-pn and 1.6 to 2.0 for the EPIC-mos
cameras (in units of 10$^{-11}$ ct cm$^{-2}$ erg$^{-1}$). Even though
the ECFs are calculated for the on-axis aim point, they can still be
used for sources detected anywhere in the field of view, by correcting
them using the exposure map.

We also calculate, for each ObsID, a further set of conversions using
the MEKAL model \citep{mewe86a}. The MEKAL model is the standard model
used to describe thermal and line emission from clusters of
galaxies. The MEKAL conversions are done over a grid of $n_{\rm H}$,
temperature and redshift, however the metal abundance is kept fixed at
$Z=0.3\times$ the Solar values in \citet{anders89a}. (This choice of
metallicity is standard in the field because previous work, such as by
\citet{maughan08a}, has shown that abundances vary little from this
value over a wide range of redshifts.) The gridded MEKAL conversions
can be used to convert count-rates to bolometric luminosities and vice
versa (and we refer to these conversions as LCFs hereafter).  The LCFs
are used to calculate synthetic cluster count-rates for the survey
selection function (Section~\ref{sec:sf_descr}) and to estimate
luminosities for XCS candidates during the literature redshift search
(Section~\ref{sec:litz}).  The LCFs, like the ECFs, are calculated for the
on-axis aim point, but can be adjusted to another location using the
exposure map.

\subsubsection{Galactic HI Column}\label{sec:nh_desc}

X-ray photons are absorbed by material along the line of sight, and in
particular by helium and oxygen for photons above $\sim 0.5$ keV
\citep{wilms00a}. One can predict the level of absorption if $n_{\rm H}$ is
known, so, for XCS, we estimated the $n_{\rm H}$ values for each source
using the compilation of \citet{dickey90a}, which combines the Bell
Labs HI Survey \citep{stark92a} data with other surveys for all sky
coverage. We use $n_{\rm H}$ to calculate ECFs and LCFs (see above),
but also at other points in the XCS pipeline, e.g. when analysing
X-ray spectra (Section~\ref{sec:Txmethod}).  We note that self-shielding of
molecular hydrogen, from ambient ultra-violet radiation, can occur
when $n_H>5\times10^{20}$ cm$^{-2}$ \citep{arabadjis99a}. This
molecular gas absorbs X-rays and thus distorts flux conversions that
are based only on $n_{\rm H}$ values. For this reason, XCS fluxes
derived when $n_H>5\times10^{20}$ cm$^{-2}$ should be regarded as
lower estimates.

\section{Generation of The XCS Source Catalogue}

\label{sec:xapa}

In this section, we provide details of our source detection algorithm,
known as the XCS Automated Pipeline Algorithm or {\sc Xapa}.  In
Section~\ref{sec:wavedetect}, we explain how {\sc Xapa} applies wavelets to
the pipeline generated images (Section~\ref{sec:images}) to generate a
source list per ObsID. In Section~\ref{sec:find_srcprop}, we describe
the parameters that are measured by {\sc Xapa} for each detected source.  In
Section~\ref{sec:points} and \ref{sec:extended} we demonstrate the
quality of the {\sc Xapa} data products for point and extended sources
respectively.  In Appendix \ref{sec:xapa_ap} we provide related flow charts.

\subsection{Source Detection }
\label{sec:wavedetect}

{\sc Xapa} source detection is based upon the mission-independent source
detection package {\sc WavDetect} \citep[][F02 hereafter]{freeman02a}, which
is available as part of the {\sc CIAO} software package.  F02 have shown
that {\sc WavDetect}'s wavelet-based algorithm is more sensitive than
standard sliding-cell algorithms (e.g.  {\sc CellDetect} from {\sc Ciao},
\citealt{fruscione06a}) and is considerably faster than
event-list-based algorithms such as {\sc CIAO}'s {\sc VTPdetect}. Before deciding
to use {\sc WavDetect} as the basis for the {\sc Xapa} algorithm, we also examined
the \emph{XMM} SAS {\sc Ewavelet} program and the {\sc SExtractor} package
\citep{bertin96a}, finding them both to be inadequate for our purposes
\citep[see][for a discussion]{davidson06a}.

The F02 version of {\sc WavDetect} consists of two components, {\tt
wtransform} and {\tt wrecon}.  The former convolves binned images with
Mexican Hat \citep{slezak90a} wavelet functions with various
user-specified scale sizes and then identifies pixels that are
significantly above the background.  In {\sc Xapa}, we use the F02 version
of {\tt wtransform} as part of an automated pipeline known as {\tt
md\_detect},\footnote{Where the {\tt md\_} prefix acknowledges the
architect of the routine, Michael Davidson.} as illustrated by the
flowchart of Fig.~\ref{fig:md_detect_flowchart}. We use a set of
nine wavelet scales, numbered according to increasing size, and
corresponding to $\sqrt{2}, ~2, ~2\sqrt{2}, ~4, ~4\sqrt{2}, ~8,
~8\sqrt{2}, ~16$ and $32$ image pixels. At each scale, the convolved
image is compared with a threshold image.  Convolved image pixels with
values greater than their corresponding threshold image pixels are
assumed to be associated with astronomical sources (`significant
pixels' hereafter). For those pixels, we reject the null hypothesis
that they are consistent with the measured background. We then
generate a set of support images, which record the significant pixels
at each wavelet scale.

In order to enhance the detectability of faint extended emission, {\tt
md\_detect} performs the wavelet analysis in two stages (or
`Runs'). In Run 1 (scales 1-2), bright compact sources are located
first. These are then masked out before performing Run 2 (scales
3-9). The masking step was found to be necessary because bright point
sources can pollute the wavelet signal on large scales, and hence
mimic extended sources. Unfortunately, this masking can occasionally
result in genuine extended sources being excluded from the candidate
list, so an extra step was added to {\sc Xapa} to mitigate this effect
(Section~\ref{sec:cuspy}).

The second component of the F02 version of {\sc WavDetect} is {\tt
  wrecon}. This generates a source list for each image, by grouping
collections of significant pixels together into source regions, or
`cells'. A drawback of the F02 version of {\tt wrecon} is that it
uses the instrument PSF to define the size of the cells. This means
that extended sources can be broken up into multiple contiguous
`sources' (because a single PSF-sized cell is not big enough to
enclose all the flux). To overcome this problem, we wrote a modified
version of {\tt wrecon}, called {\tt md\_recon}, for {\sc Xapa}. Unlike {\tt
  wrecon}, {\tt md\_recon} does not assume a priori the size of the
detected sources, and is consequently considerably better at fitting
ellipses to extended sources. The operation of {\tt md\_recon} is as
illustrated by the flowchart in Fig.~\ref{fig:md_recon_flowchart}.
At each wavelet scale, {\tt md\_recon} first combines lists of
significant source pixels into source cells. Multi-scale objects,
i.e. those detected by {\tt md\_detect} on multiple scales, are then
filtered using a `vision model' (Section~\ref{sec:xapa_vision}). The
vision model is a set of rules for combining the support images
derived for different wavelet scales. The vision model is able to
recognise when a point source is embedded in an extended source. It
also fits elliptical regions to the recovered sources (the region
enclosed by a source ellipse is referred to as $\epsilon_f$ in the
following descriptions).

\subsubsection{Extended Sources with Central Cusps}
\label{sec:cuspy}

\begin{figure*}
\includegraphics[width=0.95\textwidth]{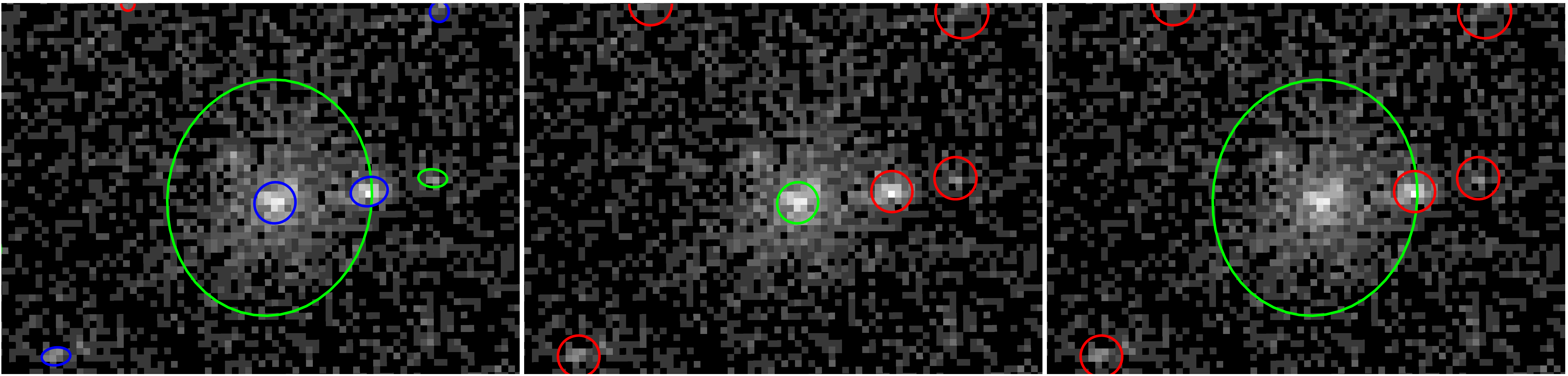}
\caption{Illustration of the effect of extended source
cuspiness. Left: The original (before the cuspiness test was
introduced) Run1 (blue) and Run2 (green) detections. Middle: The final
source list if the cuspiness test is not performed. Right: The final
source list (after the cuspiness test was introduced).  Extended and
point sources have green and red outlines
respectively.\label{fig:cuspy}}
\end{figure*}

The two step (Run 1, Run 2) procedure adopted by {\tt md\_detect} for
source detection works well, in that it prevents bright point sources
from contaminating the extended source list.  However, it has the
disadvantage that when a genuine extended source is detected in Run 1,
it will be excluded from Run 2. This means that its size will be
underestimated by the vision model, and it will not appear in an
extended source list. Extended sources with cuspy brightness profiles
will be particularly affected by this, e.g. clusters with cool cores. We
have therefore devised a `cuspiness test' that is carried out between
Run 1 and Run 2. This involves generating a grid of 5 by 5 pixels,
$Q$, centred on the position of each source detected in Run 1.  A
quantity, $C$, representing the cuspiness of the central region is
then calculated, as follows:
\begin{equation}
C=\frac{Q_{\rm max} - Q_{\rm min}}{Q_{\rm max}}
\end{equation}
Tests showed that real point sources have $C \geq 0.85$, so if a Run
1 source is found to have $C < 0.85$ --- i.e. it possesses a
flatter central profile than a real point source --- it is removed from
the list of Run 1 detections, resulting in it being available to be
detected again in Run 2. This situation is illustrated by
Fig.~\ref{fig:cuspy}.

\subsubsection{The {\sc Xapa} Vision Model}
\label{sec:xapa_vision}

Here we give more details about the vision model used to filter
sources detected at multiple scales by {\tt md\_recon}.  To describe
our vision model we introduce the following two terms: a `structure' is
a connected set of pixels in the support image for a particular scale;
and an `object' is a set of connected structures from different
scales. The steps are:

\begin{enumerate}
\item For each structure, comprising a set of pixels \{(x,y)\} in
  $S_i$ which is the support image for scale $i$, determine whether
  the structure defines the `root' of an object, i.e. whether
  $S_{j}(\{(x, y)\}) = 0$ for all $j < i$.
\item For each such root, check to see if there is a structure in the
  scale above at this position, i.e. if $\exists (x^\prime,y^\prime),
  (x^\prime,y^\prime) \in \{(x,y)\},S_{i+1}(x^\prime, y^\prime) \neq
  0$.
\item If such a structure exists, and its maximum pixel value lies
  within \{(x,y)\}, then these two structures are linked, such that
  the image pixels belonging to the object comprise the union of the
  pixels in the linked structures from scales $i$ and $i+1$.
\item The process of upward linking continues until the condition in
  step (ii) is not satisfied, at which time the object is
  terminated. When each scale has been scanned for root structures and
  they have been propagated in the `tree-like' fashion, then for each
  object created there exists a set of image pixels belonging to
  it. An ellipse can then be fitted to these regions and a source list
  created.
\end{enumerate}

\begin{figure*}
\includegraphics[width=9.5cm]{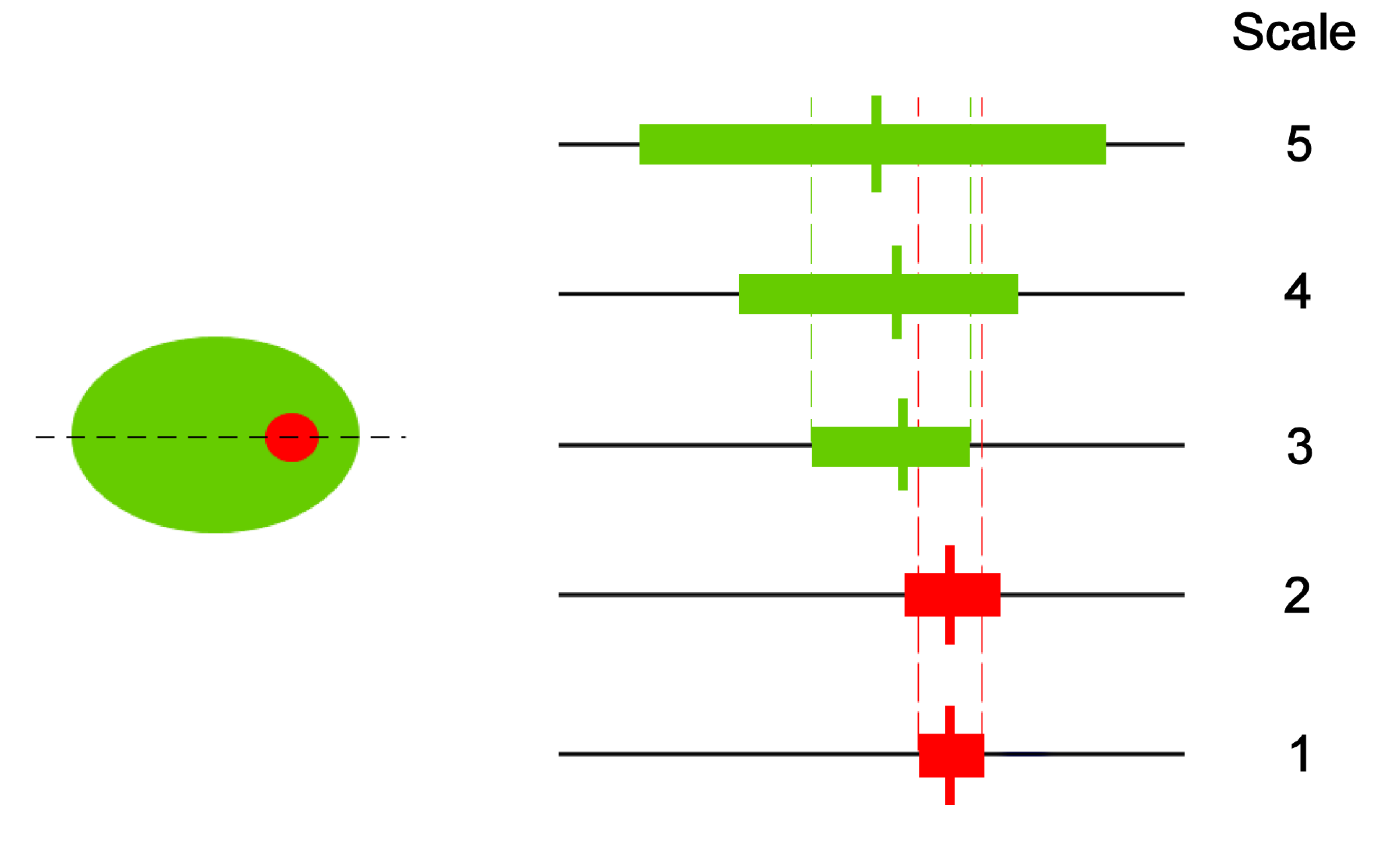}
\caption{Illustration of the `tree' vision model. Left: The source
configuration showing a point source embedded in a larger source. The
dashed line indicates a 1-d cut through the sources. Right: A schematic
of the significant pixels at each scale showing how the structures are
connected to form objects. The vertical bars denote the position of
the maximum coefficient at each scale. The maximum of scale 3 lies
outside of the structure of scale 1 hence a new object is
started. \label{fig:tree_diagram_crop}}
\end{figure*}

This vision model can handle both point and extended sources. 
Crucially, it can also cope with point sources embedded in
extended sources, and with close pairs of points which should be separated
rather than blended. A schematic to illustrate how the vision model
works when a point source is embedded in an extended source can be
seen in Fig.~\ref{fig:tree_diagram_crop}.

\subsection{Source Properties}
\label{sec:find_srcprop}

Once {\tt md\_recon} has been run on a given image, the source list is
passed on to the next part of the {\sc Xapa} pipeline, {\tt find\_srcprop}.
The two-stage operation of {\tt find\_srcprop} is illustrated by the
flowchart in Fig.~\ref{fig:find_srcprop_flowchart}. In the first
stage, {\tt find\_srcprop} determines the significance of each
detected source. In the second, a sub-routine known as {\tt
find\_srcprop\_final}, computes other source properties (such as the
count-rate and probability of extent); it is the results from the 
{\tt find\_srcprop\_final} that appear in the XCS data tables (Section~\ref{sec:tables}).

\subsubsection{Measuring Source and Background Counts}
\label{sec:fluxes}

\begin{figure}
\includegraphics[width=8.5cm]{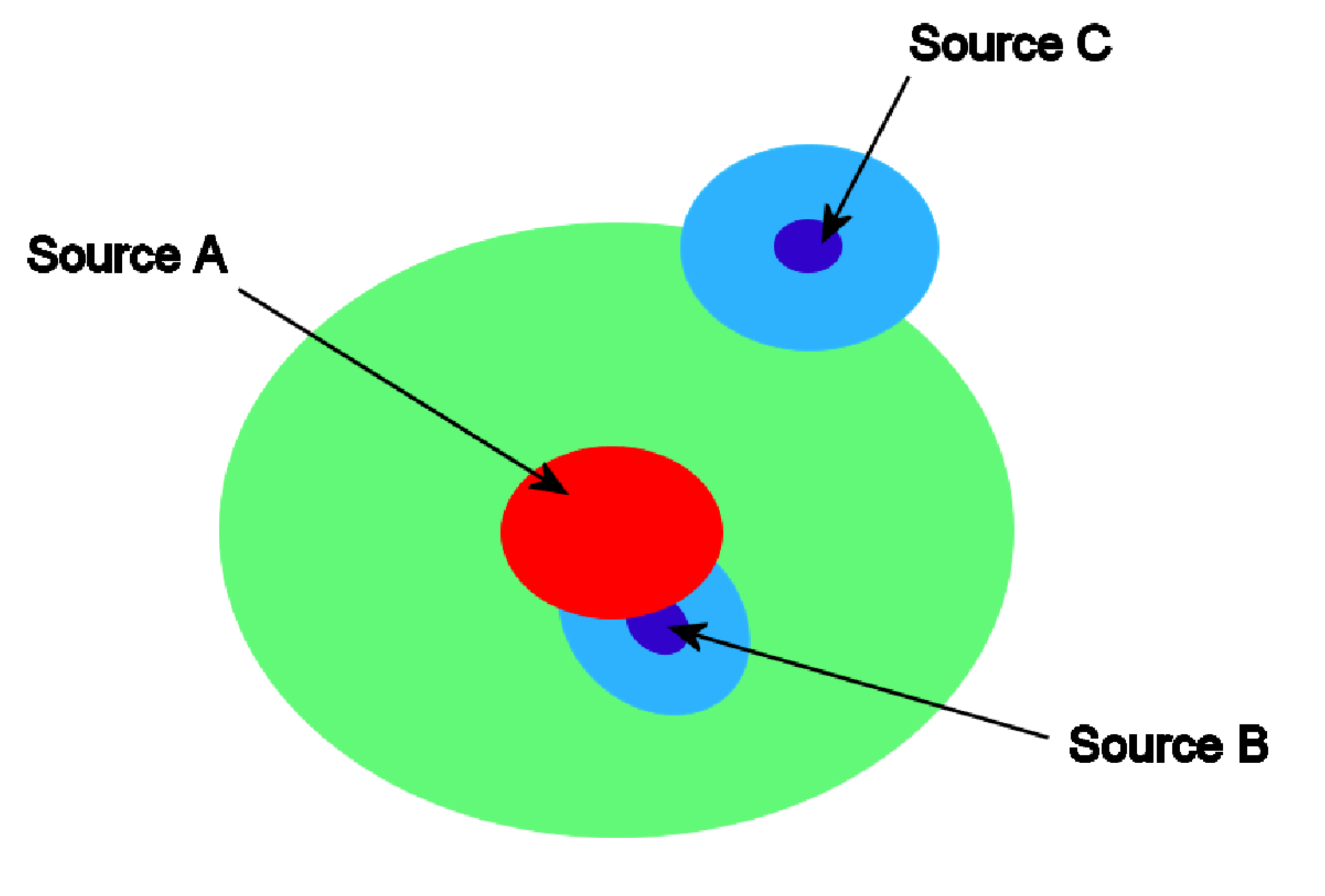}
\caption{A diagram showing how the aperture used to measure source
flux is created. The source to be measured is \emph{Source A} and
there are also two other objects nearby (\emph{Source B} and
\emph{Source C}). Both the $1*\epsilon_{\rm f}$ and $3*\epsilon_{\rm
f}$ ellipses are shown for each source (red and green respectively for
\emph{Source A} and dark blue and light blue for \emph{B} and
\emph{C}. Hence, the area used to calculate the flux for \emph{Source
A} is the red plus the green
region. \label{fig:flux_apertures_colour}}
\end{figure}

Here we describe how background corrected source counts were
calculated in {\sc Xapa} by {\tt find\_srcprop} and by its sub-routine
{\tt find\_srcprop\_final}.  Tests during the development of {\sc Xapa}
showed that the best results were obtained using different aperture
sets for each stage. The aperture set comprises the region for source
flux determination, the region for background flux determination and a
masked region (which is not used for either). In Table
\ref{apertures_find_srcprop} we note the configuration for both
aperture sets. In specifying these, we denote by $\epsilon_{\rm f}$
the ellipse as fitted to the object region, so that $3\epsilon_{\rm
f}$ is the ellipse with major and minor axes three times those fitted
to the source by the vision model. We use Uniq($X$) to denote, for a
particular source, those pixels which lie only within region $X$
defined relative to that source: e.g. Uniq($3\epsilon_{\rm f}$)
defines, for a particular source, the set of pixels which lie within
the $3 \epsilon_{\rm f}$ region for that source and for no other (as
illustrated in Fig.~\ref{fig:flux_apertures_colour}).

\begin{table}
\caption{Mask and aperture configurations for source and background
  flux determination used in  
{\tt find\_srcprop} and {\tt
  find\_srcprop\_final}.\label{apertures_find_srcprop}} 
\begin{tabular}{l|p{6.4cm}}
\hline
Type & Configuration ({\tt find\_srcprop} )\\ \hline
Run 1 & Mask: Run 1 sources masked at $2\epsilon_{\rm f}$ \\
& Flux:  $1\epsilon_{\rm f}$+Uniq($3\epsilon_{\rm f}$) \\
& Background:  Inner radius at $2\epsilon_{\rm f}$, min. area = 400 pix \\
Run 2 & Mask: All sources masked at $3\epsilon_{\rm f}$\\
&  Flux: $1\epsilon_{\rm f}$+Uniq($3\epsilon_{\rm f}$)\\
&  Background: Inner radius at $3\epsilon_{\rm f}$, min. area = 2000 pix \\
\hline
\hline
Type & Configuration ({\tt find\_srcprop\_final})\\ \hline
Point & Mask: Point sources masked at $2\epsilon_{\rm f}$ \\
& Flux:  $1\epsilon_{\rm f}$ \\
& Background:  Inner radius at $2\epsilon_{\rm f}$, min. area = 400 pix \\
Extended & Mask: All sources masked at $3\epsilon_{\rm f}$\\
&  Flux: $1\epsilon_{\rm f}$ +Uniq($3\epsilon_{\rm f}$),
 with internal point sources \\
& masked at $1\epsilon_{\rm f}$\\
&  Background: Inner radius at $3\epsilon_{\rm f}$, min. area = 2000 pix \\
\hline
\end{tabular}
\end{table}

The expected background contribution is computed locally. An
elliptical annulus is placed around the source position: the inner
edge varies but is usually at $3\epsilon_{\rm f}$ and the outer edge
is increased until there are at least 2000 background pixels, or no
more area is available. The background count-rate, $b_{\rm {pix}}$,
is then calculated as $b_{\rm {pix}}=B /\bar{E}^\prime \times
a^\prime$, where $B$ is the total number of counts in the annulus,
$\bar{E}^\prime$ the mean exposure in the annulus and $a^\prime$ is
the number of pixels in the annulus. The expected number, $B_{\rm a}$,
of background counts within the source aperture is then computed as
$B_{\rm a}=b_{\rm {pix}} \times \bar{E} \times a$, where $\bar{E}$ is
the mean exposure in the source aperture and $a$ the number of pixels
in the aperture.

\subsubsection{Removing Low-Significance Sources}

The first task is to remove any sources which are statistically of low
significance, because they will not yield accurate properties.  The
source and background apertures used to determine this significance
must be chosen carefully (Section~\ref{sec:fluxes}), but once the expected
number of background counts, $B_{\rm a}$, within the source aperture,
$\epsilon_{\rm f}$, is known, it is possible to assess the
significance of the detected source. This is done by computing the
probability that the background could, by chance, produce the detected
number of counts in the source aperture, assuming a Poisson
distribution for the background counts, with mean $B_{\rm a}$.  Those
sources with a probability higher than 0.000032 are removed from the
source list: this probability is equivalent to a 4-$\sigma$ threshold
for a Gaussian distribution. In addition, detections comprised of
only a single significant pixel are excised from the source list,
regardless of their significance. These are likely to be hot pixels or
sources that are too faint to be accurately parameterised.

\subsubsection{Measuring Source Extents}		
\label{sec:extent}

After low-significance sources have been removed, the {\tt
find\_srcprop} routine is run again on the sources above the $\geq4$-$\sigma$ 
threshold, in order to classify them as point-like or extended. For
this, we need to compare the sources to the instrument
PSF. Unfortunately, no satisfactory 2-d PSF model for \emph{XMM}
exists (Section~\ref{psf_desc}), so for XCS we adopted the best
publicly-available 1-d (radially-averaged) model --- the Extended
Accuracy Model (EAM). This, in turn, necessitated the development of a
source classification criterion based on a 1-d source property. For
XCS, we used the Encircled Energy Fraction (EEF). The EEF records the
fraction of the total energy of a source as a function of increasing
(circular) aperture size. We note that even though the shape of the
PSF changes considerably towards large off-axis angles, its radial
average, the EEF profile, is only a weak function of off-axis angle
\citep{davidson06a}, making it a good basis for a classification
criterion to be applied across the full field of view.

Our extent classification is based on testing the null hypothesis that
the measured EEF for a source is consistent with the PSF EEF, at the
appropriate off-axis angle. This is implemented using a
Kolmogorov-Smirnov (K-S) test, using the EEF profile of the source and
a model-merged PSF EEF. The PSF EEF is derived from EAM EEFs produced
by the {\sc SAS} task {\sc CALVIEW} from the Current Calibration Files
(CCF) for each camera.  This is weighted by the Energy Conversion
Factor (ECF, Section~\ref{sec:ecf_desc}) appropriate for that ObsID. We
adopt for $P(point)$, the probability that the source is point-like,
the maximum value of the probability returned by the K-S test run on a
3$\times$3 pixel grid (with spacing $\pm 0.5 \textrm{ pixels}$ in $x$
and $y$) around the source position (in Section~\ref{sec:positions} we
show that the typical positional accuracy of XCS source centroids are
good to better than 1 pixel).

The reliability of the $P(point)$ values is a function of several
factors, including the position on the field of view, the background
level, the number of source counts, and the proximity of neighbouring
sources. For that reason, choosing a fixed threshold in $P(point)$ for
our classification would be inappropriate. Instead we are forced to
conduct a series of Monte Carlo (MC) simulations for \emph{every}
source: this is computationally expensive, but it is vital to prevent
misclassification. This simulation process involves generating 200
realisations of the appropriate PSF EEF model and populating them with
the same number of counts as measured in the data. Each of the 200
realisations are compared to the model and an empirical distribution of
the K-S $d$ values is established. If none of the simulated
distributions returns a $d$ value as great as the measured value, we
classify the source as being extended.  With this procedure, the
statistical probability of misclassifying an isolated point source as
extended is 0.005 or less. However, we note that this does not take
into account systematics, such as when two or more point sources
have been blended by {\sc Xapa} into a single source profile.  These can
only be removed a posteriori, by eye-balling the extended sources that
make it through to the cluster candidate list.  This eye-balling, or
quality control (see Fig.~\ref{fig:overview_flow}), process is
described in more detail in M11.

\subsubsection{Correcting Artefacts}
\label{sec:artefacts}

After the second pass of {\tt find\_src\_prop} has been completed, we
have a preliminary list of sources (classified as extended or
point-like) for a given ObsID.  Initial tests showed that these
preliminary lists include a number of artefacts.  These must be
corrected for before inclusion in an XCS source catalogue (see
below). The corrections are not foolproof, as not all genuine clusters
make it through to the candidate list and not all contaminating
sources are excluded, but because the corrections are folded into the
survey selection function (Section~\ref{sec:sf_descr}), they should not
impact our ability to use XCS cluster catalogues for statistical
studies.

\begin{figure}
\includegraphics[width=0.5\textwidth]{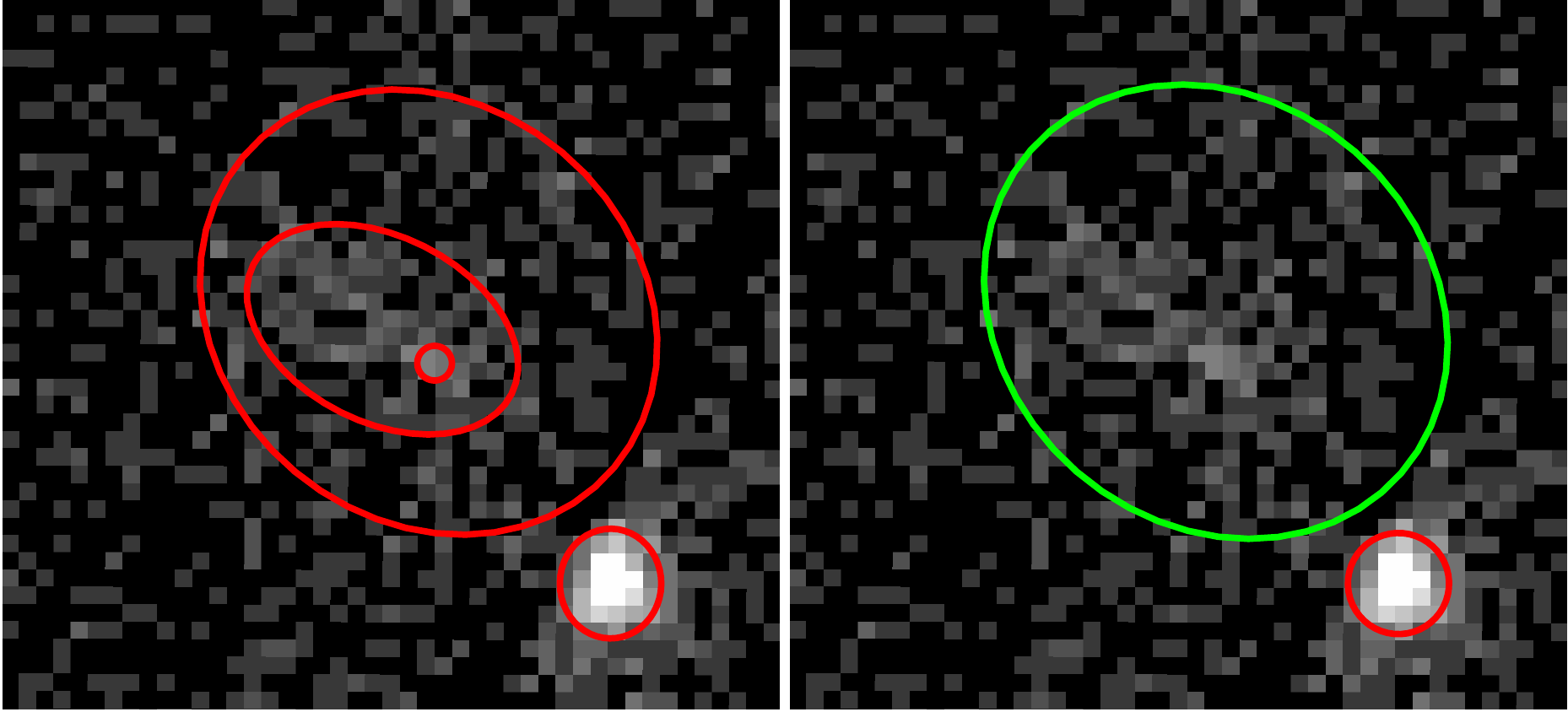}
\caption{An example where several initial detections of an
extended source are subsequently merged by {\sc Xapa} to improve the derived
properties.\label{fig:extmerge_figures}}
\end{figure}

{\sc Xapa}'s {\tt md\_recon} algorithm successfully detects sources within
sources (see Fig.~\ref{fig:tree_diagram_crop}).  However, one
unintended consequence is the occasional multiple detection of a
single source that has become split into two or more overlapping
sources.  This more often happens with extended sources, but can also
occur with point sources at the edge of the field of view. Therefore,
where there are incidences of two sources with overlapping cells, the
sources are merged and source properties recalculated by {\tt
find\_srcprop} (see Fig.~\ref{fig:extmerge_figures}).  This
refinement ensures that in most cases the source flux and morphology
are recovered well.

\begin{figure}
\includegraphics[width=0.5\textwidth]{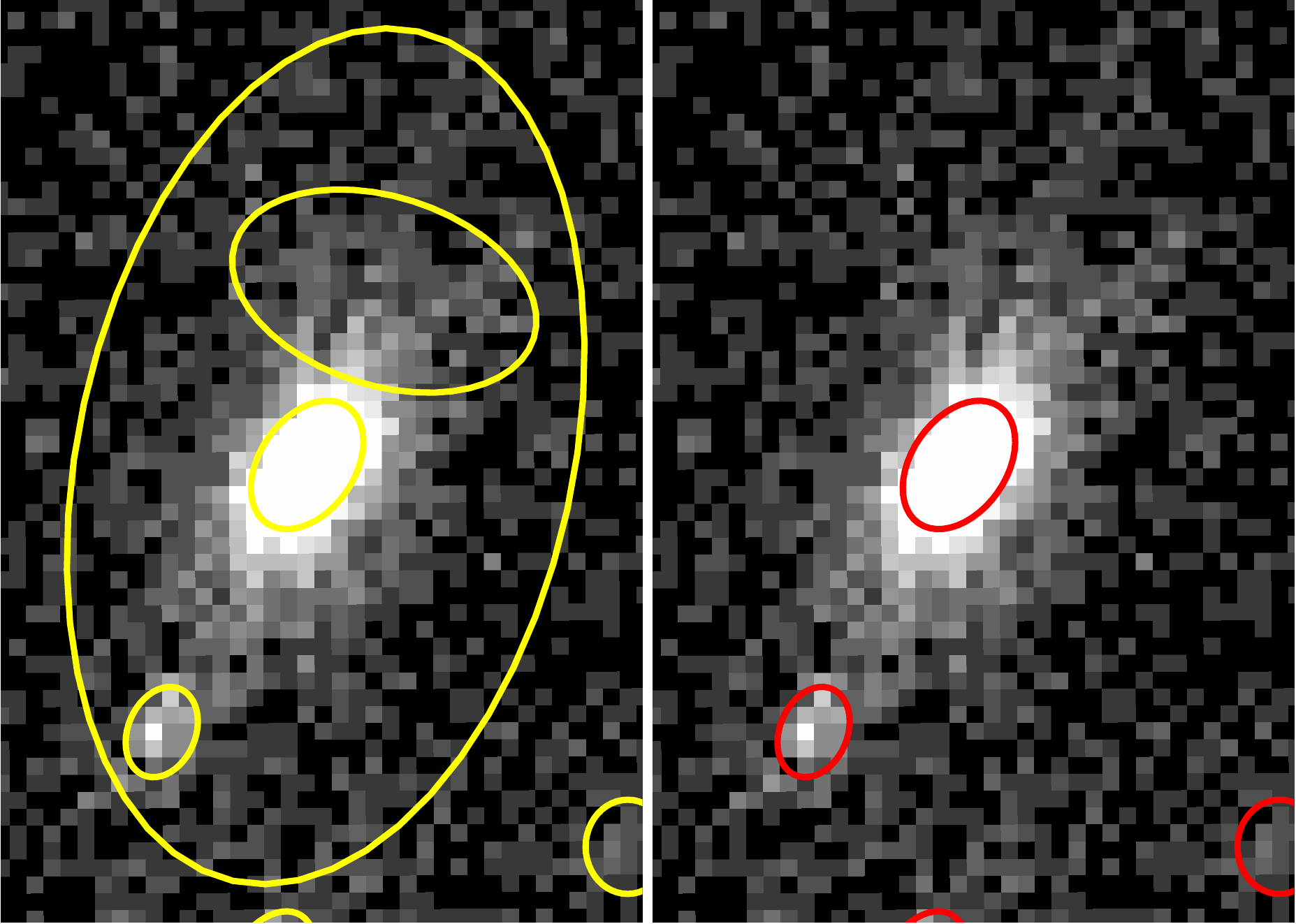}
\caption{Source ellipses defined by {\sc Xapa} for a bright, off-axis, point
source. Left: before the lobe removal step was included. Right: after
the lobe removal step was included: note that the two point sources
have still been recovered, but there is no erroneous large (extended)
ellipse enclosing both of them.\label{fig:lobe_figures}}
\end{figure}

When a bright compact source lies in the outskirts of the field of
view, it can produce a significant number of counts in the asymmetric
outer regions of the PSF. We term these objects as 
`point-sources-with-lobes'. The core of such sources are detected in
Run 1 of {\tt md\_detect}, and hence the core counts will be masked
from Run 2 (Section~\ref{sec:wavedetect}), but the remaining outer counts
might still yield a Run 2 detection (see
Fig.~\ref{fig:lobe_figures}).  Removing these
point-sources-with-lobes, without also removing clusters with cuspy
cores (Section~\ref{sec:cuspy}), proved to be one of the most difficult
problems to overcome with {\sc Xapa}. After extensive tests, we arrived at
the following compromise: an extended source is excised from the
source list, as a suspected point-source-with-lobe, if it is both
located within the 3$\epsilon_{\rm f}$ region of a Run 1 source, and
has less than one fifth of the counts of that source. This removes the
majority of the lobe artefacts, but can unfortunately also result in
some genuine faint extended sources being excluded from the XCS
cluster candidate list.

\subsubsection{Extended Source Flags}
\label{sec:flags}

When developing {\sc Xapa}, we had to find a compromise between
contamination and completeness, i.e. between effective and over
cleaning of the extended source list. Therefore, rather than removing
from the extended source list every object that could be 
erroneous, we have flagged certain sources that, conservatively, we
view as suspicious.  Our aim is to use the survey selection function
(Section~\ref{sec:sf_descr}) to help us understand whether flagged sources
should be included in statistical studies or not, but to date we have
taken a conservative approach and not included them in cluster
candidate lists, or as targets for optical follow-up (M11). The source
flags are as follows:

\begin{enumerate}
\item{{\it Extended Sources that are PSF-sized}.  At large off-axis
angles it is not infrequent for the flaws in the PSF model to cause an
obvious, bright, point source to be classified as extended. Therefore,
any source that is only just extended (i.e.  that has a size very
close to the PSF at the respective off-axis angle) is flagged as being
`PSF-sized' by {\sc Xapa}.}

\item{{\it Extended Sources with Internal Point Sources}. Even with
the inclusion of the point-source-with-lobe test
(Section~\ref{sec:artefacts}), the {\sc Xapa} vision model
(Section~\ref{sec:xapa_vision}) will occasionally misclassify flux from the
outskirts of a point source (or flux from a collection of neighbouring
point sources) as an erroneous extended source. We can mitigate 
this by flagging up likely incidences. Therefore, any extended
source region that encloses one or more point sources that contribute
$\ge 1.3$ times the extended source flux is flagged as being `point
contaminated' by {\sc Xapa}.}

\item{{\it Extended Sources with Internal Run1 Sources}. The final
flag is similar to the `point contaminated' case, but covers the
incidences of genuine point sources, detected in Run 1 by {\tt
md\_detect}, being erroneously passed on to Run 2 by the cuspiness
test (Section~\ref{sec:cuspy}).  Therefore, any extended source region that
encloses one or more Run 1 detection regions that contribute at least
half the extended source flux is flagged as being `Run 1
contaminated' by {\sc Xapa}.}
\end{enumerate}

\subsubsection{Source Parameters}
\label{sec:parameters}		

Once the source list per ObsID has been cleaned of artefacts, a file
is generated that saves all the relevant data. This file is then
interrogated when the survey-wide database is being generated
(Section~\ref{sec:tables}).  The following attributes are saved per source:

\begin{enumerate} 
\item{The centroid location in image coordinates;}
\item{The centroid location in sky coordinates (J2000);} 
\item{The centroid location in radial coordinates, i.e. the
off-axis angle (arcminutes) and the azimuthal angle (degrees);}
\item{The major axis, minor axis and orientation of the source ellipse;}
\item{The average exposure time at the source location (seconds);}
\item{The 0.5-2.0 and 2-10 keV background-subtracted source counts (in the merged image and in the individual camera exposures);}
\item{The 0.5-2.0 and 2-10 keV background-subtracted count-rates and 1-$\sigma$ count-rate uncertainties (in the merged image and in the individual camera exposures);}
\item{The source significance and extent probability;}
\item{The value of the source flags (see Section~\ref{sec:flags}).}
\end{enumerate}

\subsubsection{Master Detection List}
\label{sec:tables}

{\sc Xapa} produces a source list for each of the input ObsIDs, then
these lists are concatenated to form a Master Detection List
(MDL). Present in the \emph{XMM} archive are many areas that have been
observed multiple times. As a result, some sources will have been
detected by {\sc Xapa} multiple times.  When duplicates are found, only the
detection with the most soft-band counts is passed to the MDL.  To
remove duplicates, it is necessary to set an appropriate matching
radius. The positional accuracy of the survey is higher for point
sources than for extended sources, so it makes sense to use a
different radius for each type. The accuracy for point sources varies
as a function of off-axis and azimuthal angles (amongst other
parameters). However, for simplicity we use a single value for the
radius of 5 arcsec. The case for extended sources is less straightforward
because of the variety of source types and morphologies. The
positional accuracy for large diffuse objects, such as low-redshift
clusters, can be very poor, making it hard to pick an appropriate
radius. Fortunately, the largest diffuse sources should have already
been masked from their host ObsID. So, for XCS, we use a fixed
matching radius of 30 arcsec for extended sources. This radius is large
enough to allow reliable source matching, but small enough to minimise
removal of genuine cluster candidates.

As of May 1st 2011, {\sc Xapa} had run on \Nvi\, ObsIDs, resulting
in \Nvii\, point sources and \Nviii\, extended sources being included in
the MDL.  Of the \Nviii\, extended sources, roughly half were flagged
(\S~\ref{sec:flags}) and these were removed from the list of potential
cluster candidates (leaving \Nix\, sources). Additional cuts to this
list included the removal of sources within 20$^{\circ}$ of the
Galactic plane and 6$^{\circ}$ [3$^{\circ}$] of the Large [Small]
Magellanic Cloud. Those cuts were made because it can be hard 
to carry out effective optical follow-up in regions of high
projected stellar density. Moreover, the closer one gets to the Galactic
plane, the higher the hydrogen column (large $n_{\rm H}$ values impact
our ability to recover accurate source fluxes).  Further cuts, see
below, are then imposed to ensure that the vast majority of XCS
cluster candidates are genuinely serendipitous detections, rather than
the intended target of the ObsID. A final cut, on minimum source count
($>50$) is then applied, leaving \Niv\,  sources drawn from \Nv\, different ObsIDs;
when we use the term `candidate' hereafter, we are referring to 
these \Niv\, sources. The candidates have
a range of counts, from 50 to several thousand. Of particular interest
to the cosmology and evolution studies we plan with XCS are the \Nivb\,
with more than 300 counts, because these should deliver, once redshift
information is available, reliable temperature estimates (Fig.~\ref{comp_results_counts}).

As mentioned above, filters were applied to exclude non-serendipitous
or `target' objects from the candidate list. The targets in
question are primarily clusters, but other types of extended X-ray
sources should also be excluded (e.g. low redshift galaxies). It is also important to identify 
extended sources that are physically associated with the target, e.g. if they both 
belong to the same supercluster. The target filters involved both checks of the ObsID file 
headers and automated queries to the NASA Extragalactic Database (NED)\footnote{http://nedwww.ipac.caltech.edu}. 
The filters were run separately on each ObsID that a particular extended source 
was detected in. A given extended source (that passed the other cuts described above) 
was included in the candidate list if it was classed as being a serendipitous detection in at least one of those 
ObsID (even if it was classed as being the target of one or more others). We acknowledge that some \emph{XMM} targets, and some sources associated with targets, do make it through into our candidate list. However, 
as shown in M11, these are straight forward to remove at the quality control stage (eight
such examples were removed from XCS-DR1). 

Extended sources were excluded from the candidate list if they met one or more of the following criteria:

\begin{enumerate}

\item Their {\sc Xapa} centroid fell within $2$ arcmin of the aim point of an ObsID with an object classification (as listed in the header) of `cluster' or `group'. 

\item Their {\sc Xapa}  ellipse overlaps the aim point of an ObsID with a target name (as listed in the header) that has been associated with a cluster or group in NED. (This filter is necessary either when the pointing type is not included in the header, or is incorrect.)

\item Their {\sc Xapa} ellipse overlaps the centroid of a cluster or group in NED, when the aim point of the ObsID falls within $2$ arcmin of that cluster or group. (This filter is necessary because sometimes non standard target names are listed in the header.)

\item Their redshift is within 5000 km s$^{-1}$ of the redshift of an object in NED, when the target name (as listed in the header) has been associated with that object. Both redshifts are automatically extracted from NED. (This filter is necessary because some ObsID targets are deliberately positioned off axis. This filter also reduces the number of sources entering the candidate list that are physically associated with targets (including with non-cluster targets, such as AGN).)

\item Their {\sc Xapa} ellipse overlaps the aim point of an ObsID with a target name (as listed in the header) that has been associated with a known galaxy in NED. (This filter was found to be the most effective way to exclude low redshift galaxies from the candidate list.)

\end{enumerate}

\subsection{{\sc Xapa} Verification: Point Sources}
\label{sec:points}

As mentioned above (Section~\ref{sec:tables}), {\sc Xapa} has catalogued to
date in excess of 100,000 unique point sources.  In this section we
test {\sc Xapa} astrometry and flux measurements using these point sources,
finding both measures to be robust.

\begin{figure*}
\centering
\includegraphics[angle=90,width=0.42\textwidth]{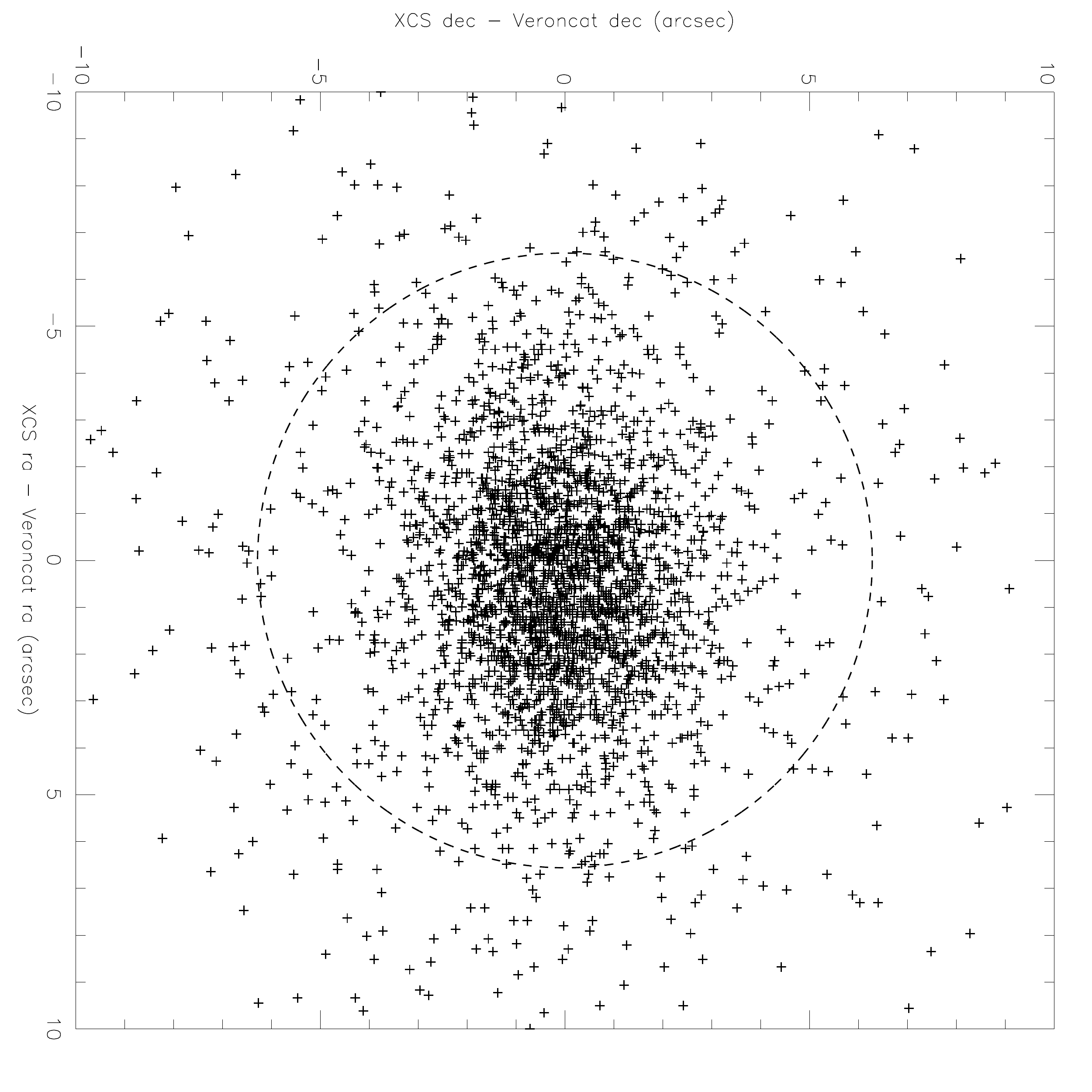}
\includegraphics[angle=90,width=0.52\textwidth]{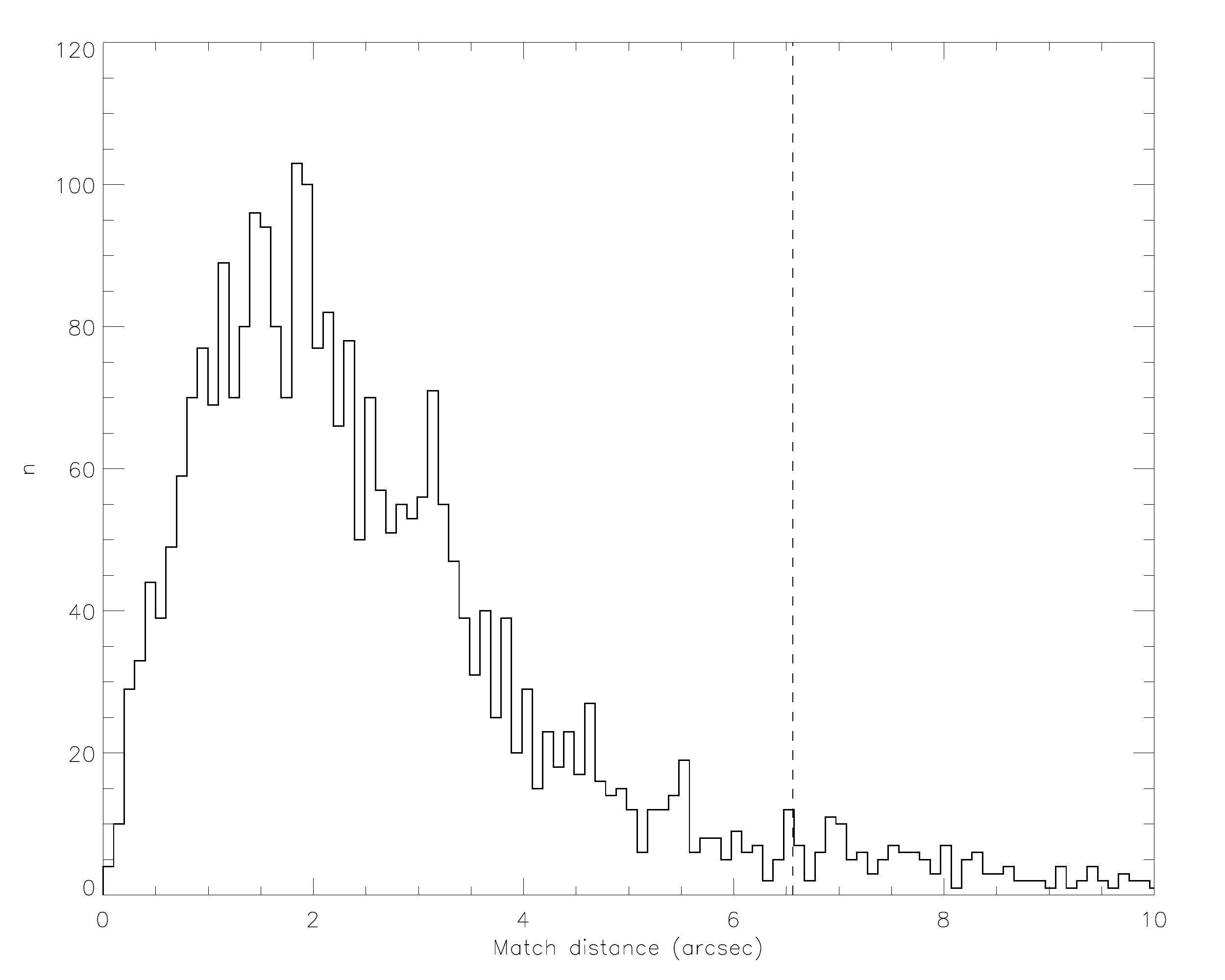}
\caption{The relative position of the matches between XCS and VeronCat 
\citep{veron06a} source positions. The dashed line represents the $95\%$ 
matching radius.}
\label{fig:VeronCat-QSO}
\end{figure*}

\subsubsection{Positions}
\label{sec:positions}

To determine the positional accuracy of the XCS point sources, it is
desirable to use a catalogue that has high spatial resolution and
astrometric precision. It would also need to have significant overlap
with the \emph{XMM} archive. A natural choice for this is the Sloan
Digital Sky Survey (SDSS\footnote{http://www.sdss.org},
\citealt{abazajian09a}); the data is of high quality and contains many
objects that would be expected to have X-ray counterparts, e.g.
quasars and AGN. A cross match of XCS point sources against the SDSS
Quasar Catalog IV \citep{schneider07a} using a radius of 10 arcsec
produces 1131 matches. This was extended further using the catalogue
of \citet[][VeronCat hereafter]{veron06a}.  VeronCat is a compilation
of all known AGNs and QSOs (including those in the SDSS). A 10 arcsec
matching radius returns 2807 matches, the distribution of which can be
seen in Fig.~\ref{fig:VeronCat-QSO}. We have determined the chance
of false association between the XCS and VeronCat with a 10 arcsec
matching radius to be 1\%.  The mean matching distance is 2.6 arcsec, and
95\% of the matches fall within 6.6 arcsec. This level of precision is
consistent with previous determinations of the positional accuracy
obtainable with XMM data \citep{watson09a}.

\subsubsection{Fluxes}
\label{sec:pointfluxes}

To assess the accuracy of the point source fluxes measured by {\sc Xapa} we
have compared the XCS point source list to the \emph{XMM}
Serendipitous Survey 2XMM catalogue \citep{watson09a}.  This catalogue
is the ideal counterpart to XCS because it is also based on automated
pipeline analysis of the entire XMM archive. A 10 arcsec matching radius
has been used to compare the samples. Fig.~\ref{fig:2xmm-xcs} shows
the flux comparison from the individual cameras aboard \emph{XMM},
using a $0.5 - 2.0$ KeV band. There is clear consistency between the
two surveys, with no significant systematic offsets. It is important to note 
that the default {\sc Xapa} fluxes for extended sources are not similarly reliable. 
This is for two reasons, first the ECFs used to generate the fluxes relate to 
power-law spectra (whereas extended sources are more likely to have thermal spectra) and
second, the fluxes have not been properly corrected for any source
flux lying outside the {\sc Xapa} defined ellipse. In
Section~\ref{sec:spatialmethod}, we describe how aperture corrected energy
fluxes are determined for the candidates.

\subsection{{\sc Xapa} Verification: Extended Sources}
\label{sec:extended}
 
As mentioned above (Section~\ref{sec:tables}), {\sc Xapa} has catalogued to
date in excess of 10,000 unique sources that have been statistically
classified as extended. {\sc Xapa} is not infallible however, and some of
the objects in the  candidates list will be erroneous --
because they are blends of point sources or other artefacts of the
data reduction -- and a small fraction will be other types of
genuinely extended X-ray sources (such as nearby galaxies or supernova
remnants). Nevertheless most of them will be clusters. In this section
we first compare the {\sc Xapa} determined extents for the clusters 
in the XCS-DR1 sample to those in the 2XMM catalogue (Section~\ref{sec:2XMMext}). 
We then compare the candidate list to the cluster sample
of the XMM-LSS survey in the same ObsIDs (Section~\ref{sec:LSS}). We then
describe how we quantify the completeness level using simulations of
our selection function (Section~\ref{sec:sf_descr}). We note that it is
harder to quantify the contamination (due to blends and artefacts) level
than the completeness level. In XCS we do not use simulations for
this, but rather examine each source (and its optical counterpart) by
eye (M11).

\subsubsection{Comparison with 2XMM}
\label{sec:2XMMext}

To investigate the quality of the {\sc Xapa}  determined source extent, we have used the \Nxix\, XCS-DR1 clusters with matches to extended sources in the 2XMM catalogue. For these clusters, we have compared the {\sc Xapa} major axis to the 2XMM extent measure. In the latter case, the quoted value is equivalent to the core radius of a $\beta$-profile (Eqn.~\ref{eq:beta}), so is always smaller than the {\sc Xapa} value, typically by a factor of 5. The two measures are correlated (correlation coefficient of 0.514), a with $\simeq 30$ per cent scatter about the best fit relation.. Therefore, despite the very different methods by which extents are measured by the two surveys, both descriptions are useful when determining source sizes. We note that only \Nxx (11 per cent) of the XCS-DR1 clusters in ObsIDs processed by 2XMM were not classified as extended sources in the 2XMM catalogue. Of course there might well be other clusters that 2XMM has detected as extended, but that {\sc Xapa} has not.

\subsubsection{Comparison with the XMM-LSS}
\label{sec:LSS}

The \emph{XMM} Large Scale Structure (XMM-LSS) Survey is reported in
\citet{pierre06a} and \citet{pacaud07a}. It covers a single contiguous
region of roughly 6 deg$^2$, comprised of 51 ObsIDs, in which the
authors have undertaken a dedicated cluster survey, accompanied by a
detailed selection function. In this region they detected 33 `Class 1'
extended objects. This class is designed to be uncontaminated by
mis-classified point sources. A more detailed examination of these
objects (including optical overlays, photometry and spectroscopy) has
confirmed 28 of these to be genuine clusters; the remaining 5 were
shown to be nearby X-ray emitting galaxies. Twenty-nine of the 33
Class 1 objects have counterparts in XCS that were classified as
extended by {\sc Xapa}. This includes 2 of the non-cluster objects. Three of
the remaining four Class 1 objects were detected by {\sc Xapa}, but classified
as point-like. The final object (XLSS J022210.7-024048) was detected
by {\sc Xapa}, but subsequently removed from the source list because it did
not meet our 4-$\sigma$ significance requirement.

The radius used in the matching of {\sc Xapa} sources to the XMM-LSS was
typically 10 arcsec. However for XLSS J022433.8-041405 a radius of 24 arcsec was
required to get a match; this source is large and elliptical, hence
there is some uncertainty in the source centre, though the extent of
the XCS source and its XMM-LSS counterpart are overlapping.
 
The XMM-LSS also have a C2 class of clusters with slightly less
conservative selection criteria. This sample has yet to be published,
but the authors report this class to contain $\sim60$ sources. Within
the XMM-LSS ObsIDs, XCS detects 82 extended objects without flags
(Section~\ref{sec:flags}), so the overlap is likely to be substantial.
 
\subsubsection{Selection Function: Method}
\label{sec:sf_descr}

Pioneering work by \citet{adami00a}, and later by \citet{burenin07a},
demonstrated the impact of complex selection effects on cluster
samples derived from X-ray surveys. \citet{pacaud07a} have shown,
using the XMM-LSS Class 1 sample described above, that the measured
evolution in the normalisation of the $L_{\rm X}-T_{\rm X}$ relation
is significantly affected by selection biases. In another X-ray study,
\citet{mantz10a} provide an in-depth discussion of Malmquist and
Eddington biases and their effect on measurements of scaling
relations. Optical and SZ cluster surveys are also increasingly supported
by selection function simulations \citep{melin05a,koester07a}.

The ability to measure selection functions for XCS was embedded at the
outset in {\sc Xapa}. Indeed, one of the driving reasons behind us designing
our own source detection pipeline, rather than using the excellent
data products available from the \emph{XMM} Survey Science Centre
\citep{watson09a}, was the requirement that we needed to be able to
quantify the extended source selection function using synthetic
clusters. In the following, we describe how the selection functions
are carried out and present some results.

Our approach follows a general method in which synthetic
cluster profiles are added to EPIC merged images, which are
then run through {\sc Xapa}. 
The angular size of the synthetic cluster profile is determined from
the angular diameter distance at the chosen input redshift. The
profile is then randomly positioned into a blank \emph{XMM} `image', with a
uniform probability across the field of view, and then convolved with
the appropriate PSF model. For this purpose we use the two-dimensional
Medium Accuracy Model (MAM, Section~\ref{psf_desc}).  This is a natural
choice of PSF model for the selection functions because it accounts
for the azimuthal variation in the PSF, and also because the
alternative  model (EAM, Section~\ref{psf_desc}) is implemented in {\sc Xapa} for
source classification (Section~\ref{sec:extent}): to keep the simulations
fair, we cannot use the same model for blurring as we do for extent
classification. The convolution with the PSF creates a probability
density function (PDF) for the synthetic cluster profile.  We note
that the shape of the synthetic cluster profile depends on the user's
specific requirements, and we will discuss some examples in
Section~\ref{sec:sf_beta} \& \ref{sec:sf_clef}.

Next, an ObsID is chosen for the synthetic cluster to be placed in.
The choice of ObsID will depend on the particular test being
undertaken.  For example, one might want to know the detection
sensitivity in a particular ObsID, or one might want to know the
detection sensitivity for a set of ObsIDs, e.g. those with similar
$n_{\rm H}$ or exposure times.  The synthetic cluster is added to
the chosen ObsID as follows:
\begin{enumerate}
\item The absorbed count-rate of the cluster profile is determined
  from the gridded LCFs (Section~\ref{sec:ecf_desc}) for that ObsID, so
  that it matches the synthetic cluster's luminosity, temperature and
  redshift.
\item The cluster PDF is normalised to the LCF predicted
  count-rate, thus creating a count-rate image.
\item The count-rate-image is converted into a count-image by
  multiplying by the appropriate exposure map.
\item The synthetic count-images for the individual cameras are then
  added to the respective real images
  (Section~\ref{sec:images}).
\item The individual images are then added to make a merged image.
\end{enumerate}

The resulting merged image, containing the synthetic cluster, is then
processed by {\sc Xapa} in the standard way. There are two criteria that
must be met in order for an input synthetic cluster to be deemed
successfully `recovered' by {\sc Xapa}: the detection software must identify a
source at the synthetic cluster location, and that source must be
classified as extended. This has to be a new source; if the synthetic
cluster happens to have been placed at random close to a previously
detected real extended source, then the synthetic cluster is not
classed as having being recovered (even if its `counts' dominate
those from the real source).  Depending on the application, we might
further require that the new detection not be flagged (see
Section~\ref{sec:flags}). It is not sufficient to perform the synthetic
cluster recovery test only once, rather one must perform it multiple
times to ensure an accurate measurement of the recovery probability
for a given set of input parameters. There is so much parameter space
to be tested (see below) that the number of selection function tests
can run into the millions for certain applications. Determining the
survey selection function is by far the most computationally demanding
part of XCS.

\subsubsection{Selection Function: Results (Analytical Models)}
\label{sec:sf_beta}

\begin{figure}
\centering
\includegraphics[angle=90,width=0.5\textwidth]{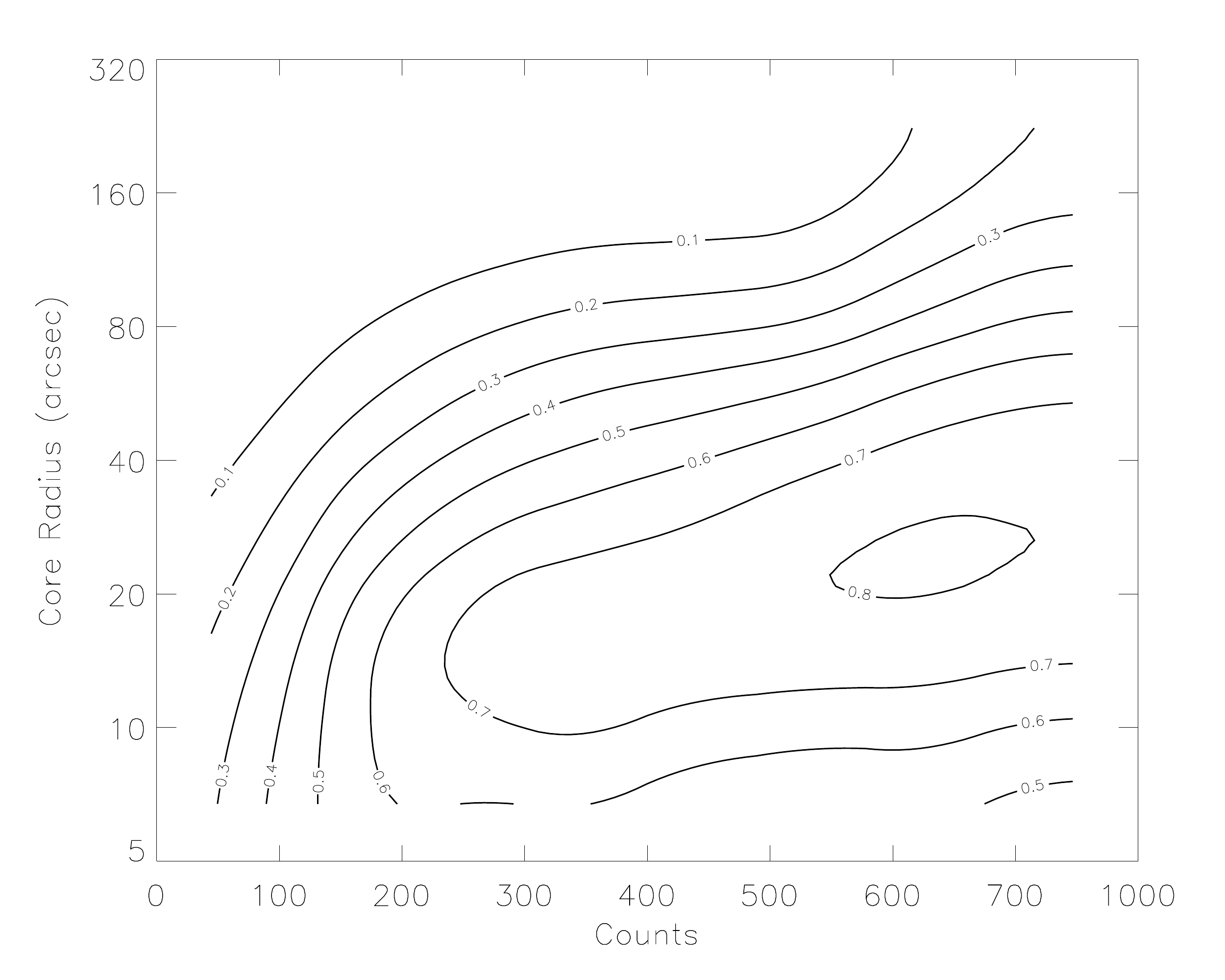}
\caption{Predicted recovery efficiency for 3 keV clusters as a  function of core radius and recovered counts. The synthetic clusters used for this test had circularly symmetric $\beta$-profiles  ($\beta=2/3$).}
\label{fig:betafun2}
\end{figure}

The simplest profile type that we have studied is that of an
isothermal $\beta$-profile cluster (Eqn.~\ref{eq:beta}). 
Using this profile we have tested the selection function
dependency on cluster parameters (e.g.  redshift, temperature,
luminosity, core radius, profile slope and ellipticity); on image
parameters (e.g. exposure time, off-axis angle, azimuthal angle); and
on cosmological parameters (e.g.  $k$ and $\Omega_{\rm m}$, the
curvature and present mean mass density of the Universe
respectively). Some results from the $\beta$-profile selection function
runs have already been published \citep{sahlen09a}. In Fig.~\ref{fig:betafun13}, \ref{fig:betafun16} and \ref{fig:betafun2} we
show some additional results.

Figures~\ref{fig:betafun13} and \ref{fig:betafun16} show how the
selection function depends on cluster luminosity and redshift.  We
show results for 3 keV and 6 keV clusters with a range of
luminosities. In all cases, the input profiles were spherically
symmetric, with $r_c=160$ kpc and $\beta= 2/3$. The profiles were
placed randomly in a subset of ObsIDs chosen to be a representative
sample of the whole archive, i.e. to have a similar distribution of
exposure times and Galactic latitudes. This simple test confirms that
bright clusters can be consistently detected out to redshifts of at
least 1, whilst fainter clusters can only be found with reasonable
certainty at lower redshifts. We note that typical $L_{\rm X}$ values
at 3 keV and 6 keV are roughly 1 and 10$\times10^{44}$ erg~s$^{-1}$
respectively, based on the low-redshift $L_{\rm X}-T_{\rm X}$ relation
of \citet{arnaud99a}. Therefore, we can expect to detect roughly 60\%
and 85\% of 3 and 6 keV clusters respectively at $z=0.6$, but only
10\% and 75\% at $z=0.9$ (assuming no evolution in the $L_{\rm
X}-T_{\rm X}$ relation).

Fig.~\ref{fig:betafun2} shows how the selection function depends on
cluster angular size.  Here we have run a set of 3 keV clusters
through the selection function process with physical core radii
varying according to the findings of \cite{jones84a}, i.e. in the
range of 50 kpc to 400 kpc.  Over the range of redshifts probed by XCS
($0.1\leq$ z $\leq 1.0$), these core radii have an angular size in the
range 217 to 7 arcsec. Fig.~\ref{fig:betafun2} shows the
fraction of clusters recovered by {\sc Xapa} as a function of both angular
size and the number of input synthetic cluster source counts. For
clusters with more than 300 counts, the cluster recovery rate is good
($\geq$70\%) when the extent is in the range $\simeq 10 - 20$ arcsec. 
These limits roughly translate to $0.1<z<0.6$ for $r_c=50$ kpc
and $z>0.3$ for (more typically) $r_c=160$ kpc.

\subsubsection{Selection Function: Results (Numerical Models)}
\label{sec:sf_clef}

We have also investigated the recovery of clusters with profiles drawn
from cosmological hydrodynamic simulations. For this purpose we have
used clusters from the CLEF (CLuster Evolution and Formation in
Supercomputer Simulations with Hydro-dynamics) simulation
\citep{kay07a}. The use of the CLEF clusters has enabled an
investigation into the effects of more realistic (than
$\beta$-profiles) cluster shapes on the XCS selection function,
because CLEF includes clusters with cool cores and substructure.

The process by which the CLEF profiles are input into the \emph{XMM} images
is the same as for the analytical models. For simplicity, we use
the CLEF catalogued mean cluster temperature, rather than the full temperature 
map, when calculating the
total count-rates using the gridded LCFs (Section~\ref{sec:ecf_desc}).
These count-rates are then distributed using the emission measure profile
(see \citealt{onuora03a} for details) as a probability map. The
emission measure maps fully encode variations in temperature and
density and so this approach will preserve any substructure in the
surface brightness and also the presence of any central peak caused by
a cool core.

The selection function work using CLEF has shown that strong peaks in
the surface brightness profile, either due to substructure, or to a
cool core, make clusters easier to detect than $\beta$-profiles.
However, these clusters do not always make it into the `recovered list',
because they are either misclassified as point sources or flagged as
being PSF-sized extended sources (Section~\ref{sec:flags}). This
misclassification trend is mitigated if the total number of detected
counts is large enough to sample more of the extended profile, and is
almost completely resolved at the 500-counts-per-source level
\citep{hosmer10a}.

\begin{figure}
\centering
\includegraphics[angle=90,width=0.5\textwidth]{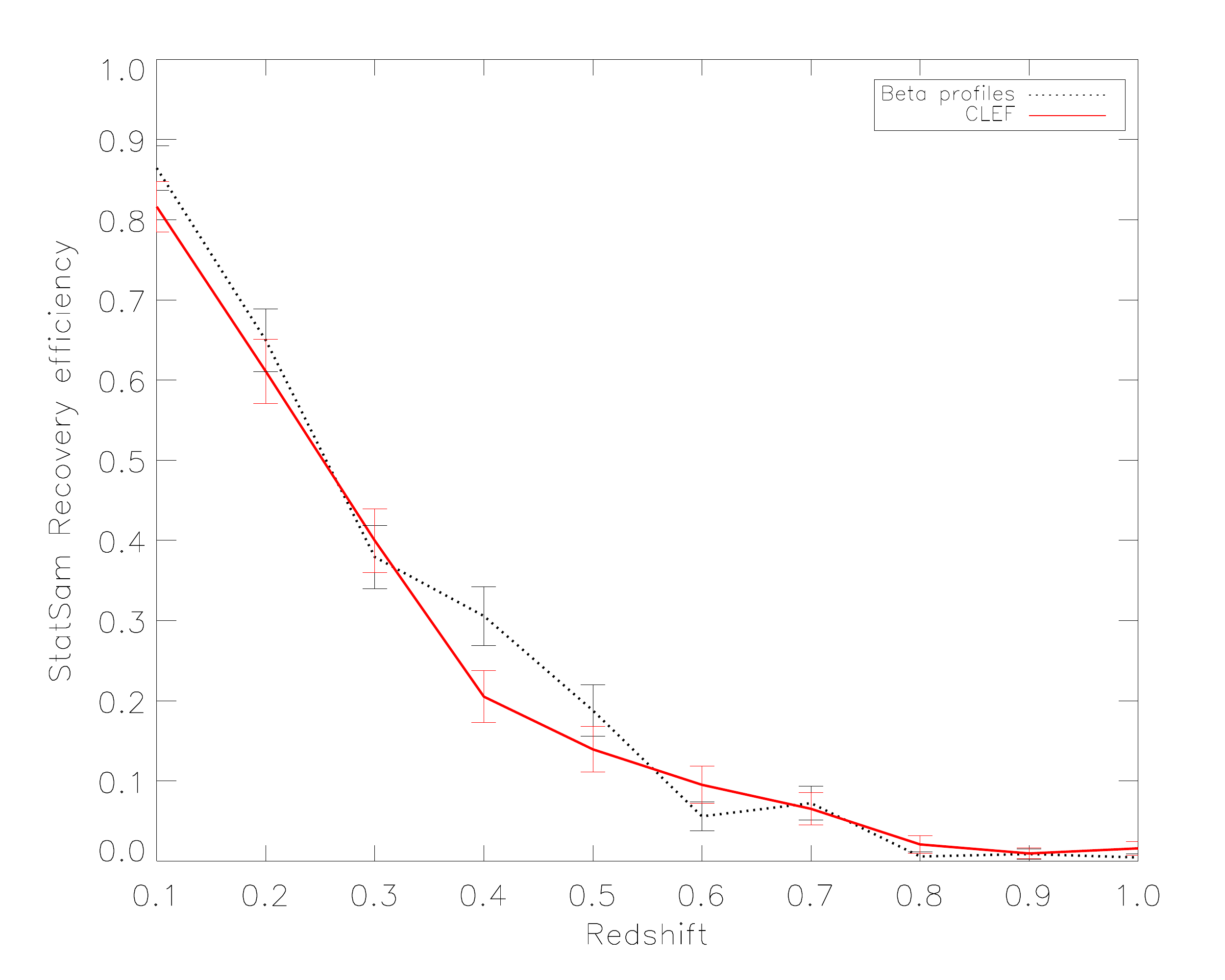}
\caption{Predicted recovery efficiency of CLEF \citep{kay07a} and
$\beta$-profile clusters as a function of redshift. A 500-count cut
has been imposed, where the counts are as measured by the {\sc Xapa}
pipeline. The $\beta$-profile clusters are paired with a CLEF
counterpart, in that they have the same redshift, temperature,
luminosity and location in the respective ObsID.
\label{fig:BetavsCLEF500}}
\end{figure}

The CLEF investigation has further shown that symmetrical $\beta$-profiles are an acceptable approximation to the XCS selection
function. This is important because CLEF, and most other
hydro-dynamical samples, are only available for a single assumed
underlying cosmology. In order to use XCS to measure the underlying
cosmology, we need to know the selection function across a broad range
of cosmological parameters. The suitability of the $\beta$-profiles
was demonstrated by comparing the results of two duplicate selection
function runs. The first used the CLEF cluster profiles, each input
multiple times to determine recovery efficiencies. The second run
replaced the CLEF clusters with isothermal $\beta$-profiles
($r_c=160$kpc, $\beta=2/3$), whilst keeping all other aspects the same
(ObsID, location, luminosity, temperature, redshift and input
cosmology).  The results, after a 500-count detection limit has been
imposed, are shown in Fig.~\ref{fig:BetavsCLEF500}. Plotted in red
is the average recovery efficiency obtained using the CLEF cluster
sample, and over-plotted is the data from using the $\beta$-profiles
(dotted-black line).  

\section{Analysis of XCS Cluster Candidates}
\label{sec:postproc}

In this section we describe a further set of XCS analysis
pipelines. These re-examine the \emph{XMM} observations of the 
candidates delivered by {\sc Xapa}. The first pipeline examines each of the candidates in turn, jointly interrogating the respective ObsID
and NED in the search for published redshifts (Section~\ref{sec:litz}). The
second pipeline carries out batch X-ray spectroscopy on candidates
with redshift measurements and delivers measurements of the X-ray
temperature (Section~\ref{sec:Txmethod}). The third outputs total
luminosities, by fitting spatial profiles to those candidates with
$T_{\rm X}$ measurements (Section~\ref{sec:spatialmethod}). The fourth
pipeline has been designed to estimate redshifts directly from the
X-ray data (Section~\ref{sec:zxmethod}).  The methodology of all of these
pipelines is described below, together with a range of verification
tests. Here, all quantities are calculated assuming a standard
concordance cosmology ($H_0$ = 70 km s$^{-1}$ Mpc$^{-1}$, $\Omega_{\rm
m}$ = 0.27 and $\Omega_\Lambda$ = 0.73) and all quoted luminosities
are bolometric and within radii where the over density is 500 relative
to the critical density ($R_{500}$).

\subsection{Automated NED Queries: Literature Redshifts}
\label{sec:litz}

Redshift measurements are essential if we are to use the 
candidates for science applications. However, with \Niv\, candidates in
our current catalogue (May 1st 2011), we need to find automated
ways to derive as much redshift information as possible. In order to
automatically identify redshifts that are already in the literature,
we have constructed an algorithm that searches the NASA Extragalactic
Database (NED). These `literature redshifts', or $z_{\rm lit}$, are
only available for a small fraction of the candidates, but
they are still extremely important, in that they allow us to check our
other redshift estimation techniques, in particular the X-ray
redshifts described below (Section~\ref{sec:zxmethod}) and the red-sequence
redshifts described in our companion catalogue paper (M11).

The NED search was carried out for all the candidates. 
An initial search extracts all sources, classified as either a galaxy or a cluster, within a
$30$ arcmin search radius of the candidate centroid. Then, for every
extracted object with a catalogued redshift, we calculate a crude
placeholder luminosity, $L_{\rm X, ph}$. The $L_{\rm X, ph}$ is
derived using the gridded LCFs (Section~\ref{sec:ecf_desc}) and the
soft-band {\sc Xapa} count-rate.  From the $L_{\rm X, ph}$, we then estimate
a corresponding placeholder temperature, $T_{\rm X, ph}$, for the
candidate using the $L_{\rm X}-T_{\rm X}$ relation of
\citet{arnaud99a}. From the $T_{\rm X, ph}$ we can then estimate a
placeholder $R_{500, {\rm ph}}$ value ($R_{500}$ is the radius from
the cluster centre that represents an over density of 500 times the
critical density), using the prescription in \citet{arnaud05a}, and a
corresponding redshift appropriate angular search radius,
$\theta_{500, {\rm ph}}$. The velocity dispersion--temperature
relation of \citet{bird95a} is used in a similar way to estimate a
placeholder velocity dispersion $\sigma_{v, {\rm ph}}$ for the
candidate.

Any NED objects that lie outside their respective $\theta_{500, {\rm
ph}}$ are discounted as a true match. If any lie inside, then
those classified in NED as clusters are then considered as a potential
match.  Should there be only one such object, then that is chosen as
the best match. If there is more than one, then the object with the
smallest positional offset is chosen. If no objects classified in NED
as clusters fall inside the search radius, but some galaxies do, we
then look for groupings of galaxy redshifts within the 
($\theta_{500,  {\rm ph}}$, $\sigma_{v, {\rm ph}}$) volume. If more than one
grouping of galaxies is found, then the one with the smallest
positional offset is chosen as the best match. When the query was run
(May 1st 2011), a total of \Nx\, candidates were associated with
published redshifts for clusters (\Nxa) or galaxy groupings (\Nxb) in
NED. 

The NED redshifts were then passed to the `Redshift Followup (Archive)' stage of the 
XCS pipeline (Fig.~\ref{fig:overview_flow}) and individually checked. In doing 
so, it was discovered that some matches are wrong, i.e. the XCS source is not
associated with the selected NED cluster. This is especially true at low NED 
redshifts (where the allowed matching radius is large). We found that imposing 
a redshift limit of $z=0.08$ was effective at removing the erroneous matches, although 
this reduced the number of NED redshifts available to \Nxc. Of these \Nxc\, candidates, 
\Nxd\, passed the quality control stage and made it into the XCS-DR1 sample. That
is not to say that the remainder are not clusters, but rather that they cannot be 
confirmed as being so using the currently available optical and X-ray data (see M11). 
Of these \Nxd\, only \Nxe\, are listed in XCS-DR1 with the automatically selected 
NED redshift. This is because, in the other cases, alternative redshifts were 
available. The alternative redshifts mostly came from either our own observations 
or from our analysis of the  SDSS archive, but in eight cases they came from 
the literature. In those eight cases, the NED redshift listed for the cluster had 
not been updated to reflect more 
recent optical follow-up. The tendency for NED to retain outdated redshift information
first became apparent to us when we compared the NED redshift ($z=1.2$) for the highest 
redshift XCS cluster (XMMXCS J2215.9-1738) to the value we published ($z=1.46$) 
based on 31 secure spectroscopic redshifts \citep{hilton10a}. The $z=1.2$ value had been taken from \citep{olsen08a} and was based on single ($i$) band photometry. 

\subsection{Spectral Fitting: X-ray Temperatures}
\label{sec:Txmethod}

In this section we describe our pipeline to measure X-ray temperatures
for candidates with a secure redshift measurement; the $T_{\rm
  X}$-pipeline hereafter.  In Section~\ref{sec:Xspectra} we explain
how spectra are extracted and corrected for background
contamination. Next we describe how these spectra are fitted to X-ray
models and how parameter uncertainties are calculated
(Section~\ref{sec:spec_fitting}). Both of these tasks are carried out in an
automated fashion so, to assess their efficacy, we have carried out a
series of tests. These tests are described in
Section~\ref{sec:spec_sim}.

\subsubsection{Generating the Spectra}
\label{sec:Xspectra}

Spectra are generated for every candidate with an associated
redshift measurement.  The first step is to establish all the ObsIDs
in which the candidate was observed and all the exposures within
them. We need to do that to ensure we have the maximum number of
source counts available to carry out the fit. In the simplest case,
the candidate will have only been observed in the ObsID listed in the
MDL (Section~\ref{sec:tables}), and there will be only three sub-exposures to
exploit (one each for EPIC-mos1, EPIC-mos2 and EPIC-pn). However, in
other cases the candidate might be covered by multiple ObsIDs (only
the one generating the most soft counts is listed in the MDL). 
Moreover, there can also be multiple exposures within an
ObsID, especially if the exposure time is long and had to be broken up
over several satellite orbits. Finally, in some cases, one or more of
the cameras might have been turned off, so fewer than 3 exposures are
available.

When all the exposures have been gathered then the cleaned event
lists, described earlier (Section~\ref{sec:flares}), are used to generate
spectra. Only photons in the 0.3 keV to 7.9 keV band are used for this
(the telescope is poorly calibrated at softer energies and the spectra
are background dominated at higher energies). The regions used to
extract the source spectra are the ellipsoidal regions, $\epsilon_f$,
that {\sc Xapa} defined for the respective candidate, although, if
other {\sc Xapa} sources overlap with any part of $\epsilon_f$, then events from
those pixels are not included when the spectra are produced.  The
redistribution matrices and area response files necessary for spectral
analysis are then created, using the \emph{XMM} SAS
package. These files are ObsID, camera and position dependent and so
one needs multiple sets for each candidate.

Every source spectrum generated needs an associated field
spectrum for the purposes of background subtraction.  
The background subtraction in the $T_{\rm X}$-pipeline was
done using an in-field method, since XCS clusters do not generally
have large angular sizes.  The background spectra were usually taken
from a circular annulus around the source, although in the case of
sources very near the edge of the field of view, an ellipse
perpendicular to the off-axis direction, with a circular region centred on
the cluster excluded, was used instead. The outer radius of the
background annulus is 1.5 times the {\sc Xapa} defined major axis of the
respective candidate.  The inner edge varies depending on the
exposure, but is no less than 1.05 times the major axis. Any pixels
within the background region that overlap with other {\sc Xapa} sources are
excluded from the background spectrum. The normalisation of the
background is performed within XSPEC and reflects the ratio of the
number of pixels in the source and background extraction regions.

\subsubsection{Spectral fitting}
\label{sec:spec_fitting}

The spectral fitting was carried out using XSPEC. The fitting was done
using the maximum likelihood Cash statistic \citep{cash79a}. As
mentioned above, there can be multiple spectra per candidate and these
were usually all fitted simultaneously.  The only exception were very
low-count spectra, i.e. those with either less than 10 soft-band
counts in total or those with less than 10\% of the soft-band counts
of the spectrum with the most counts. These spectra were excluded from
the simultaneous fit because it was found that they degraded the fits.

In XSPEC the photons within each spectrum are grouped into bins before
fitting.  For the $T_{\rm X}$-pipeline we varied the minimum number of
counts per bin according to the total number of counts in the
spectrum. That way, higher signal-to-noise spectra could be fitted to
higher spectral resolution (and vice versa). For spectra with fewer
than 250 counts, the minimum was set at one count per bin. For spectra
with more than 850 counts, the minimum was five. In between those
limits, the minimum was scaled between 1 and 5 counts using a
power-law with an index of 0.75.  This particular scaling of the
minimum number of counts per bin was chosen after carrying out
spectral simulations. It was designed to minimise the bias in the
derived parameters while also minimising the statistical uncertainties.

Four different models are fitted to the data. All of the models include a
photoelectric absorption component (WABS; \citealt{morrison83a}) to
simulate the $n_{\rm H}$ absorption and a hot plasma component (MEKAL;
\citealt{mewe86a}) to simulate the X-ray emission from the ICM. 
The different models are:
\begin{enumerate}
\item WABS*MEKAL with the hydrogen column, $n_{\rm H}$, frozen at the
  \citet{dickey90a} value and the metallicity, $Z$, frozen at the
  canonical, $0.3Z_{\odot}$, value.
\item WABS*MEKAL but with $n_{\rm H}$ and $Z$ allowed to vary.
\item WABS*(MEKAL+POWERLAW), as (ii), but including an extra
  power-law component to simulate a potential contaminating point
  source.
\item WABS*(MEKAL+MEKAL), as (ii), but with two MEKAL components
  rather than one, in order to simulate the case where there is a
  significant cool core in the cluster.
\end{enumerate}

The best-fitting model of these four is usually used to derive the
luminosity and temperature of the cluster, but if the best-fitting
model does not give sensible parameters, then the next best model will
be selected, and so on. The accepted ranges are $0.3 \, {\rm keV} \, <
T_{\rm X} < 17.0 \, {\rm keV}$ and luminosity less than $
5\times10^{46}$ erg s$^{-1}$.  It is important to note, however, that these
luminosity values are not aperture corrected and only relate to the
luminosity originating from the $\epsilon_f$ region defined by
{\sc Xapa}. In general, a cluster will be more extended than this ellipse,
and so these aperture luminosities, $L_{\rm X, ap}$, need to be
corrected for missing flux using a spatial model. We describe how such
models are fit to XCS candidates in Section~\ref{sec:spatial_models}.

The 68\% uncertainty bounds on the best-fit $L_{\rm X, ap}$ and
$T_{\rm X}$ values are provided as part of the standard XSPEC
fitting process:  the parameter in question is
stepped from its best-fit value until the fit statistic increases by
the amount required for the confidence region needed (at each step
point the other free parameters are refit).  This stepping is done
in both the positive and negative directions to obtain a confidence
region.

\subsubsection{$T_{\rm X}$-Pipeline Validation}
\label{sec:spec_sim}

\begin{figure}
\includegraphics[width=9.5cm]{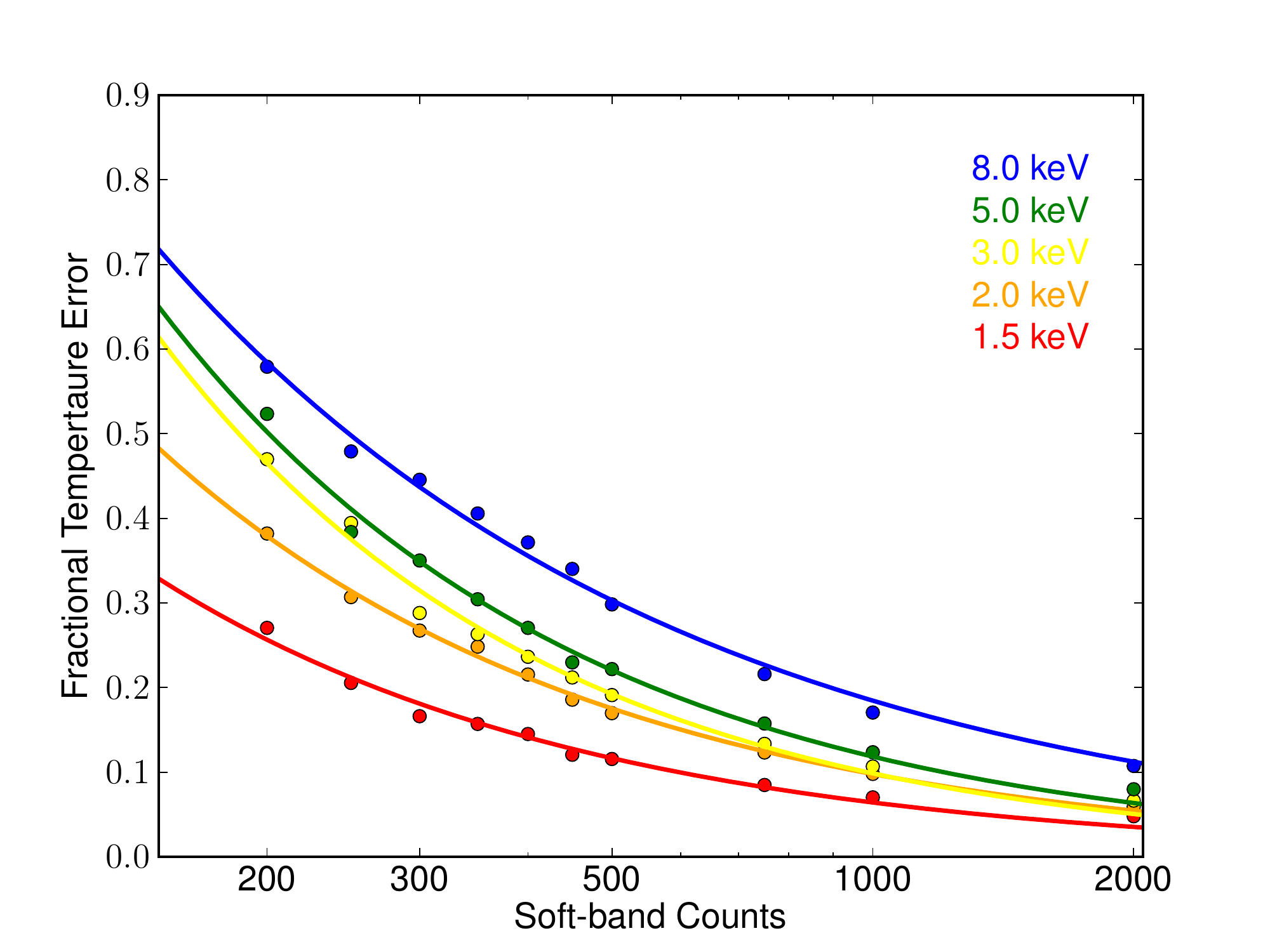}
\caption{Fractional temperature uncertainty as a function of number of
  soft-band counts as a result of fitting simulated $z=0.5$ MEKAL
  spectra with different temperatures, going from cool to hot
  clusters. The red, orange, yellow, green and blue points represent
  spectra of 1.5, 2, 3, 5 \& 8 keV respectively. \label{spec_sim_cts}}
\end{figure}

The spectral pipeline is fully automated and so it is important to
check the reliability of the results it produces before using them for
scientific studies (such as the measurement of cosmological
parameters). We have performed these checks using both XSPEC
simulations and actual data. The results of the first check are
presented in Fig.~\ref{spec_sim_cts}, where temperature uncertainty
is plotted against the number of counts in the fitted spectrum. For
this test we have simulated cluster spectra (all at $z=0.5$), using
the MEKAL model, with a range of temperatures ($1.5 \, {\rm keV} \,
<T_{\rm X}<8\,{\rm keV}$).  It can be seen that it is much easier to
constrain the temperatures of cool systems (red) than it is for the
hottest systems (blue). The constraints also become worse as the
number of source counts decreases. It can be seen that below 300
counts the temperature uncertainties exceed 50 per cent for the 8 keV
systems, though they are considerably smaller than that for lower
temperature systems. This test has informed our decision as to what
count limit we should impose on the candidate list when
defining a sample for cosmological tests. We have set this limit at
300 because we use $T_{\rm X}$ values as a mass proxy when measuring
cosmological parameters \citep{sahlen09a}.

\begin{figure}
\includegraphics[width=9.5cm]{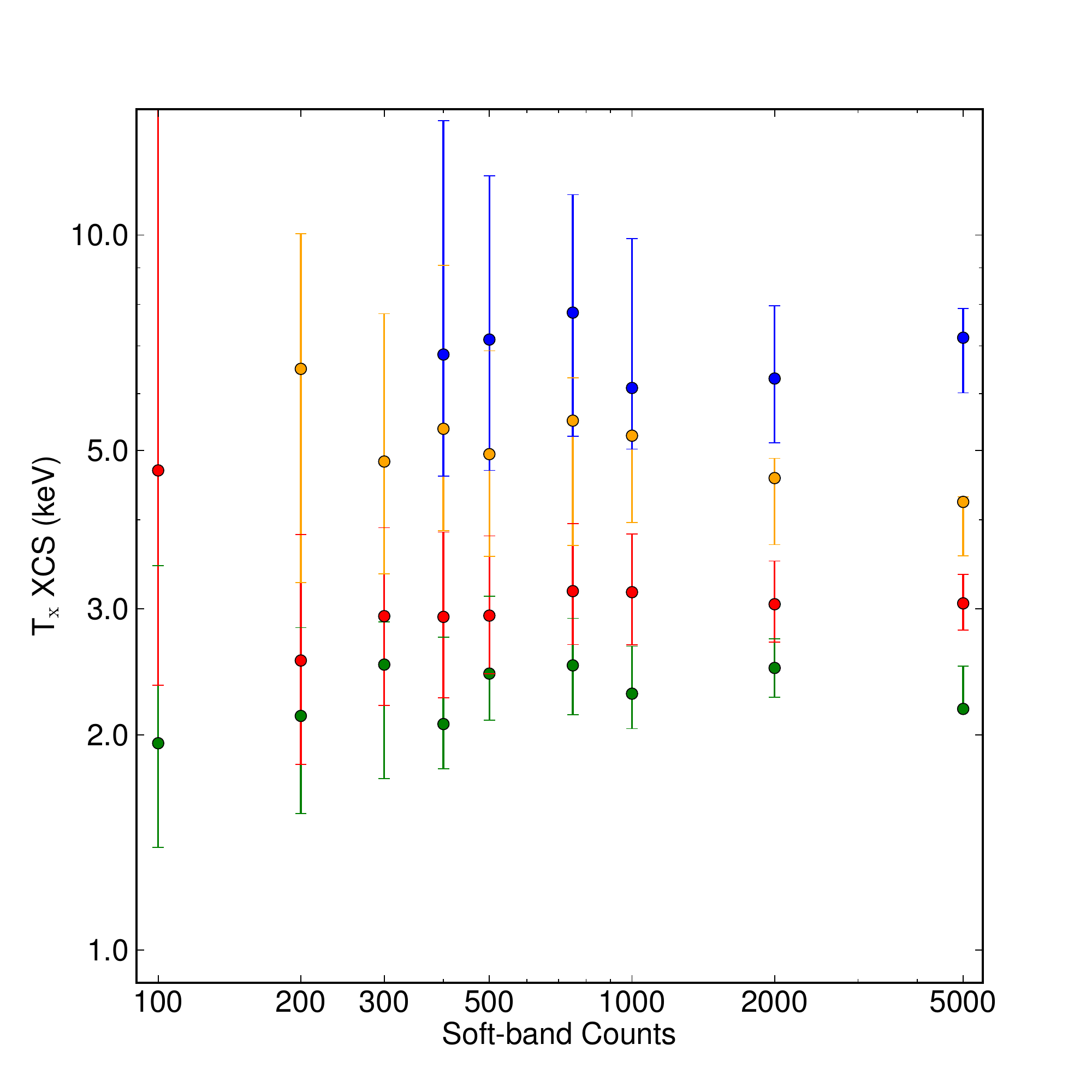}
\caption{The XCS determined X-ray temperatures (and uncertainty) as a
function of the number of counts in the fitted spectrum.  Each colour
represents a cluster that was detected with more than 5000 counts.  For details
of the four clusters used in this plot, see Table \ref{tab:counts}. The
respective exposures were then subdivided to generate lower count
spectra. Note that the higher temperature systems do not yield fits at
the low-count end. The 1-$\sigma$ error bars come from the {XSPEC}
fitting software (see Section~\ref{sec:spec_fitting} for details).
\label{comp_results_counts}}
\end{figure}

The results in Fig.~\ref{spec_sim_cts} were based on model spectra
(albeit with actual \emph{XMM} background contamination) and so should
be seen as a best-case scenario: real clusters do not have a perfectly
isothermal ICM, nor have zero contamination from point
sources. Therefore we have carried out a related test using four real
clusters (Table~\ref{tab:counts}), the results of which can be seen in
Fig.~\ref{comp_results_counts}. Here the best-fit $T_{\rm X}$ value
(and its 1-$\sigma$ uncertainty from XSPEC) are plotted against the
total number of counts in the spectra. This was achieved by
artificially reducing the exposure time of the respective event
file. We note that only one realisation of this proceedure was
performed for each total number of soft-band counts and that the error
bars are the standard XSPEC generated values.  From
Fig.~\ref{comp_results_counts} it is clear that there are no
systematic biases in the derived values of $T_{\rm X}$ as the number
of counts decreases.  The error bars increase, with decreasing counts,
in line with the expectation from Fig.~\ref{spec_sim_cts}. The fit
failed to converge at low counts for the hotter clusters, but in
general Fig.~\ref{comp_results_counts} supports our decision to cut
the candidate list at 300 counts for cosmological studies. This test
also demonstrates that it is still worth fitting candidates with fewer
counts, since we can derive reliable $T_{\rm X}$ values in the galaxy
groups regime down to 100 counts.

\begin{figure}
\includegraphics[width=9.5cm]{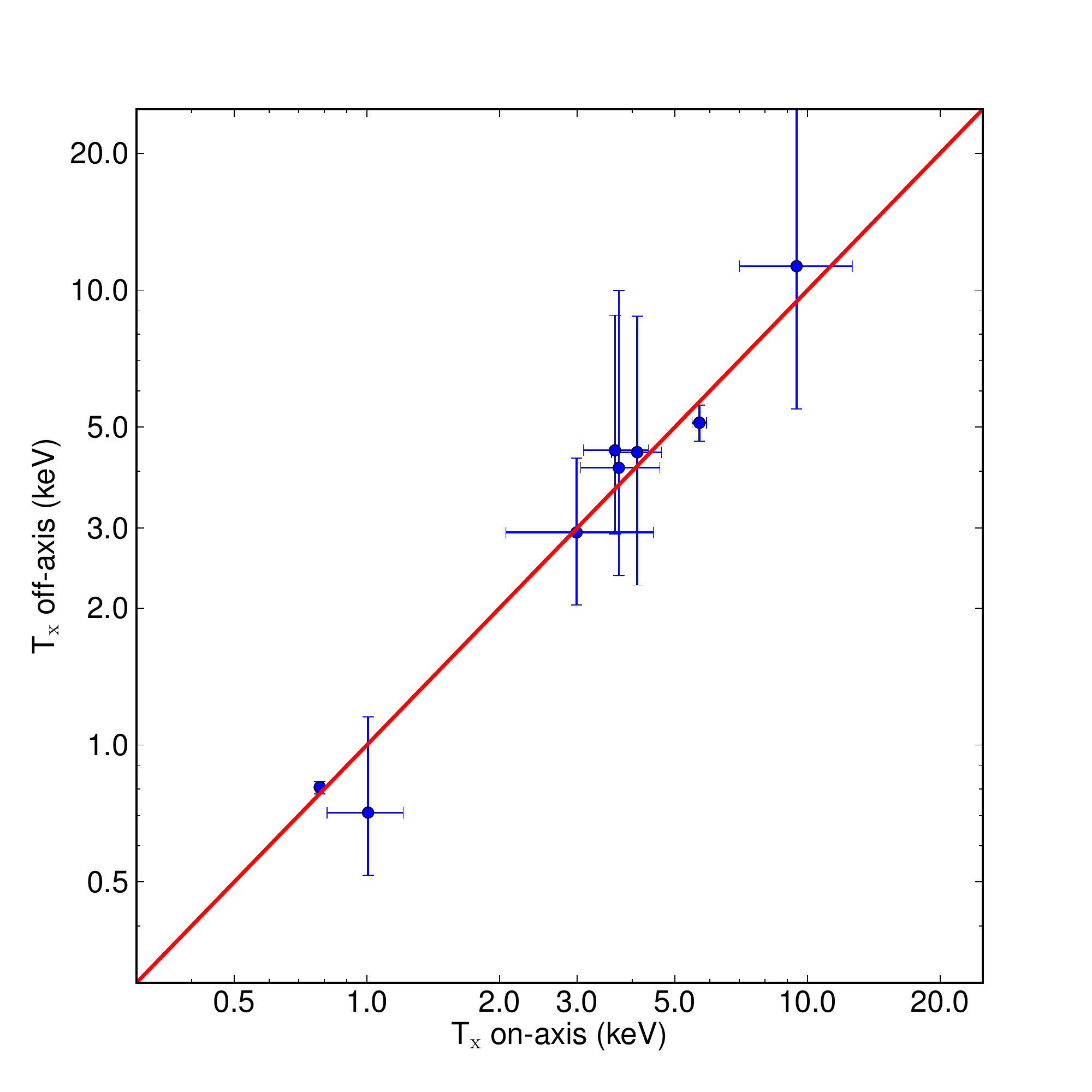}
\caption{Comparison of XCS determined X-ray temperatures when the
cluster is observed off-axis (y-axis) or on-axis (x-axis). For details
of the eight clusters used in this plot, see Table \ref{tab:axis}. The
solid line shows the one-to-one relationship. The error bars are
1-$\sigma$. Both x and y-errors come from XSPEC (see
Section~\ref{sec:spec_fitting} for details).\label{comp_results_angle}}
\end{figure}

We have carried out a test to see if the $T_{\rm X}$-pipeline works at
large off-axis angles, since candidates are located across the \emph{XMM}
field of view.  There are not many clusters to choose from for this
test, but we did identify eight systems that have been observed by \emph{XMM}
both on and off-axis.  For this purpose we define off-axis [on-axis]
to mean a source centroid more than $10$ [less than $3$] arcmin from the
ObsID aim point. The standard XCS spectral reduction was undertaken and
the results can be seen in Fig.~\ref{comp_results_angle}. It can be
seen that the fits to spectra taken off-axis, while in general having larger
uncertainties due to having a lower signal-to-noise, are
consistent with the corresponding on-axis results. We can
therefore be confident that the pipeline produced XCS $T_{\rm X}$
values that are not biased in cases where the objects are located on the
outskirts of the field of view.

\begin{figure}
\includegraphics[width=9.5cm]{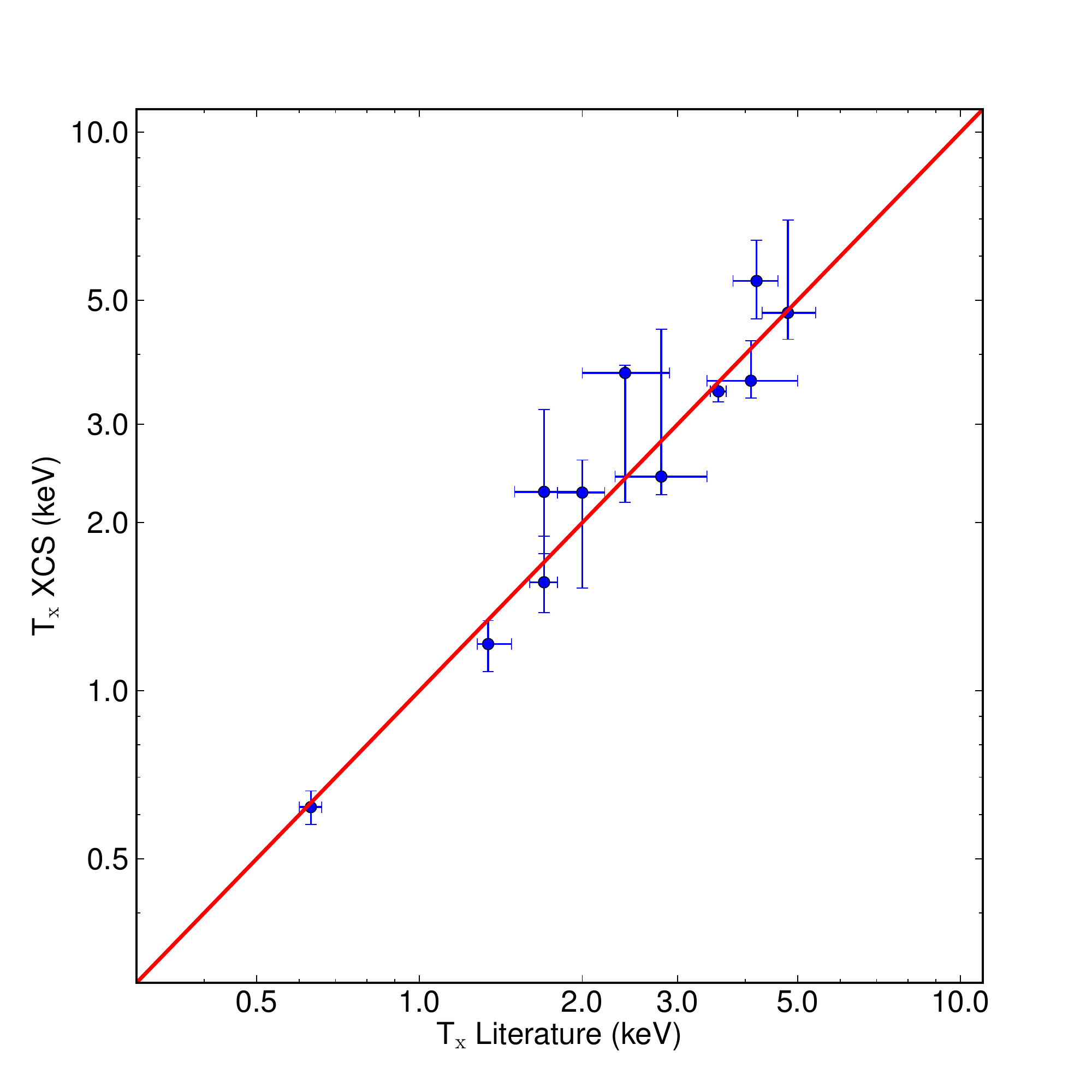}
\caption{Comparison of XCS determined X-ray temperatures with values
  determined by other authors. For details of the eleven clusters used
  in this plot, see Table \ref{tab:lit}. The solid line shows the
  one-to-one relationship.  The error bars are 1-$\sigma$. The
  x-errors are as quoted in the literature. The y-errors come from
  XSPEC (Section~\ref{sec:spec_fitting} for details).\label{comp_lit}}
\end{figure}

The test in Fig.~\ref{comp_results_angle} shows that the XCS
pipeline is internally consistent, but it is also important to compare
XCS parameters to those derived externally, i.e. by other authors,
since they will use different approaches. In particular, most cluster
spectral fitting is done on an object by object basis, with the
background regions and the light curve cleaning being adjusted by
hand. By contrast the XCS pipeline is completely automated because we
do not have the resources to fit hundreds of candidates
individually.  We have therefore tested the quality of the results
from our pipeline using previously published results. We have
constructed a sample of 11 XCS clusters which have previously
published temperatures measured with \emph{XMM}
\citep{pacaud07a,gastaldello07a,hoeft08a}.  The results can be
seen in Fig.~\ref{comp_lit}, where the temperatures derived from the
XCS pipeline are plotted against those measured by other authors. It
can be seen that there does not appear to be any systematic offset
and the XCS temperatures are consistent with the literature
values. This final test demonstrates that the XCS $T_{\rm X}$ values
are reliable and hence suitable for science applications without the
need for a further `hands on' analysis stage.

\subsection{Spatial Fitting: X-ray Luminosities}
\label{sec:spatialmethod}

As mentioned above (Section~\ref{sec:spec_fitting}), the spectral pipeline
produces both luminosity and temperature fits, but the
luminosities, $L_{\rm X, ap}$, are within an aperture and are not
corrected for missing flux. In order to extrapolate the cluster
emission to $R_{500}$, so that the total cluster luminosity can be
calculated, it is necessary to measure the surface brightness
profile.
This is achieved in the XCS spatial pipeline, the $L_{\rm
  X}$-pipeline hereafter, by fitting an analytical function to the cluster
and then using this to extrapolate to $R_{500}$. We decided against
using the alternative, non-parametric approach that produces
de-projected gas densities, e.g.  \citet{croston08a}, because it is
complex and, importantly, does not allow us to extrapolate
fluxes to $R_{500}$.

\subsubsection{Spatial Models}
\label{sec:spatial_models}

Surface brightness fits are performed for every candidate that
passes the spectroscopic pipeline. The main function used to
characterise the shape of the clusters was a simple one-dimensional,
spherically-symmetric, $\beta$-profile model \citep{cavaliere76a}:

\begin{equation}
S(r)=S(0)\left[1+\left(\frac{r}{r_{c}}\right)^{2}\right]^{-3\beta+\frac{1}{2}},
\label{eq:beta}
\end{equation}

\noindent where $r_{c}$ is the core radius and $\beta$ is the density index parameter, which encodes the power-law decline. Three different types of $\beta$-profile models were fitted to the data:
\begin{enumerate}
\item One with $\beta$ frozen at the canonical
value of 2/3. The free parameters are the normalisation and the R.A. and Dec. of the centroid.
\item One as (i), but with $\beta$ allowed to vary. 
\item One with an inner power-law cusp inside a certain parameterised radius (usually of the order of the core radius). This gives us a crude description of clusters with cool cores or AGN contamination. The free parameters are as (ii), plus both an extra normalisation and an extra power law index.
\end{enumerate}

The same background regions were used for the surface brightness
fitting as were used for the spectral fitting. However, in addition to
knowing the total number of counts in the background region, it is
also necessary to know how those counts are distributed. The total \emph{XMM}
background varies considerably across the field of view and so for
extended sources, such as clusters, one cannot assume that the
background counts are divided equally between all the pixels. The
background can be considered as having two components, an `X-ray
  component' that is focused (and so vignetted) by the telescope
mirrors and a  `particle component' that is not. In reality, these
terms are not particularly accurate, since the X-ray component
includes soft protons that are focused by the mirrors and the particle
component includes high-energy photons that are created as the result
of particle collisions with the telescope structure.)

The X-ray and particle components need to be treated separately during
spatial fitting because their spatial variation is different: the
X-ray component is assumed to vary in the same way as the exposure map,
because it is vignetted, whereas the particle component should show no
positional dependence.  To determine the particle background
count-rate per pixel, we have selected, from the respective ObsID, two
or more source regions that are at significantly different off-axis
angles.  We then compare the ratio of the normalised counts within
those regions to the ratio of the same regions in the exposure
map. The difference between those ratios tells us at what level the
counts are contaminated by the particle background. This process is
illustrated in Fig.~\ref{background_flow}.

\subsubsection{Spatial Fitting}
\label{sec:spatial_fits}

As with the spectroscopic pipeline, the fit is performed
simultaneously on all ObsIDs, and sub-exposures, in which the
candidate was observed (barring those with very few counts), see
Section~\ref{sec:Xspectra}. For the spatial fitting we generated new,
4.35 arcsec pixel, image files in the same 0.3 keV to 7.9 keV band as was
used for the $T_{\rm X}$-pipeline.  The three spatial models (see
above) were convolved with the 1-d EAM (Section~\ref{psf_desc})
XMM point-spread function model before the fitting took place.  They
were then multiplied by the exposure map at the respective location,
in order to add observational effects such as vignetting and chip
gaps.  The background was accounted for as described above.

The maximum-likelihood Cash statistic was used for the comparison
between the model and the data. The MINUIT package \citep{james75a} of
minimisation algorithms was used to find the best fitting of the three
models.  The best-fitting model was then used to calculate the scaling
of the luminosity from the spectral extraction region to
$R_{500}$. This was achieved by calculating the ratio of the summed
emission from the spectral extraction region (i.e. $\epsilon_f$) to
that from a circular region, radius $R_{500}$.  The ratio was then
used to scale $L_{\rm X, ap}$ to $L_{\rm X, 500}^{\rm fit}$. This luminosity
scaling value is typically in the range 0.9 to 3.0, depending on the
complex interplay between the cluster size and redshift, the location
on the field of view, and the depth of the exposure.

\begin{figure}
\includegraphics[width=8cm]{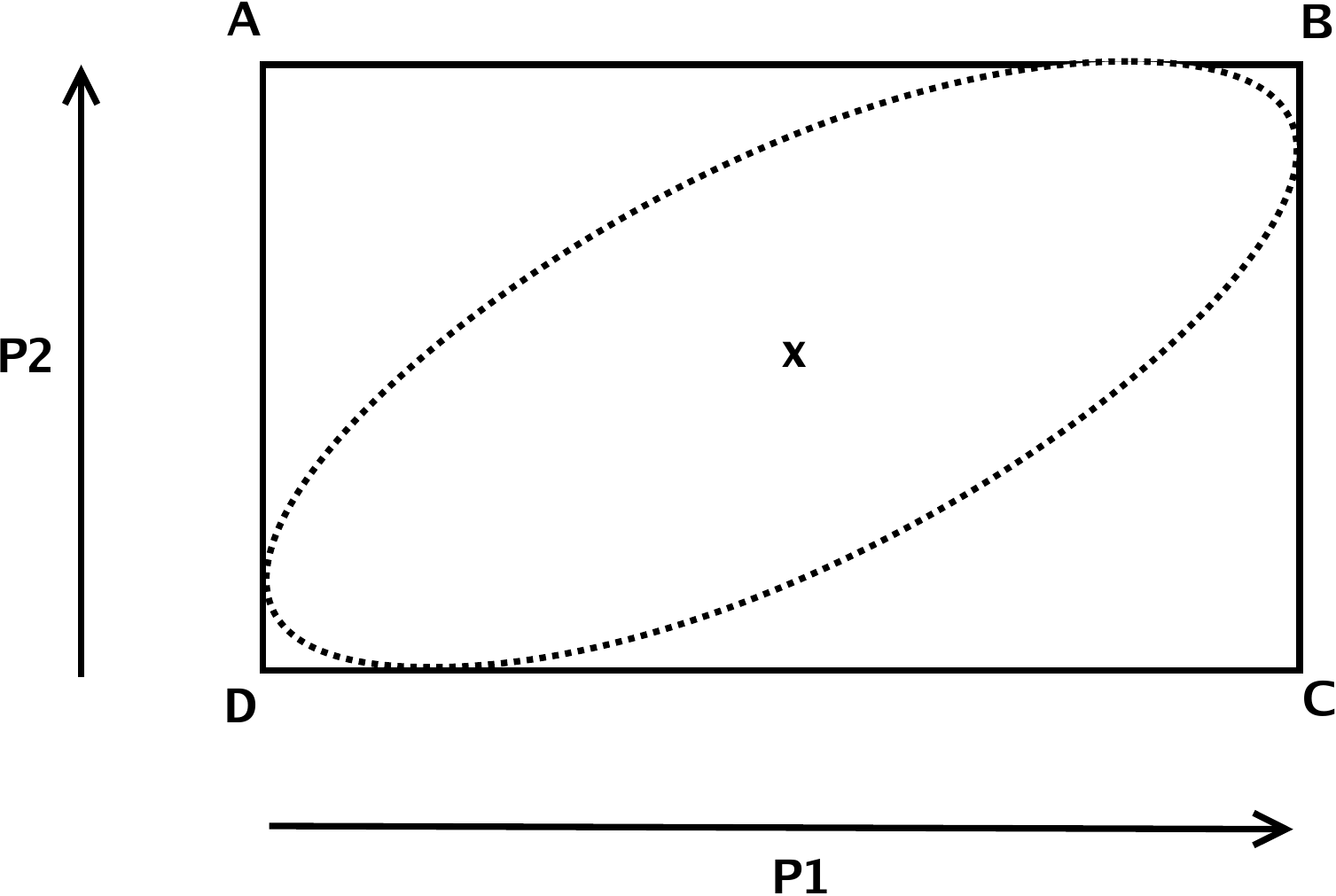}
\caption{Schematic of how the luminosity uncertainties are calculated
by combining the uncertainties on the two quantities (P1,P2). The
cross represents the best-fit point and the dotted line represents the
(unknown) 1-$\sigma$ confidence contour. The model luminosity is
evaluated at A, B, C and D, and the maximum and minimum values are
used as upper and lower uncertainty bounds. \label{errors_dia}}
\end{figure}

The 1-$\sigma$ uncertainty bounds on the free parameters in the
spatial fits were generated in a similar way to that used in the
$T_{\rm X}$-pipeline (Section~\ref{sec:spec_fitting}), i.e. by stepping,
and fixing, the parameter of interest. This was done separately for
each of the three models used in the spatial fitting. The uncertainty
bounds on the $L_{\rm X, 500}^{\rm fit}$ were not so straightforward
to calculate, however. This is because the $L_{\rm X, 500}^{\rm fit}$
calculation involves both the $T_{\rm X}$-pipeline and the $L_{\rm
X}$-pipeline and, since they are performed separately, there is no
information on the correlations between them.  Ideally one would carry
out a simultaneous fitting of a spectral and spatial model to a data
cube (X,Y and energy), as was demonstrated by \citet{lloyd-davies00a},
but this process would be too complex and CPU intensive for the batch
processing required by XCS. Therefore, we adopt the conservative
approach of taking the uncertainty bounds for the two quantities
(i.e. on the luminosity scaling value and on $L_{\rm X, ap}$) and
calculating the luminosities for the four most extreme combinations
(Fig.~\ref{errors_dia}). The highest and lowest luminosities are
then used as the uncertainty bounds of the $L_{\rm X, 500}^{\rm fit}$
measurement, although this will almost certainly be an overestimate.

We note that not all of the candidates that are passed to the spatial
fitting generate acceptable fits. When the spatial fitting fails for a
candidate, we then estimate the luminosity, $L^{\rm est}_{\rm X, 500}$
by extrapolating the $L_{\rm X, ap}$ value assuming a standard cluster
profile. For this, we fix the power law slope to be $\beta=2/3$ and
use a core radius appropriate to the $T_{\rm X}$. Thus all candidates
that pass the spectroscopic pipeline will have either an associated
$L_{\rm X, 500}^{\rm fit}$ measurement or $L^{\rm est}_{\rm X, 500}$
estimate. 

\subsubsection{$L_{\rm X}$-Pipeline Validation}

\begin{figure}
\includegraphics[width=9cm]{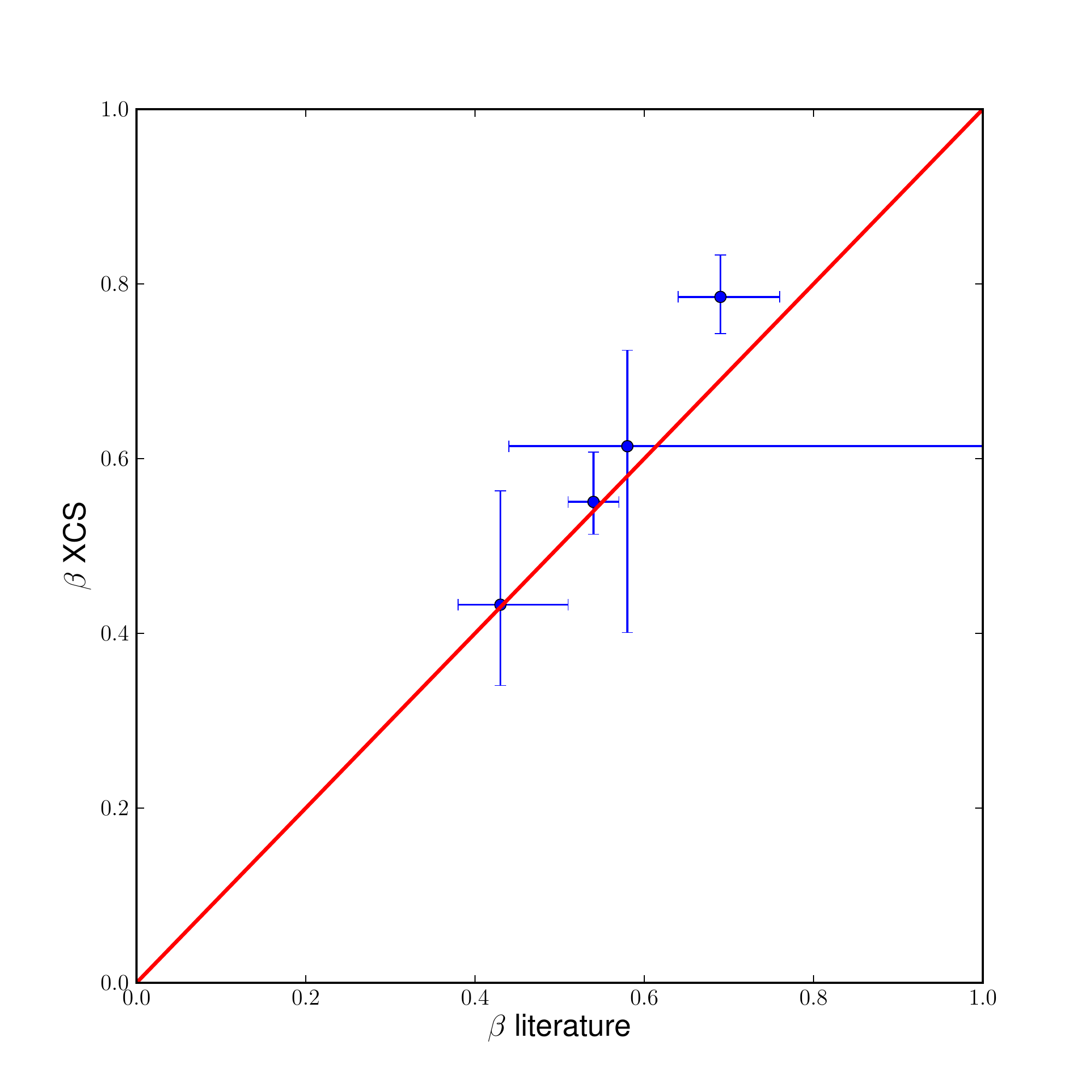}
\caption{Comparison of the XCS determined outer slope (or $\beta$)
  with that derived by other authors. In both cases, the surface
  brightness of the clusters was fit using a circularly-symmetric King
  profile (without a central cusp, see Equation \ref{eq:beta}). For
  details of the four clusters used in this plot, see Table
  \ref{tab:beta}. The solid line shows the one-to-one relationship.
  The error bars are 1-$\sigma$. The x-errors are as quoted in the
  literature. The y-errors come from the XCS fitting
  software (see Section~\ref{sec:spatial_fits} for
  details). \label{comp_beta}}
\end{figure}

The $L_{\rm X}$-pipeline is fully automated and so it is important to
check the reliability of the results it produces before using them for
scientific studies (such as the study of the evolution of the $L_{\rm
X}-T_{\rm X}$ relation). We have done this by comparing the XCS
derived results for $\beta$ and $L_{\rm X, 500}^{\rm fit}$ with those
derived by other authors. First we have examined clusters in common
with the sample of \citet{alshino10a}. This should be a fair
comparison since this sample is a subset of the XMM-LSS
\citep{pacaud07a} and is therefore, like XCS, drawn from an \emph{XMM}
survey.  Fig.~\ref{comp_beta} shows XCS $\beta$ for 4 clusters
plotted against the $\beta$ values taken from \citet{alshino10a}.  It
can be seen that there is no systematic bias between the values.  In
addition, the scatter about the one-to-one relation (solid line) is
consistent with the measurement uncertainties. We can therefore be
reassured that the spatial parameters we obtain from the spatial
pipeline are reliable.

\begin{figure}
\includegraphics[width=9cm]{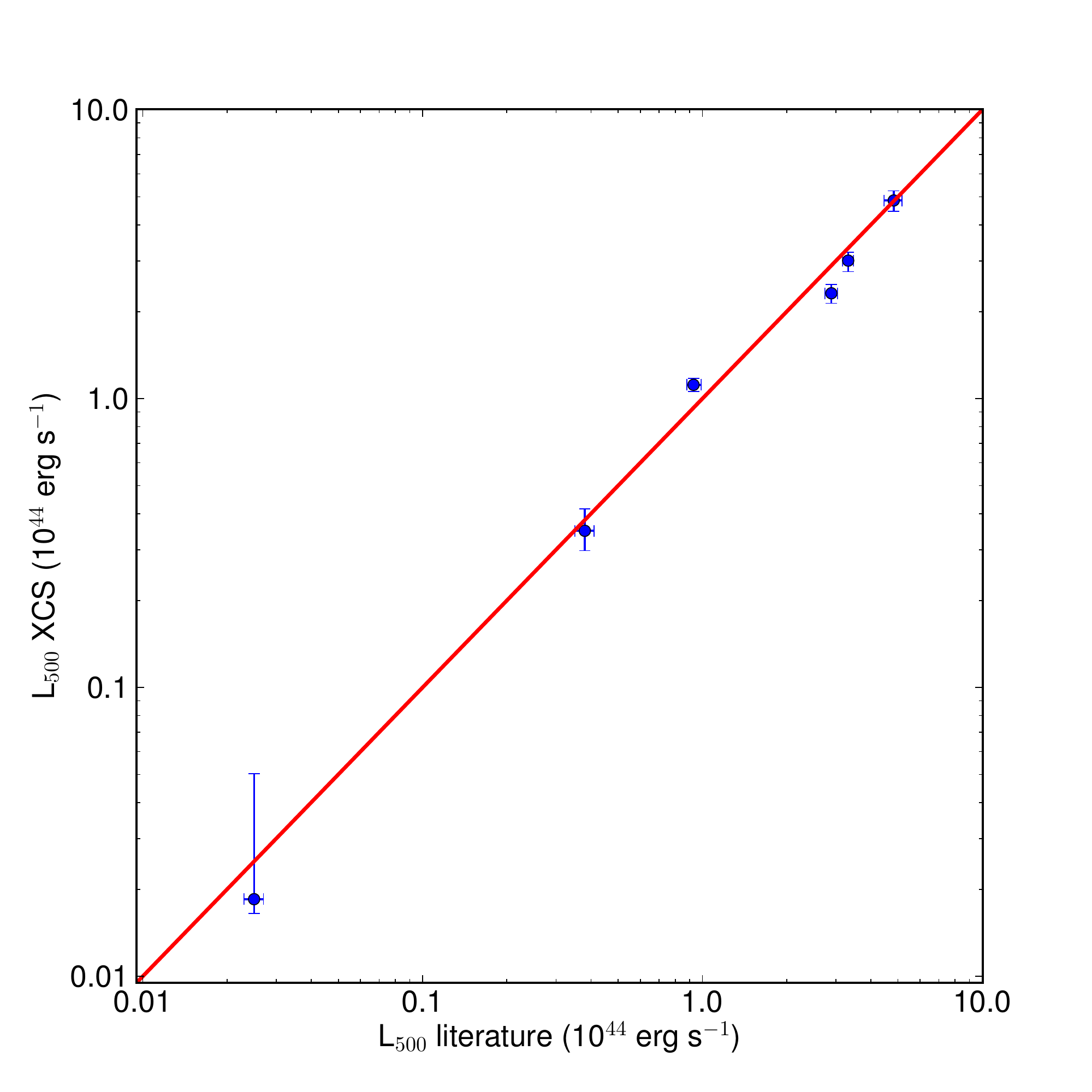}
\caption{Comparison of XCS determined bolometric luminosities within
$R_{500}$ with values determined by other authors. For details of the
six clusters used in this plot, see Table \ref{tab:lum}. The solid line shows the
one-to-one relationship.  The error bars are 1-$\sigma$. The x-errors are
as quoted in the literature. The y-errors
come from the XCS fitting software (see Section~\ref{sec:spatial_fits} for
details).
\label{comp_lum}}
\end{figure}

We have also compared the XCS $L_{\rm X, 500}^{\rm fit}$ values against
published values obtained from clusters with 300 or more counts in the
XMM-LSS sample \citet[][note that this is 300 XMM-LSS, not {\sc Xapa},
  counts]{pacaud07a}.  The XCS $L_{\rm X, 500}^{\rm fit}$ values are plotted
against those of \citet{pacaud07a} in Fig.~\ref{comp_lum}. It can be
seen that they closely follow the one-to-one relation (solid
line). This test demonstrates that the XCS $L_{\rm X, 500}^{\rm fit}$ values are
reliable and hence suitable for science applications without the need
for a further Òhands onÓ analysis stage.

\subsection{Spectral Fitting: X-ray redshifts}
\label{sec:zxmethod}

We cannot exploit the thousands of candidates that {\sc Xapa}
has produced without first determining their redshifts. As mentioned
above (Section~\ref{sec:litz}), only a small fraction have redshifts
available from the literature, so we have carried out both an
intensive optical follow-up campaign, and exploited the SDSS archive, 
to gather more redshifts. This effort has yielded redshift information for 
\Nxvi\, additional candidates to date (M11), but the redshift follow-up of the XCS is still far from
complete. We therefore decided to use the \emph{XMM} data itself to
constrain candidate redshifts. This process, of measuring `X-ray
redshifts' or $z_{\rm X}$, has been demonstrated by several authors
for individual clusters \citep{hashimoto04a, werner07a, lamar08a,
rosati09a} and recently on a sample of \emph{Chandra} clusters by
\citet{yu11a}, and has even been used to study bulk motions of the gas
within the bright, nearby clusters \citep{dupke01a,dupke01b}, but has
never been used on the industrial scale we need for XCS. In the
following we describe the X-ray redshift pipeline, $z_{\rm
X}$-pipeline hereafter, and its verification using XCS clusters with
known redshifts. 

\subsubsection{Generating and Fitting the Spectra}

Similar to the $T_{\rm X}$-pipeline (Section~\ref{sec:Txmethod}), all
exposures that overlapped with a particular candidate were used and a 
simultaneous fit was carried across all the respective spectra. Because this pipeline will be run
on the many thousands of candidates that {\sc Xapa} produces, we needed to
keep the processing time per candidate to a minimum.  We therefore
chose a single-temperature MEKAL model, convolved with a photoelectric
absorption model. Moreover, during the fitting, only the spectral
normalisation was left free.  By design, we do not want to assume
the redshift, so we ran a series of fits stepping from $z=0.01$ to
$z=2$, in steps of 0.01. At all of these steps, the metallicity was
fixed at Z=0.3$\times$Solar and $n_{\rm H}$ at the \citet{dickey90a}
value. The $T_{\rm X}$ was not free either, but rather calculated (via
the \citealt{arnaud99a} $L_{\rm X}-T_{\rm X}$ relation) from the
best-fit normalisation at that redshift step (assuming no scatter in
the $L_{\rm X}-T_{\rm X}$ relation).

At each redshift step, the Cash statistic was recorded, as
demonstrated in Fig.~\ref{example_xrz}. The $z_{\rm X}$ for the
candidates is then chosen by searching for minima in the distribution
of Cash statistic values.  Usually the redshift corresponding to the
lowest Cash statistic was used, but if the corresponding temperature
was $T_{\rm X}>8$ keV then the next deepest minimum was
chosen, and so on.  This limit was placed on the allowed temperature
because very few $T_{\rm X}>8$ keV are expected to be detected by XCS
\citep{sahlen09a}. The 1-$\sigma$ uncertainty on $z_{\rm X}$, $\sigma_{z_{\rm
    X}}$, was also determined from the Cash statistic distribution. We
note that in the following we refer to statistical uncertainties
expressed as a percentage, and by this we mean
$100\times \sigma_{z_{\rm X}}/z_{\rm X}$. 

\begin{figure}
\includegraphics[width=8.5cm]{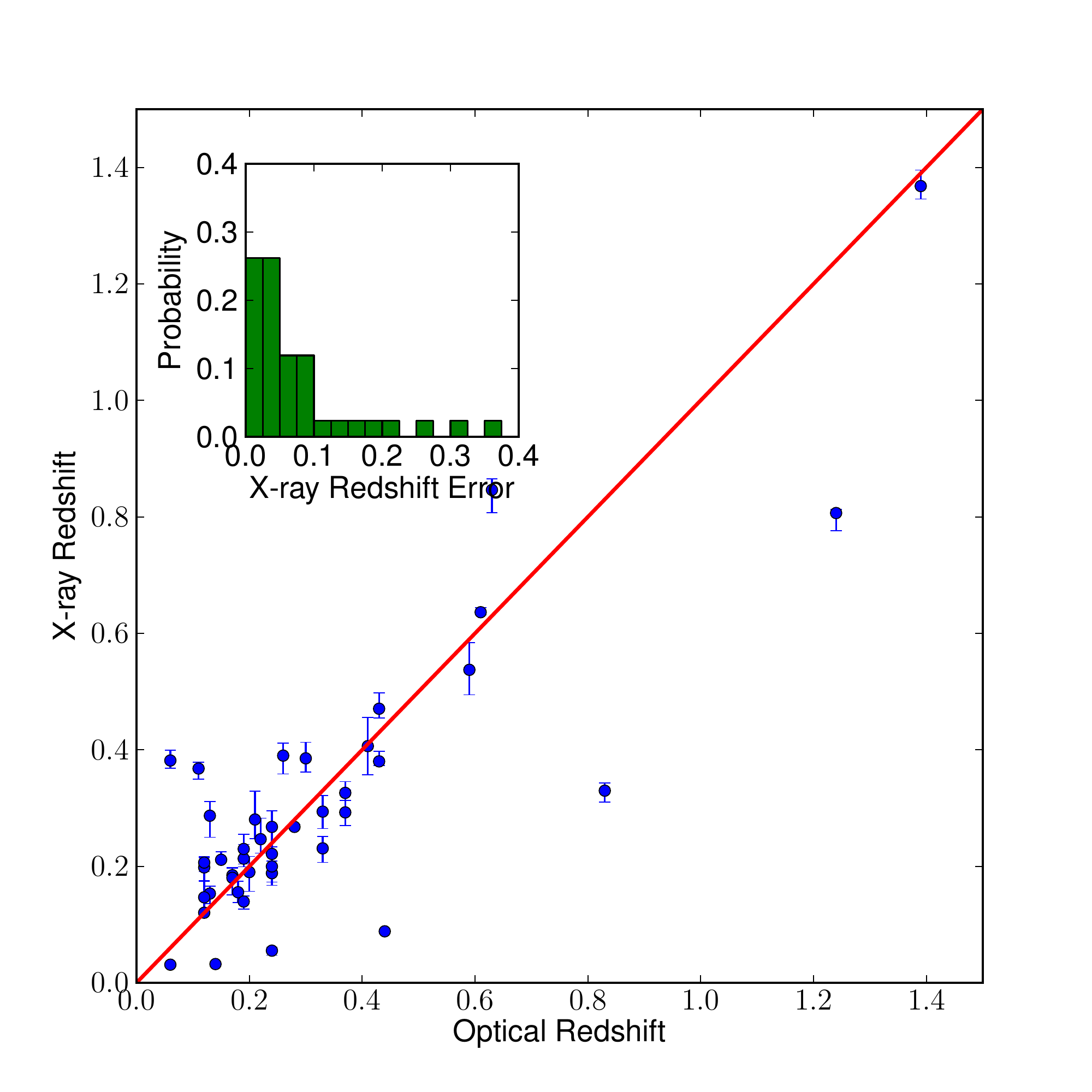}
\caption{Measured X-ray redshifts plotted against optically determined
  redshifts for clusters in XCS-DR1 (M11). The solid line shows the
  one-to-one relationship. Only X-ray redshifts with statistical uncertainties of 20 per cent or
  less are shown.  The insert shows a histogram of the difference between the X-ray
  redshifts and optically determined redshifts. \label{comp_z}}
\end{figure}

\subsubsection{$z_{\rm X}$-Pipeline Validation}
\label{sec:zxvalid}

X-ray redshift measurements have not been attempted with this level of
automation before. We checked our results to see if our $z_{\rm X}$
values are suitable for science applications. We did this using clusters
in XCS-DR1 that had optically determined redshifts  Under the assumption that the
optical redshift is correct, we have compared the $z_{\rm X}$ to the $z_{\rm true}$ for
\Nxviii\, XCS-DR1 clusters where the $z_{\rm X}$ fit yields a statistical uncertainty of
$<20$\% and $\sigma_{z_{\rm X}}<0.05$ (Fig.~\ref{comp_z}). We chose
$\sigma_{z_{\rm X}}=0.05$ as the upper limit for this comparison
because this is the typical error on the single colour photometric redshifts 
presented in M11. As shown in Fig.~\ref{comp_z}, the $z_{\rm X}$ fits are 
usually good (to within the errors), but the level of catastrophic failures, 
i.e. where $z_{\rm  true} - z_{\rm X}=\Delta_z \gg \sigma_{z_{\rm X}}$, is high. 
The failure rate for X-ray redshifts is 24 per cent, compared to $\leq 7$ per 
cent for the photometric redshifts presented in M11.  
 
These catastrophic failures are not unexpected, even for high signal-to-noise spectra, 
when the single-temperature spectral model is too
simple, e.g. if there is AGN contamination or a cool core. Similarly,
if the $n_{\rm H}$ and/or abundance was fixed at the wrong value, or
the cluster is not close enough to the \citet{arnaud99a} $L_{\rm
  X}-T_{\rm X}$ relation, then one might obtain $z_{\rm X}$ values
with small errors, but that are not physically realistic.  We have
investigated the possibility of making cuts on the sample to
objectively weed out the catastrophic failures. However, it does not
seem to be possible to predict, a priori, that a given $z_{\rm X}$
estimate would be unreliable from any combination of number of counts,
cluster temperature or $n_{\rm H}$.  We discuss in
Section~\ref{sec:discussion} how we plan to use the $z_{\rm X}$ values, 
despite their high failure rate.

\section{Discussion}
\label{sec:discussion}

\defcitealias{romer01a}{R01}

In Fig.~\ref{fig:overview_flow} we introduced the complex
methodology associated with the generation of a cluster catalogue
based on serendipitous detections in the \emph{XMM} archive. We went
on to describe, and verify, all the steps in the methodology that
involve \emph{XMM} data (other steps are described in our companion
paper, M11).  In this section we will discuss each
step again, making reference where appropriate to predictions made in
our pre-launch paper \citet[][R01 hereafter]{romer01a}, and in our
cosmology forecasting paper \citep{sahlen09a}. We also highlight areas
for improvement.

In Section~\ref{sec:downloads} we described the download of data from
the archive and showed how the area covered by the archive has grown
over the last 10 years. By now there are over 600 deg$^2$ of the sky
covered by {\emph XMM}, and 51~deg$^2$, 276~deg$^2$ and 410~deg$^2$
(at $>40$~ks, $>10$~ks and $>0$~ks depths respectively) are in regions
suitable for cluster searching. We note that the exposure times used
in Fig.~\ref{fig:area-time} are after flare correction (Section~\ref
{sec:flares}), and that flares typically affect one-third of the exposure. 
The rate of addition of new area is slowing over
time, reflecting the trend towards repeated observations and fewer,
but longer, exposures. It is, therefore, almost certain that XCS will
not reach the 800 deg$^2$ target set in \citetalias{romer01a}. The
revised target of 500 deg$^2$ target set in \citet{sahlen09a} does
remain achievable though (as long as no minimum exposure time cut
is applied).

The distribution and average (requested) exposure times of
ObsIDs in the public archive is close to what was anticipated in
\citetalias{romer01a}, but due to the unanticipated need for flare
correction, the average usable exposure is only 13~ks (compared to a
requested average of 20~ks and a predicted, in \citetalias{romer01a},
average of 22~ks). These decreases, in exposure time and areal
coverage, will certainly impact the size of the final XCS cluster
catalogue. However, we have been able to use a lower minimum
acceptable source significance (4-$\sigma$ rather than the 8-$\sigma$
used in \citetalias{romer01a}) -- because the {\sc Xapa} extent
determination is more effective at low signal-to-noise than expected
from previous experience with \emph{ROSAT} -- and this will help to
keep the cluster numbers up.

In Sections~\ref{sec:reduction} and \ref{sec:images} we described the
reduction of the downloaded data, including mitigation of time periods
affected by background flares, and the production of images. This was
done in a fairly standard way, albeit on a much larger, and more
automated, scale than is typical. In \citetalias{romer01a} we expected
that XCS source detection would be carried out only in EPIC-pn images, 
because we assumed that it would be too complicated to carry out
selection function tests on merged images. However, in practice we
have been able to run source searching and selection functions on
merged images without any difficulty (Section~\ref{sec:sf_descr}). This has
helped compensate for the decreased sensitivity, and increased
background levels, of the EPIC-pn CCDs compared to pre-launch
predictions. One thing that was not anticipated in \citetalias
{romer01a} was the need to create mask files by hand for about a third
of the ObsIDs (Section~\ref{sec:masks}). This tedious process has
been carried out by student volunteers and has not actually held up
the processing of the archive significantly. 

Overall, we are satisfied 
by the performance of the procedures described in Section~\ref{sec:xmmdata}
and do not plan any major modifications in future. That said, we did uncover 
during the quality control stage that some of the masks were too small 
and also that a small fraction of the reduced image had an atypically high 
background (see M11). These two factors have resulted in contamination in 
the candidate list at the $\simeq 7$ per cent level. To avoid such 
contamination in future, we have improved the way that eye-ball checks 
of reduced images will be carried out.

In Section~\ref{sec:xapa} we described the generation of the XCS
source catalogue using the XCS Automated Pipeline Algorithm ({\sc Xapa}),
and the tests we have carried out to demonstrate its efficacy. In
Section~\ref{sec:wavedetect}, we explained how {\sc Xapa} applies
multi-scale wavelets to generate a source list per ObsID, and
discussed some of the successes of {\sc Xapa}, including the ability to
detect sources over a wide range of sizes and signal to noise: only
very rarely does one look at an image of an ObsID, with {\sc Xapa} ellipses
overlaid, and see real sources that have been missed or artefacts
(e.g. chip edges, where there can be discontinuities in the background
level) misidentified as sources. We are especially pleased with the
{\sc Xapa} vision model (Section~\ref{sec:xapa_vision}) because of its ability to
detect sources within sources and to fit source ellipses. During the
development of {\sc Xapa}, it was found that vagaries of the \emph{XMM}
optics could result in false source detections (e.g. when point
sources had extended lobes, due to the complex off-axis PSF), or
incorrect size measurements (e.g. when an extended source had a cuspy
core). These issues were addressed with additional
sub-algorithms. 

As shown in Section~\ref{sec:points}, the parameterisation of point
sources (fluxes, positions etc.)  is very good. Not only does this
give us confidence that the extended source centroids are suitable for
cluster searching, it also demonstrates that the point source
catalogue itself can be used for science applications. XCS members and
collaborators are using the data products in the point source
catalogue in a variety of ways, including a study of the evolution of
quasar X-ray spectra and a search for X-ray Dim Isolated Neutron Stars
(or XDINS, see \citealt{haberl07a} for a review of XDINS).

We do have some concerns, however, about the parameterisation of
extent by {\sc Xapa} (Section~\ref{sec:extent}) because the available PSF models
are known to have deficiencies, especially off-axis. Occasionally,
during the quality control stage (M11), we see sources
that are obviously (from the X-ray data themselves and/or from the
related optical image) point-like but that have been classified as
extended and erroneously entered into the candidate
list. Likewise there are likely to be incidences of extended sources
that are detected but falsely classified as point-like (or flagged as
PSF-like). The latter effect was indicated by the selection function
test using numerically generated synthetic clusters (Section~\ref{sec:sf_clef}; see below).  For
these reasons we plan to adapt {\sc Xapa} to use a new 2-d PSF model that is currently 
under development by \citet{read10a}. The implementation of this new 2-d PSF 
model would be a major undertaking because it would require the extent 
determination sub-algorithm of {\sc Xapa} to be rewritten and also
necessitate the recalculation of the survey selection functions.  It
is also worth pointing out that even a perfect PSF model cannot
prevent very nearby (on the sky) sources from becoming blended into a
single source, especially when the signal-to-noise is low. These
blends will always affect our candidate list at some level (as
they will any cluster searching project based on \emph{XMM}
detections); some will be obvious from the optical follow-up (see M11 for examples), 
but some might well require higher resolution imaging, e.g. from
\emph{Chandra}, to be identified.

The collation of a Master Detection List (MDL) for the survey
(Section~\ref{sec:tables}) has been fairly straightforward, despite the
fact that so many ObsIDs overlap (and so many sources are detected
multiple times in the archive). However, we have found that the
process by which duplicate extended sources are identified (via a
fixed matching radius of 30 arcsec) does not always work at low redshifts
($z\la 0.2$), so we are in the process of improving this aspect of
{\sc Xapa}. As of May 1st 2010, the MDL contained \Nvii\, point sources
and \Nviii\, extended sources, although, as just noted, a small number
of the extended sources will be duplicate entries. We have selected
\Niv\, of the extended sources as cluster candidates, after making a
series of cuts to the extended source list, and these have then been
passed onto the post processing and optical follow-up steps described
in Section~\ref{sec:postproc} and in M11.

We stress that the MDL, and hence the candidate list, was derived from the analysis 
of individual ObsIDs, even in regions where different ObsIDs overlap. In fact, approximately 
40 per cent of ObsIDs in the XMM public archive have significant overlap with 
other ObsIDs, with a median additional exposure time of 70 percent. 
Therefore, it would be possible to increase the number of sources, 
and hence candidates, detected by {\sc Xapa} using co-adding ObsIDs.
However, this would require a major overhaul of both {\sc Xapa} and 
the selection function methodology (in the latter case because the 
point spread function would be significantly more complicated), and we have no 
plans to use co-added ObsIDs in XCS. That said, we do take advantage of 
multiple exposure when running the $T_{\rm  X}$ and $L_{\rm  X}$-pipelines.

We note that the XCS is not the largest compilation of \emph{XMM}
detections; the XMM-Newton Serendipitous Survey 2XMM catalogue
\citep{watson09a} contains 191,870 sources discovered in
3,491 \emph{XMM} ObsIDs. We make comparisons between XCS 
and 2XMM point and extended sources in Sections~\ref{sec:points} 
and \ref{sec:2XMMext}, finding them to be in good
agreement. We have compared XCS to another sample of \emph{XMM}
selected clusters (the XMM-LSS, Section~\ref{sec:LSS}), and find them to be
in good agreement also: only four of the 33 XMM-LSS `class 1' extended
sources did not make it into the candidate list (because
they did not meet the extent and/or signal-to-noise criteria).

Selection functions are very important to any cluster survey that
plans to carry out statistical studies, such as the measurement of
scaling relations or cosmological parameters. The XCS selection
functions need to describe survey completeness as a function of a wide
number of parameters, and are thus very CPU intensive.  Examples of our
selection function work so far are given in Sections~\ref{sec:sf_beta}
and \ref{sec:sf_clef}. We have demonstrated, using simple analytical
models for the ICM distribution, that XCS can detect typical (for the
local $L_{\rm X}-T_{\rm X}$ relation), 3 and 6 keV clusters to high
redshifts, but that the percentage recovery of the cooler
(i.e. fainter) clusters drops off rapidly, e.g. from roughly 60\% at
$z=0.6$ to 10\% at $z=0.9$ for 3 keV clusters
(Fig.~\ref{fig:betafun13}).  These predictions of the selection
function redshift dependence were based on the assumption that all
clusters have core radii of $r_c=160$ kpc, so we have also
investigated our sensitivity to smaller and larger clusters
(Fig.~\ref{fig:betafun2}).  We found that for clusters with more
than 300 counts, the cluster recovery rate is good ($\geq$70\%) when
the extent is in the range $\simeq 10 - 20$ arcsec. These limits roughly
translate to $0.1<z<0.6$ for $r_c=50$ kpc and $z>0.3$ for, more
typical, $r_c=160$ kpc.  XCS is not as sensitive to clusters with core
radii at the top end of the \cite{jones84a} range; roughly only 20\%
[40\%] of 300-count clusters with $r_c=400$ kpc are recovered at
$z>0.3$ [$z>1$], although this rises to 60\% [75\%] for 1000-count
clusters. This insensitivity was not anticipated in
\citetalias{romer01a} (i.e.  before we had access to realistic
selection functions); we claimed therein that all clusters with core
radii larger than $20$ arcsec would be flagged as extended sources. It may
seem counter intuitive that clusters of larger angular extent are
harder to recover, but this is due to two factors; first, more
extended clusters have lower surface brightnesses and correspondingly
lower contrast against the background, making them harder to detect,
and second, our wavelet scales were chosen with more compact clusters
($<250$kpc, \citetalias{romer01a}) in mind.

In Section~\ref {sec:sf_clef} we have used numerically-generated
`clusters', from the CLEF hydrodynamical simulation of
\citet{kay07a}, to investigate whether factors such as cool cores
(which result in luminosity enhancements at small radii), or recent
merging activity, might impact the ability of XCS to detect
clusters. We found that the numerical `clusters' are easier to
detect than the analytical $\beta$-profile `clusters', but that they
are more likely to be misclassified (as point-like) when they contain
cool cores. This effect is reduced as the number of `source' counts
increases, and above 500 counts no longer occurs
(Fig.~\ref{fig:BetavsCLEF500}). This test justifies the use of
selection functions based on simple analytical cluster profiles in the
XCS cosmology forecasting paper \citep[][which was based on a minimum
count threshold of 500]{sahlen09a}.  This is important because CLEF,
and most other hydrodynamical samples, are only available for a single
assumed underlying cosmology and, in order to use XCS to measure the
underlying cosmology, we need to know the selection function across a
range of cosmological parameters.  However, this test further suggests
that it may not be appropriate to use only simple analytical profiles
when establishing selection functions for a minimum count threshold of
300 (which we have determined is the limit to which we can expect to
recover reliable $T_{\rm X}$ measurements, see below), and this is
something we plan to investigate further.

\begin{figure*}
\includegraphics[width=8.5cm]{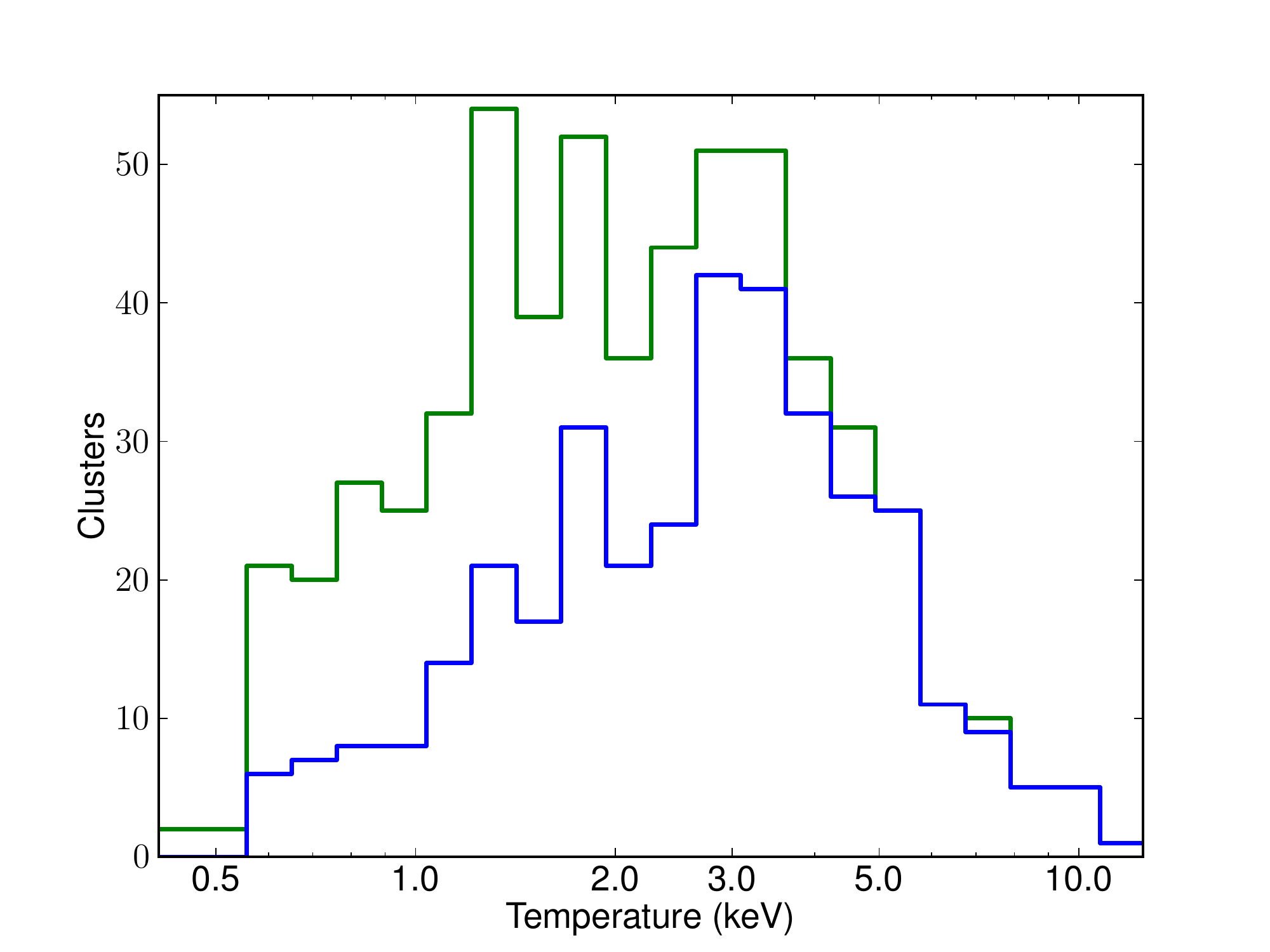}
\includegraphics[width=8.5cm]{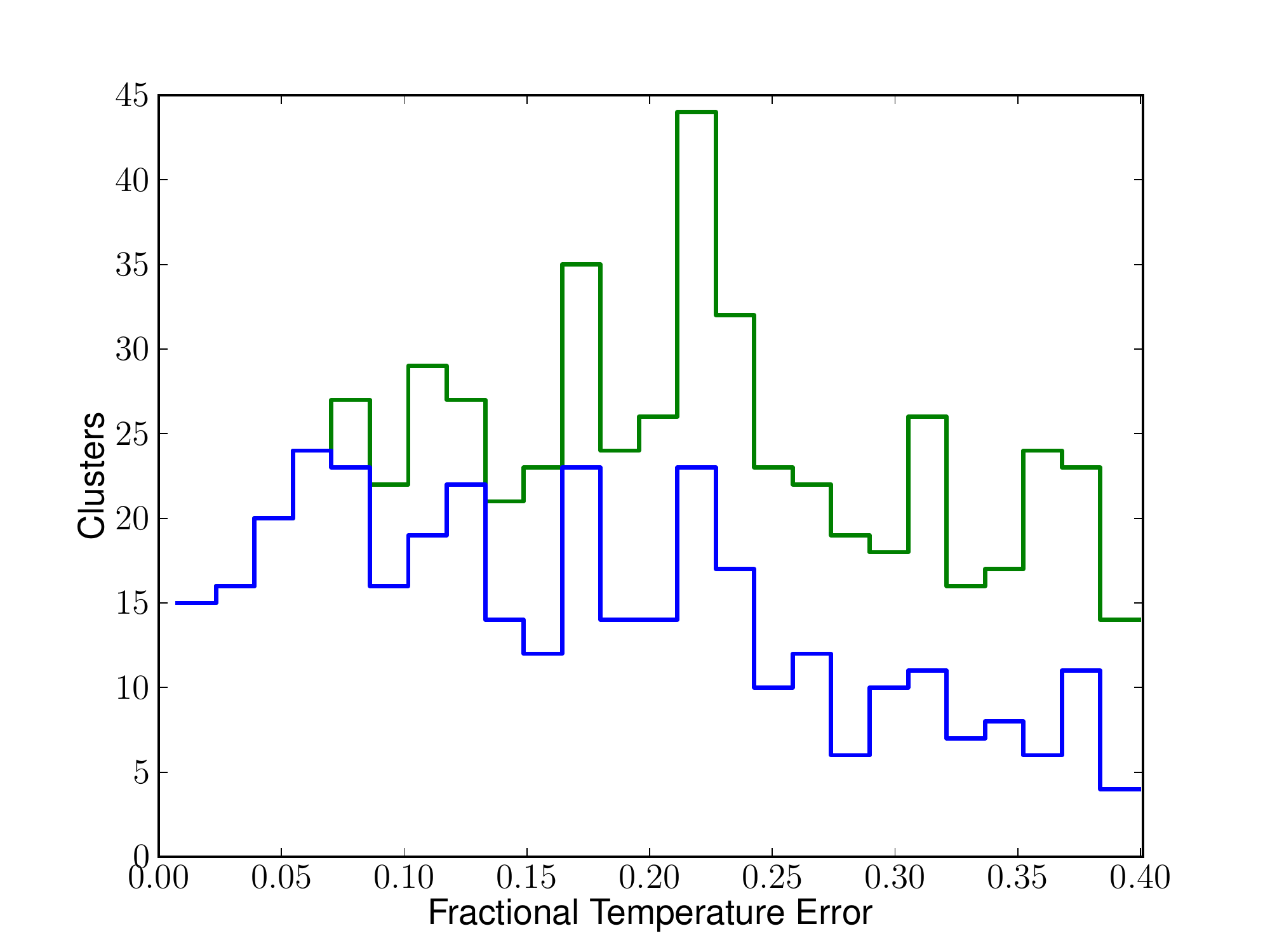}
\caption{Number of clusters with less than 40 per cent temperature
errors and with more than 300 soft-band counts (blue) and with no
count cut (green) plotted against measured temperature ({\it Left
panel}) and fractional temperature error ({\it Right
panel}). \label{fig:TxHist}}
\end{figure*}

In Section~\ref{sec:litz}, we described an automated search for
redshifts available in the literature. When this search was run in May 2011, a total of \Nx\, candidates were associated
with published redshifts using NED. We have found, however, during the
preparation of the first XCS data release (XCS-DR1, M11) that the NED
redshifts are not always appropriate for their respective candidate:
the match to a given NED cluster might be erroneous, especially at low
redshifts (where the allowed matching radius is large). 
We have therefore imposed a default\footnote{In principle we would have been
prepared to assign $z_{lit}<0.08$ values to XCS-DR1 clusters if they
had a measured photometric redshift of $z_{phot}\leq0.1$. However, in
practice there were no such cases, see M11.} minimum redshift limit
of $z=0.08$ when using literature redshifts. Moreover, even if
the match to the NED cluster is correct, the default NED
redshift for that cluster might not be the best one available in the literature.
Therefore, of the \Nx\, redshifts automatically selected from NED, only \Nxe\, were
included in XCS-DR1. That said, NED still contributed more redshifts to the 
catalogue than any of the other optical follow-up methods used in M11.

In Section~\ref{sec:Txmethod} we have described and verified an
automated method to derive X-ray temperatures from the XMM archive. We
have shown that reliable $T_{\rm X}$ values can be obtained for most
clusters if more than 300 soft-band counts are available in the
background-subtracted spectrum (Figures~\ref{spec_sim_cts} and
\ref{comp_results_counts}). We have further shown that our technique
works well even at large off-axis angles
(Fig.~\ref{comp_results_angle}) and that our automatically generated
results are consistent with those derived by other authors using more
traditional spectral fitting methods (Fig.~\ref{comp_lit}).  We note
that being able to fit $T_{\rm X}$ down to 300 counts was not
anticipated in \citetalias{romer01a}, where we assumed the minimum
counts threshold for $T_{\rm X}$ measurement would be 1000. This
decrease can be attributed to our adaptive spectral binning technique
and our use of Cash (rather than Gaussian) statistics in the fitting.

In \citetalias{romer01a} we predicted that up to 1,800 \emph{XMM}
clusters might yield temperatures (with $<20$ per cent errors). 
By comparison (by May 1st 2011), we had made only \Nii\, $T_{\rm X}$ 
measurements (with $<20$ per cent errors) for candidates
with optically determined redshifts, although, when the error threshold is
relaxed to $<40$ per cent, the number rises to \Ni. Of these \Ni, \Nib\, 
were determined from candidates detected with 300 or more background-subtracted counts (Fig.~\ref{fig:TxHist}). Even when we set the error threshold at 10\% (the
calibration uncertainty for the satellite), we still have \Niii\, clusters 
(\Niiib\, with over 300 counts) remaining.  For these \Niii, it would not
be worth doing further \emph{XMM} follow-up, although some high
resolution \emph{Chandra} imaging would be worthwhile to elucidate the
impact of point source contamination on derived temperatures (as we
have done successfully for XMMXCS J2215.9-1738, see
\citealt{hilton10a}). Errors on $T_{\rm X}$ of 40\% are too large for
some of the science applications we have in mind for XCS, e.g.
studies of the evolution in the scatter on the $L_{\rm X}-T_{\rm X}$
relation (since the intrinsic scatter is $<40\%$). Therefore, we have
made requests for additional \emph{XMM} follow-up of certain XCS
clusters. We note that only \Nic\, of the \Ni\, candidates with $T_{\rm X}$ 
measurements ($<40$ per cent errors) are included in 
XCS-DR1\footnote{An additional \Nicc\, more $T_{\rm X}$ measurements 
with larger errors are also included in XCS-DR1}. That is not to say that the 
remainder are not clusters, but rather that they cannot be confirmed as being 
so using the currently available optical and X-ray data (see M11). Even so, 
the size of the XCS-DR1 $T_{\rm X}$ sample is still much larger than any 
previous published compilations of cluster $T_{\rm X}$ measurements 
(with either heterogeneous or homogeneous selection). 

In Section~\ref{sec:spatialmethod} we described and verified an
automated method to derive X-ray luminosities from the XMM archive,
the $L_{\rm X}$-pipeline.  This pipeline is run on any candidate for
which a $T_{\rm X}$ measurement has been made via the $T_{\rm
X}$-pipeline. We demonstrated that the parameters that come out of the
spatial fitting are robust, as compared to previously published work
(Figures~\ref{comp_beta} and \ref{comp_lum}). Limitations of the
current method include the reliance on circularly-symmetric models and
the lack of covariance information between the $T_{\rm X}$-pipeline
and $L_{\rm X}$-pipelines.  Addressing these two issues is possible,
but given that we are often fitting to only a few hundred counts, and
currently using a circularly-symmetric PSF, we have no plans to adjust
the $L_{\rm X}$-pipeline accordingly (because it would increase the
computational complexity significantly).  A further limitation of the
current $L_{\rm X}$-pipeline is that the error on the input redshift
is assumed to be zero. This simplification is justified for
spectroscopically determined redshifts, but not for photometric, or
X-ray (see below), redshifts. This issue will be addressed before the
$L_{\rm X}$ values are used in a future study of the evolution of the
$L_{\rm X}-T_{\rm X}$ relation.

The impact of redshift errors notwithstanding, the uncertainty on the
$L_{\rm X}$ value for a given candidate will be much smaller than the
associated error on $T_{\rm X}$ (because the ICM emission is only a
weak function of $T_{\rm X}$). Therefore, we do not think it is
necessary to carry out a large-scale \emph{XMM} follow-up campaign in
order to improve the $L_{\rm X}$ measurements.  That said, we are
planning to request \emph{XMM} snapshots of clusters that were
discovered so close to the edge of the field of view that a large
fraction of their flux was not captured in their respective EPIC
images. These snapshots will allow us to get a better estimate of their
total flux.  We also plan to make \emph{Chandra} snapshot requests of
a representative subsample of XCS clusters, in order to gauge, in a
statistical sense, the impact of point source contamination on XCS
$L_{\rm X}$ values (although this test may be possible using the
existing \emph{Chandra} archive and we will explore that avenue
first).

In Section~\ref{sec:zxmethod} we described a method to extract X-ray 
redshifts directly from the discovery data to supplement the XCS optical follow efforts.
As shown in Fig.~\ref{comp_z}, acceptable ($\Delta_z <0.1$) redshift measurements
are made in $\simeq75$ per cent of the cases when thresholds on the $z_{\rm X}$-pipeline 
errors are set at $<20$\% and $\sigma_{z_{\rm X}}<0.05$. To date, \Nxvii\, candidates have yielded
$z_{\rm X}$ measurements that meet these criteria.
We have used $z_{\rm X}$ estimates to preselect candidates for optical
follow-up and this approach has been successful, e.g. one cluster
with $z_{\rm X}=0.84$ was demonstrated to have a true redshift of
$z=0.83$ based on subsequent Gemini GMOS spectroscopic observations (M11).
The level of catastrophic redshift errors is much higher for X-ray redshifts than for
optical photometric techniques, so all clusters with only $z_{\rm X}$ values 
will ultimately have to be followed up with optical photometry or spectroscopy. 

\begin{figure}
\includegraphics[width=9.5cm]{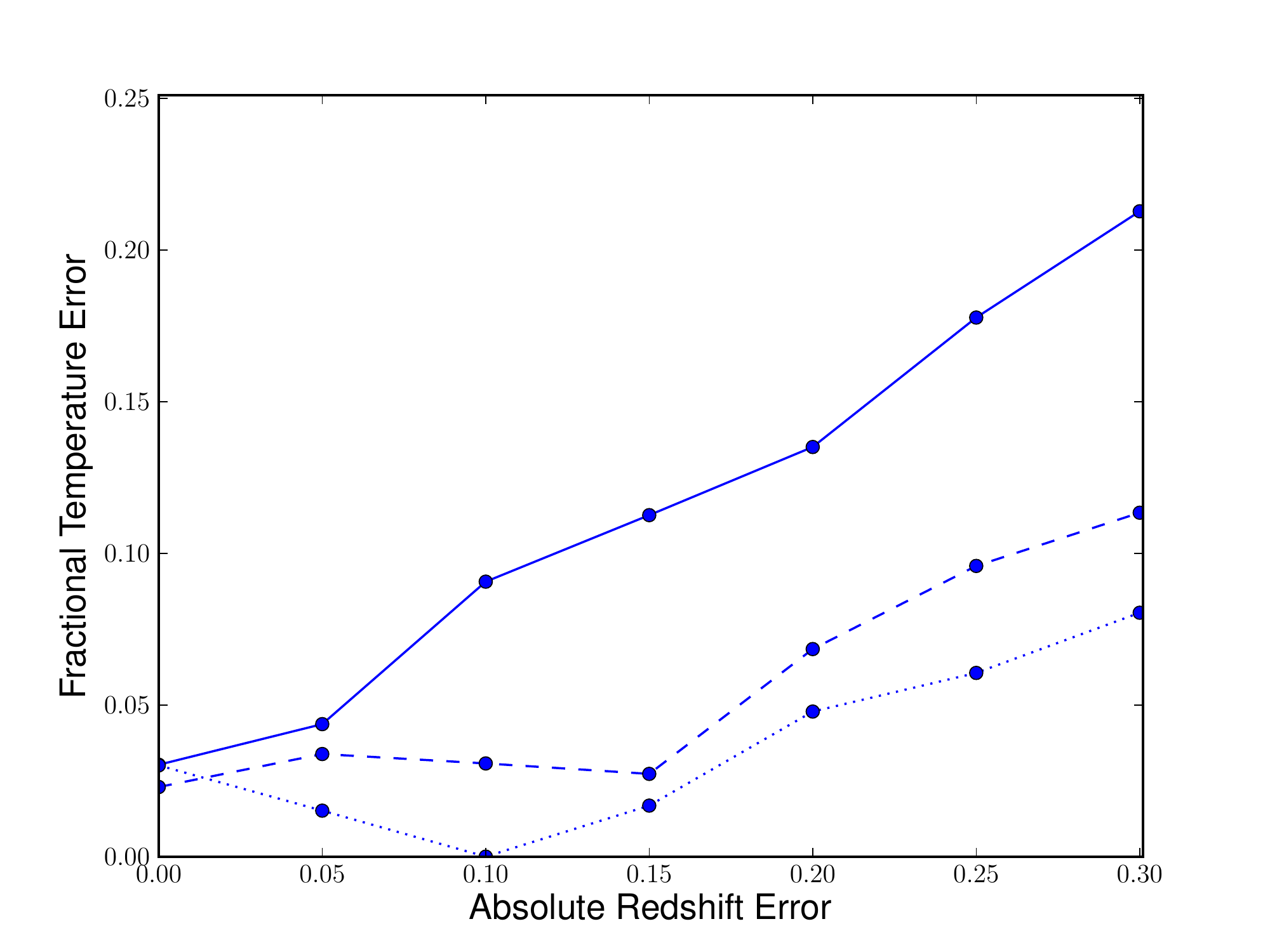}
\caption{The results of fitting simulated MEKAL spectra with a
  temperature of 3.0 keV and redshifts of 0.2 (solid), 0.6 (dashed)
  and 1.0 (dotted) showing fractional temperature uncertainty as a function
  of absolute redshift uncertainty.
\label{spec_sim_zerror}}
\end{figure}

The impact of $z_{\rm X}$ errors on $L_{\rm X}$ measurements is
significant, and so $L_{\rm X}$ values that rely on $z_{\rm X}$ will
not be used for science applications. However, we have determined that
the impact of $z_{\rm X}$ errors on $T_{\rm X}$ measurements is not
significant: as shown in Fig.~\ref{spec_sim_zerror}, an absolute
redshift uncertainty of $\Delta_z$=0.3 induces a less than 30 per cent
$T_{\rm X}$ uncertainty. For this reason, it will be possible to use
$z_{\rm X}$ determined $T_{\rm X}$ values to select XCS clusters for
SZ follow-up, because the SZ effect is (roughly speaking) redshift
independent. Most current SZ instruments are only sensitive to $T_{\rm
X} >5$ keV clusters, but few of those have been catalogued yet, 
especially at $z>0.5$: in the BAX\footnote{http://bax.ast.obs-mip.fr/} 
database there are only 39 such clusters listed. In XCS-DR1 there are \Nxiv\,
such clusters (of which \Nxivb\, are in addition to the BAX sample). By comparison, 
using the X-ray redshift technique we have identified \Nxiii\, more 
candidates (without other redshift information) meeting those criteria. 
In summary, X-ray redshifts are not a `magic bullet' and optical follow-up 
is still required in order to secure redshift measurements; however, they 
do provide some useful information, as long as they are used judiciously.

In summary, we have demonstrated that the X-ray algorithms developed for 
XCS are suitable for the compilation and analysis of large samples of clusters
detected serendipitously by \emph{XMM}.  In our companion paper (M11) 
we discuss the optical follow-up of candidates and present the 
first XCS data release (XCS-DR1), including \Nxvb\, $T_{\rm X}$ measurements.
On going science exploitation of XCS-DR1 includes projects related to
cluster scaling relations, fossil groups, the SZ effect and the derivation of 
cosmological parameters. We also plan to apply our post-processing pipelines, 
that were designed with serendipitous clusters in mind, to the many hundreds of target
clusters in the \emph{XMM} archive, so that they too can benefit from
a uniform set of $T_{\rm X}$ and $L_{\rm X}$ measurements. Even though
these target clusters cannot be used for XCS statistical studies, we
think this will be a valuable resource for the community, especially
now that \emph{Planck} is in full operation.

\section{Conclusions}
\begin{enumerate}
\item{We have demonstrated that the \emph{XMM} archive is a rich
resource for serendipitous cluster detection out to redshifts of at
least $z=1.5$.}
\item{The archive now covers over 600 square degrees 
that can be used for serendipitous source detection and, of
this, 51~deg$^2$, 276~deg$^2$ and 410~deg$^2$ (at $>$40~ks, $>$10~ks
and $>$0~ks depths respectively) are available for cluster detection.}
\item{We have shown that typically one-third of a given \emph{XMM}
exposure is rendered unusuable due to background flares.}
\item{We have shown that it is possible to exploit the whole
\emph{XMM} archive in a uniform and reproducible way.}
\item{We have developed a source detection pipeline that operates
across the entire \emph{XMM} field of view, and is effective over a
wide range of angular scales and signal-to-noises. It has many
features, including the ability to determine which sources are extended
beyond the PSF model and to detect point-like sources
that lie along the line of sight to extended sources.}
\item{We have developed a pipeline that can measure reliable X-ray
cluster temperatures. This pipeline has been shown to work well even
when the cluster is discovered on the outskirts of the field of view. We have 
demonstrated that with 300 or more background-subtracted counts, 
one can measure robust, unbiased, temperatures for most clusters.}
\item{We have developed a pipeline that can measure reliable X-ray
luminosities by making spatial fits to \emph{XMM} images. The derived
luminosity values have been shown to be robust, as have the fitted
spatial parameters.}
\item{We have developed a pipeline that can measure `X-ray
redshifts' for clusters using \emph{XMM} spectra. These redshifts can help increase the number
of clusters with X-ray temperature measurements; acceptable ($\Delta_z <0.1$) redshift measurements are made in $\simeq75$ per cent of the cases (once errors thresholds have been imposed).}
\item{To date (May 1st 2011), some key statistics for XCS
are as follows: \Nxi\, ObsIDs have been downloaded from the \emph{XMM}
archive; \Nxii\, ObsIDs have run through the event list cleaning
pipeline; \Nvi\, ObsIDs have been processed by the source detection
pipeline; \Nvii\, point sources and \Nviii\, extended sources have been
catalogued; \Niv\, cluster candidates have been selected, of which
\Nivb\, were detected with more than 300 background-subtracted counts; \Ni\, (\Niii) X-ray temperatures have been measured 
with $<40$ ($<10$) per cent errors.}
\end{enumerate}

\section*{Acknowledgments}

This work was made possible by the ESA \emph{XMM-Newton} mission, and we thank everyone who was involved in making that mission such a success.  We also acknowledge the following public archives, surveys and analysis tools: The HEASOFT analysis packages provided by NASA's Goddard Space Flight Center. The NASA/IPAC Extragalactic Database ({\bf NED}) which is operated by the Jet Propulsion Laboratory, California Institute of Technology, under contract with the National Aeronautics and Space Administration. The X-Ray Clusters Database ({\bf BAX}) which is operated by the Laboratoire d'Astrophysique de Tarbes-Toulouse (LATT), under contract with the Centre National d'Etudes Spatiales (CNES). 

Financial support for this project includes: 
The Science and Technology Facilities Council (STFC) through grants ST/F002858/1 and/or ST/I000976/1 (for ELD, AKR, NM, MHo, ARL and MS), ST/H002391/1 and PP/E001149/1 (for CAC and JPS), ST/G002592/1 (for STK) and through studentships (for NM, HCC). 
The RAS Hosie Bequest and the University of Edinburgh (for MD). 
Carnegie Mellon University (KS). 
The University of Sussex (MHo, EK, HCC). 
The University of KwaZulu-Natal (for MHi). 
The Leverhulme Trust (for MHi).
Funda\c{c}\~{a}o para a Ci\^{e}ncia e a Tecnologia through the project PTDC/CTE-AST/64711/2006 (for PTPV).
The South East Physics Network (for END, RCN). 
FP7-PEOPLE- 2007-4Ð3-IRG n 20218 (for BH). 
The Swedish Research Council (VR) through the Oskar Klein Centre for Cosmoparticle Physics (for MS).  
The  U.S. Department of Energy, National Nuclear Security Administration by the University of California, Lawrence Livermore National Laboratory under contract No. W-7405-Eng-48 (for SAS).  

Parts of the manuscript were written during a visit by AKR to the Aspen Physics Center.  The authors thank both Peter Thomas, for his encouragement and advice during the preparation of this manuscript, and an anonymous referee for their careful reading and insightful comments.
 
\bibliography{paper}
\appendix

\onecolumn

\FloatBarrier
\include{appendix2}

\FloatBarrier
\include{appendix3}

\FloatBarrier
\include{appendix1}


\end{document}

%% file: appendix2.tex
\section{Additional Supporting Figures}

\begin{figure*}
\centering
\includegraphics[width=0.35\textwidth]{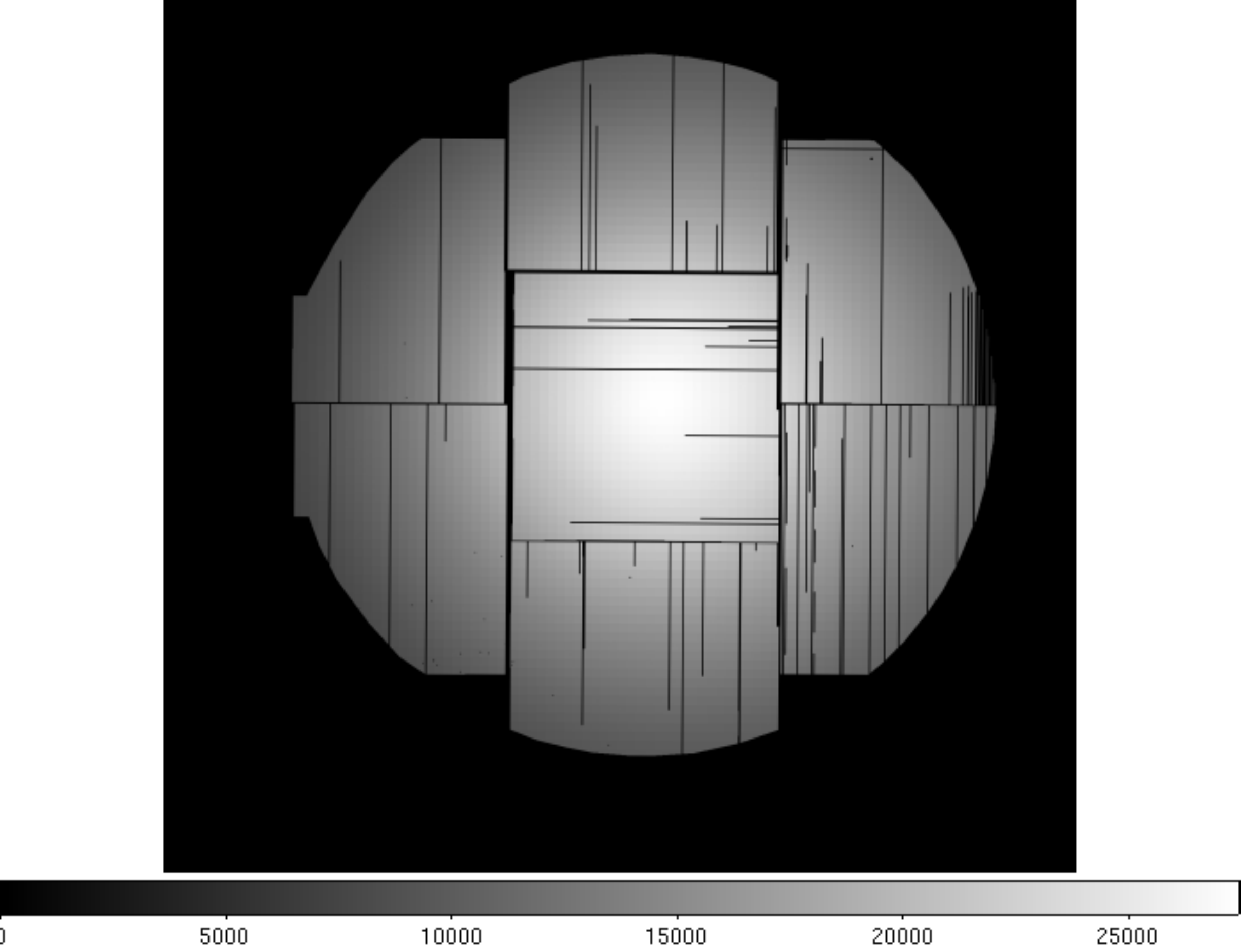}
\includegraphics[width=0.35\textwidth]{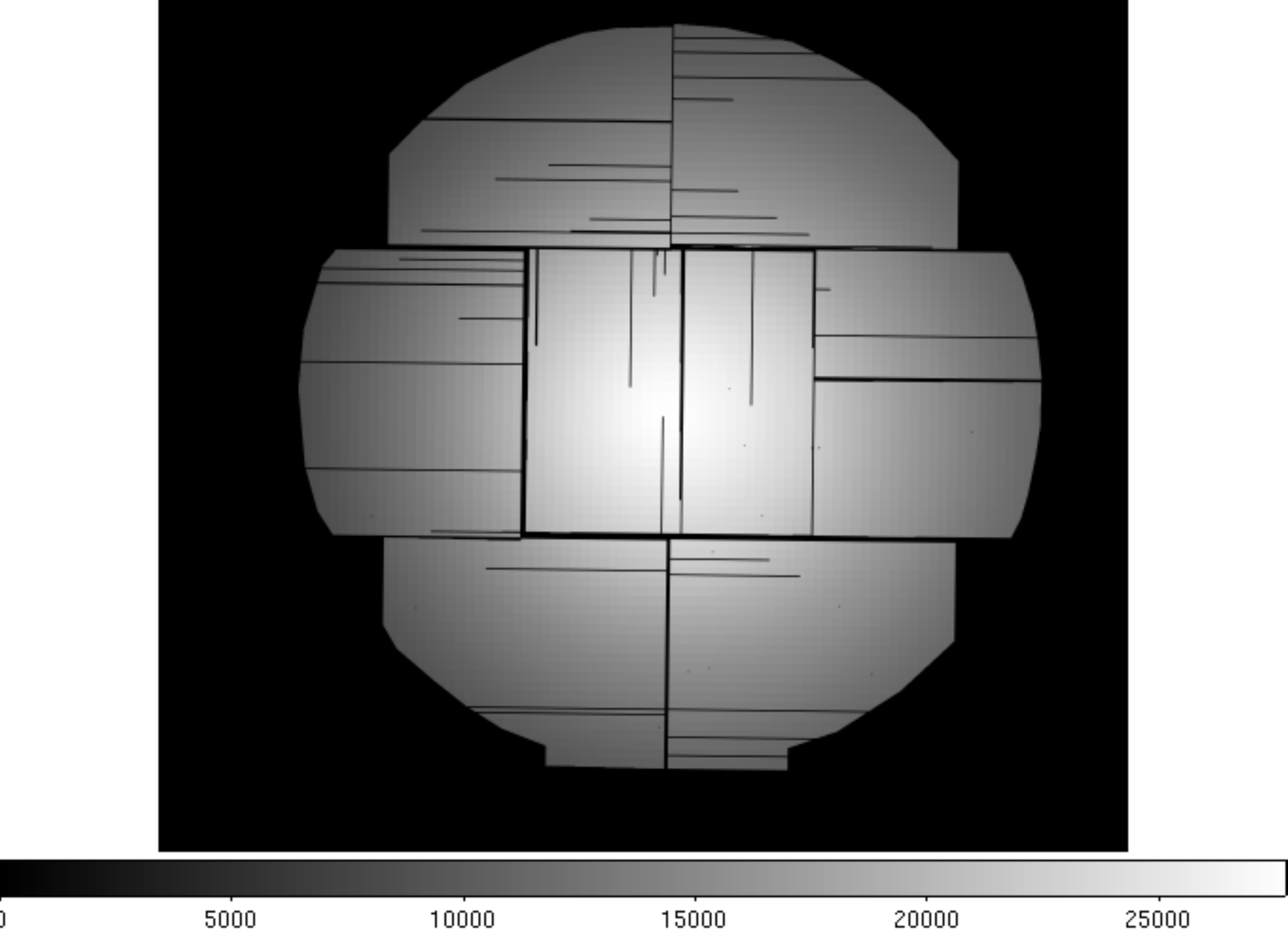}
\\
\includegraphics[width=0.35\textwidth]{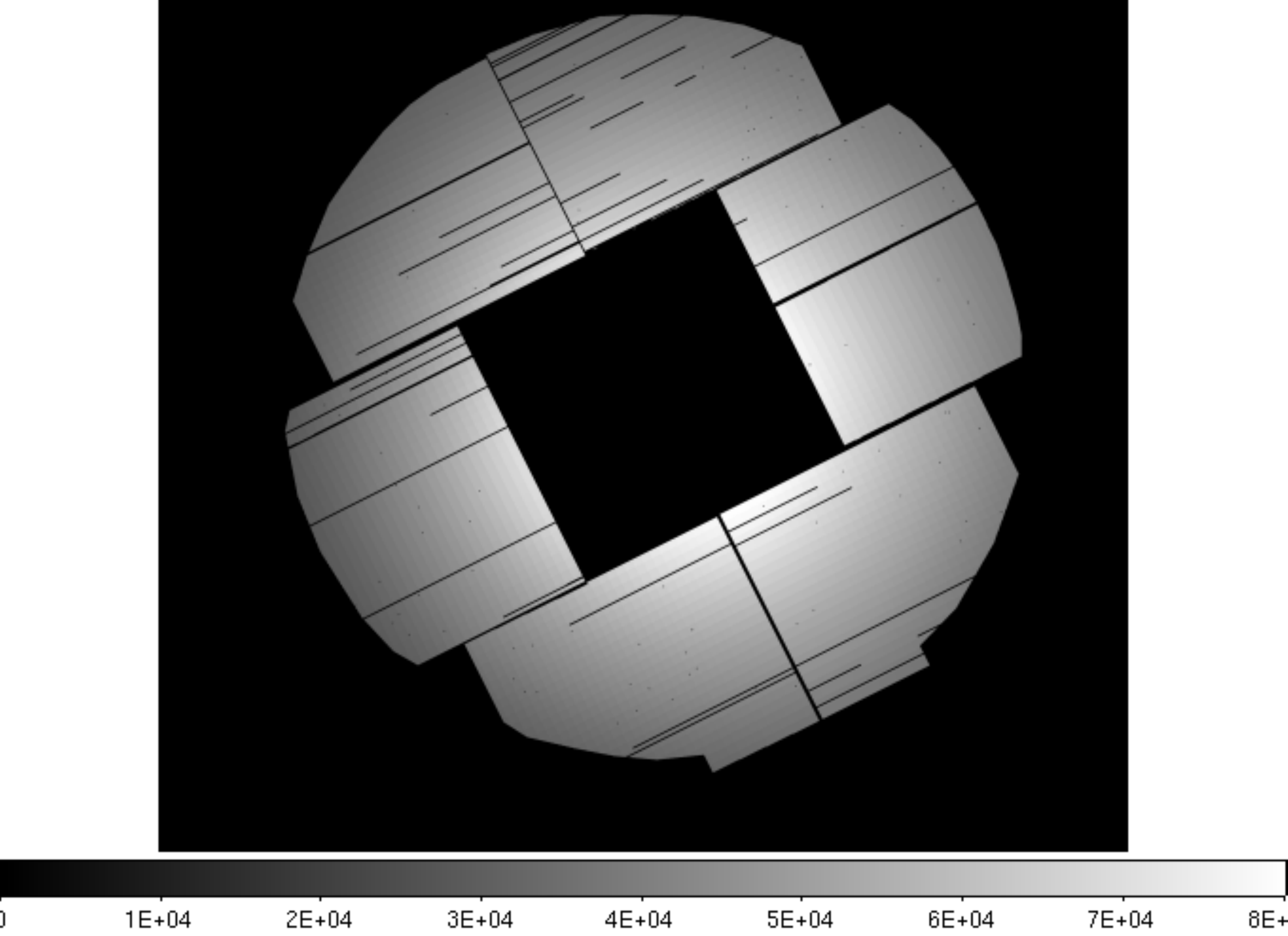}
\includegraphics[width=0.35\textwidth]{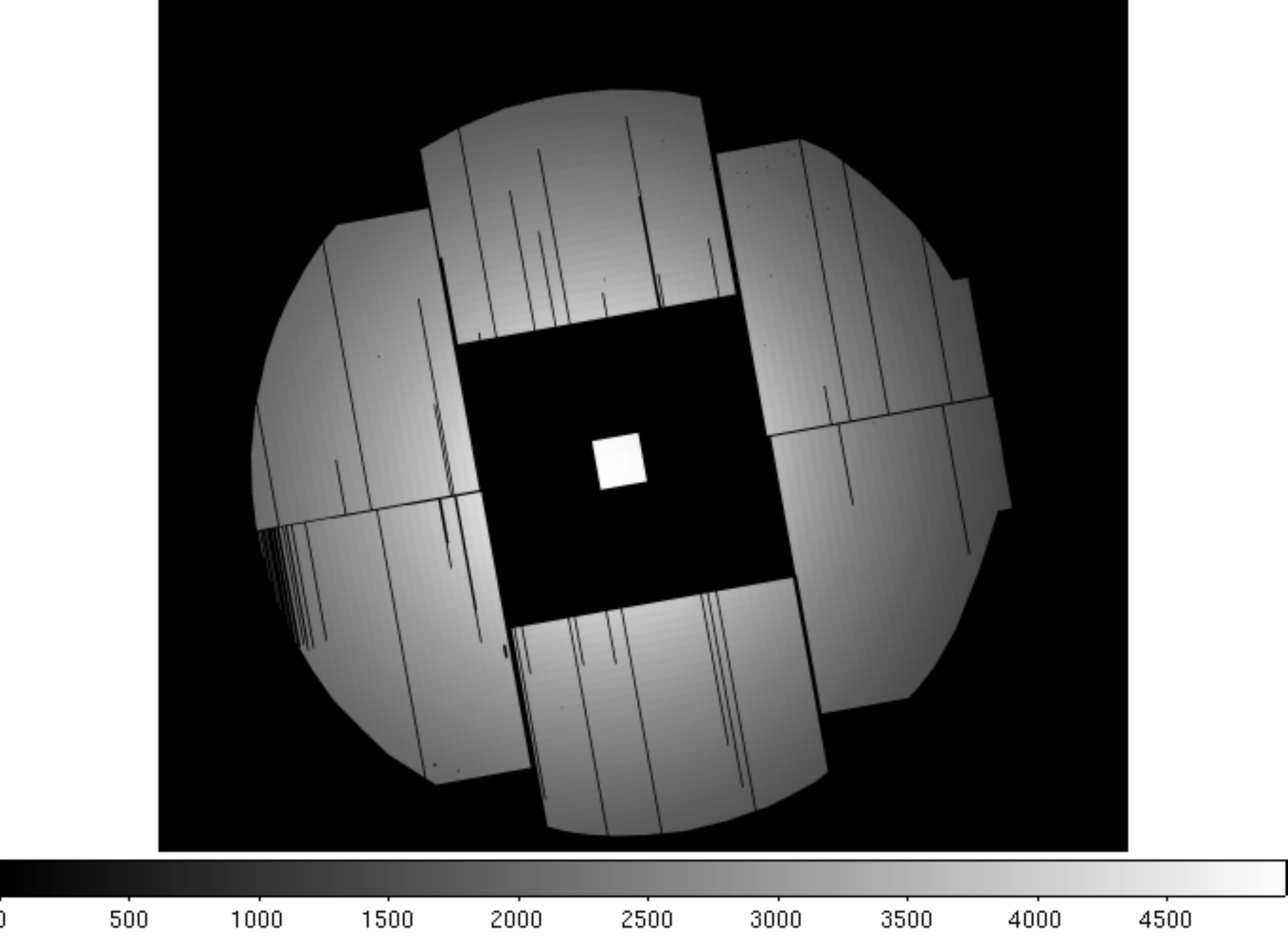}
\\
\includegraphics[width=0.35\textwidth]{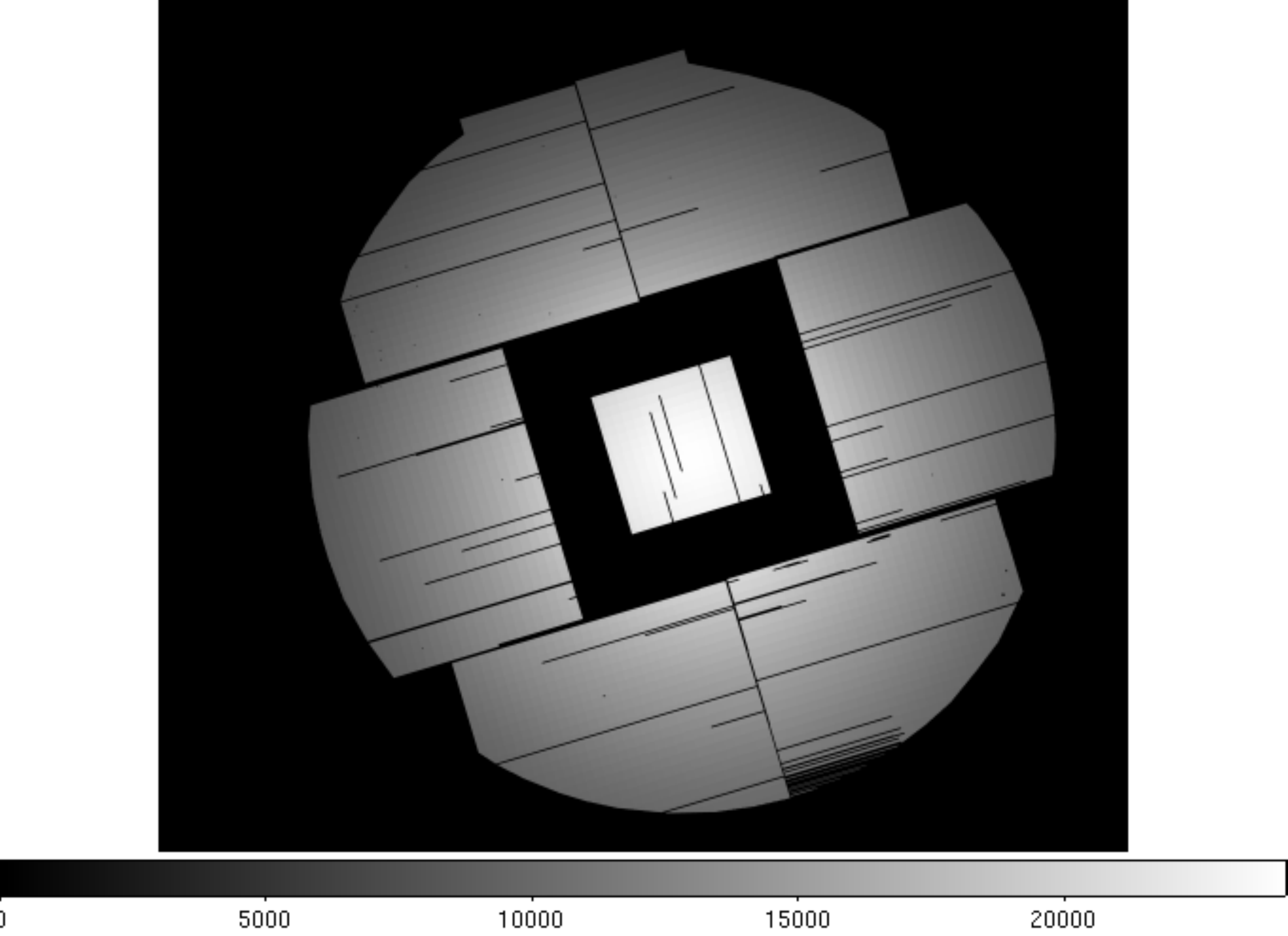}
\includegraphics[width=0.35\textwidth]{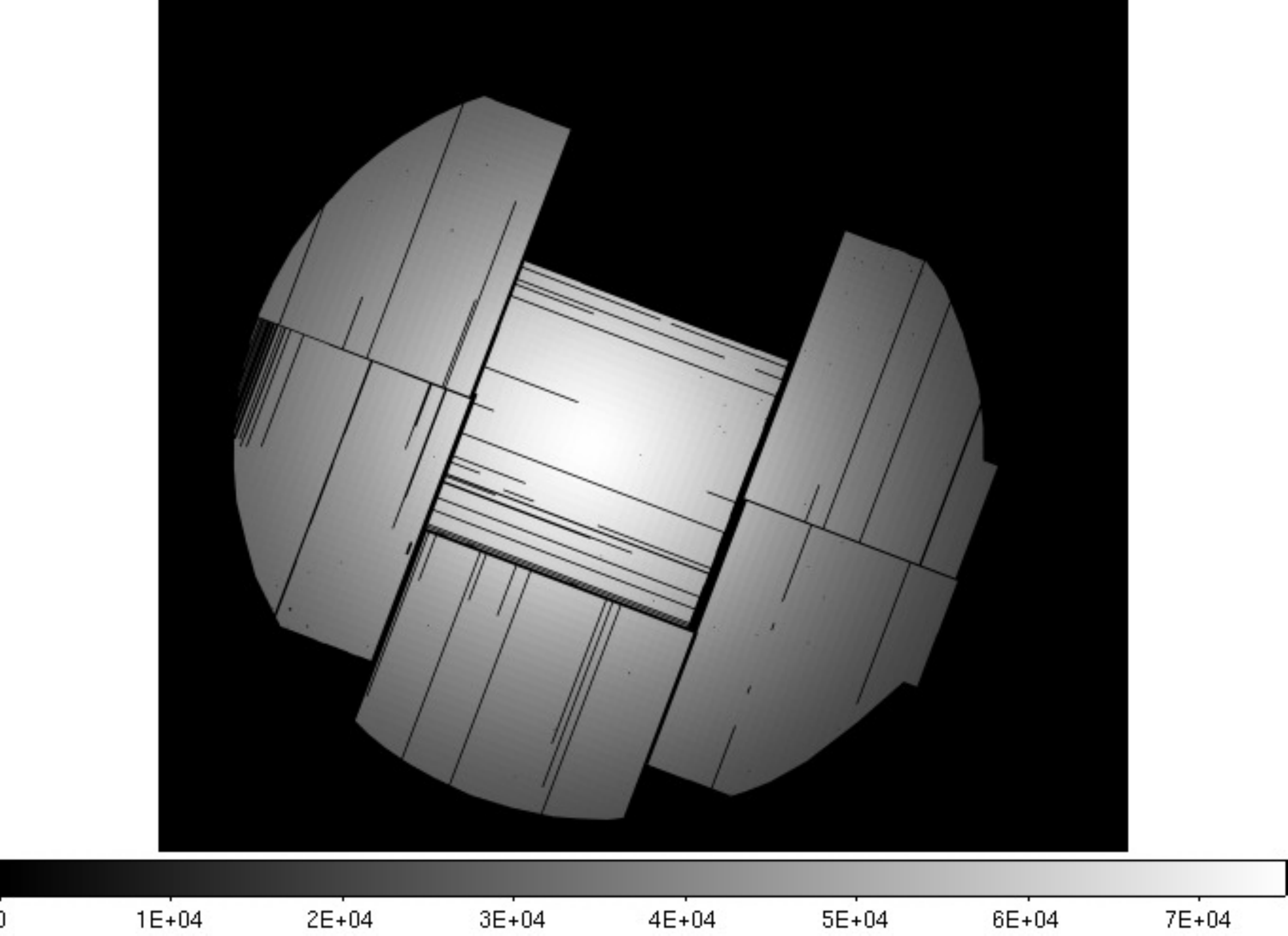}
\\
\includegraphics[width=0.35\textwidth]{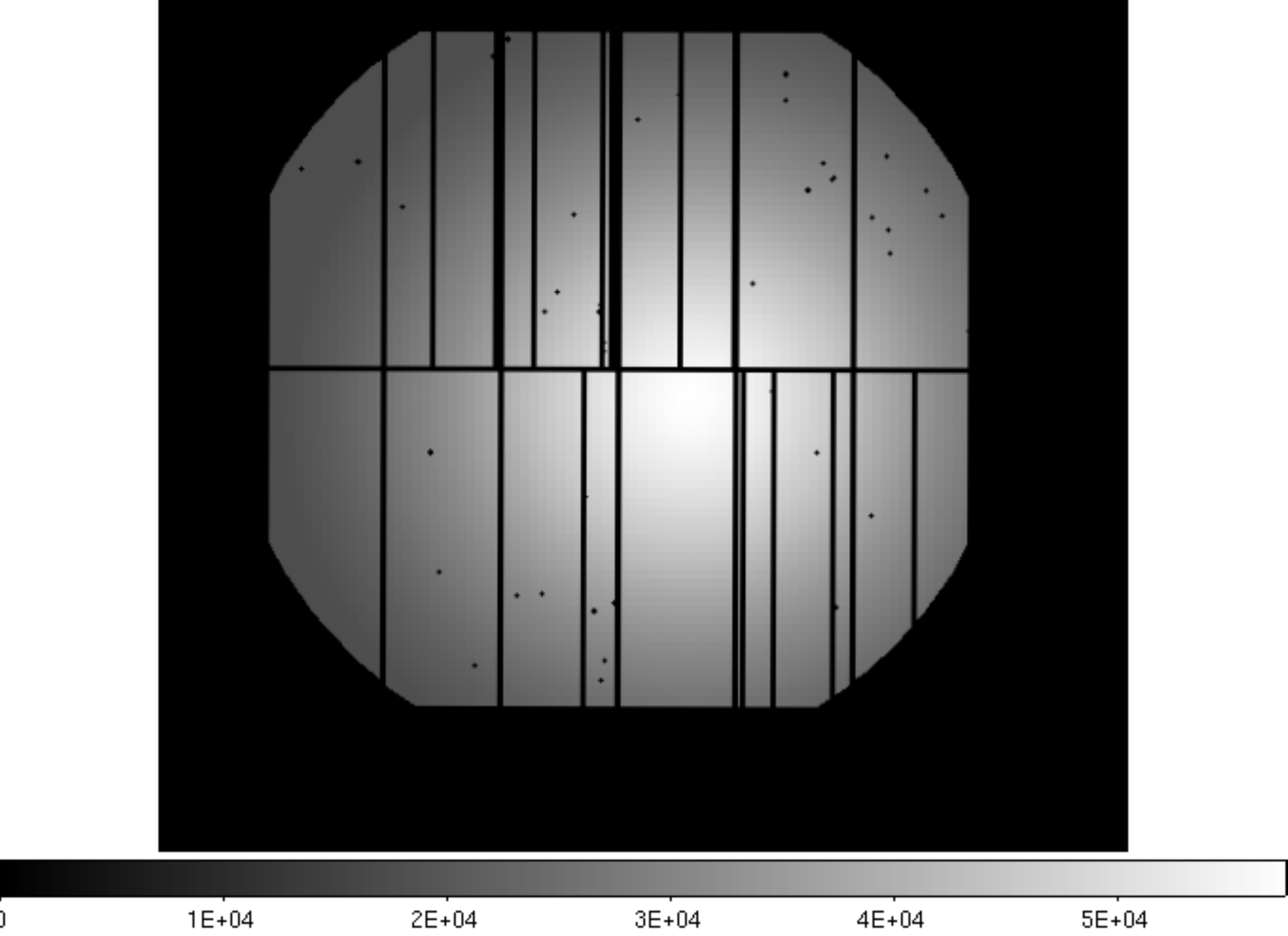}
\includegraphics[width=0.35\textwidth]{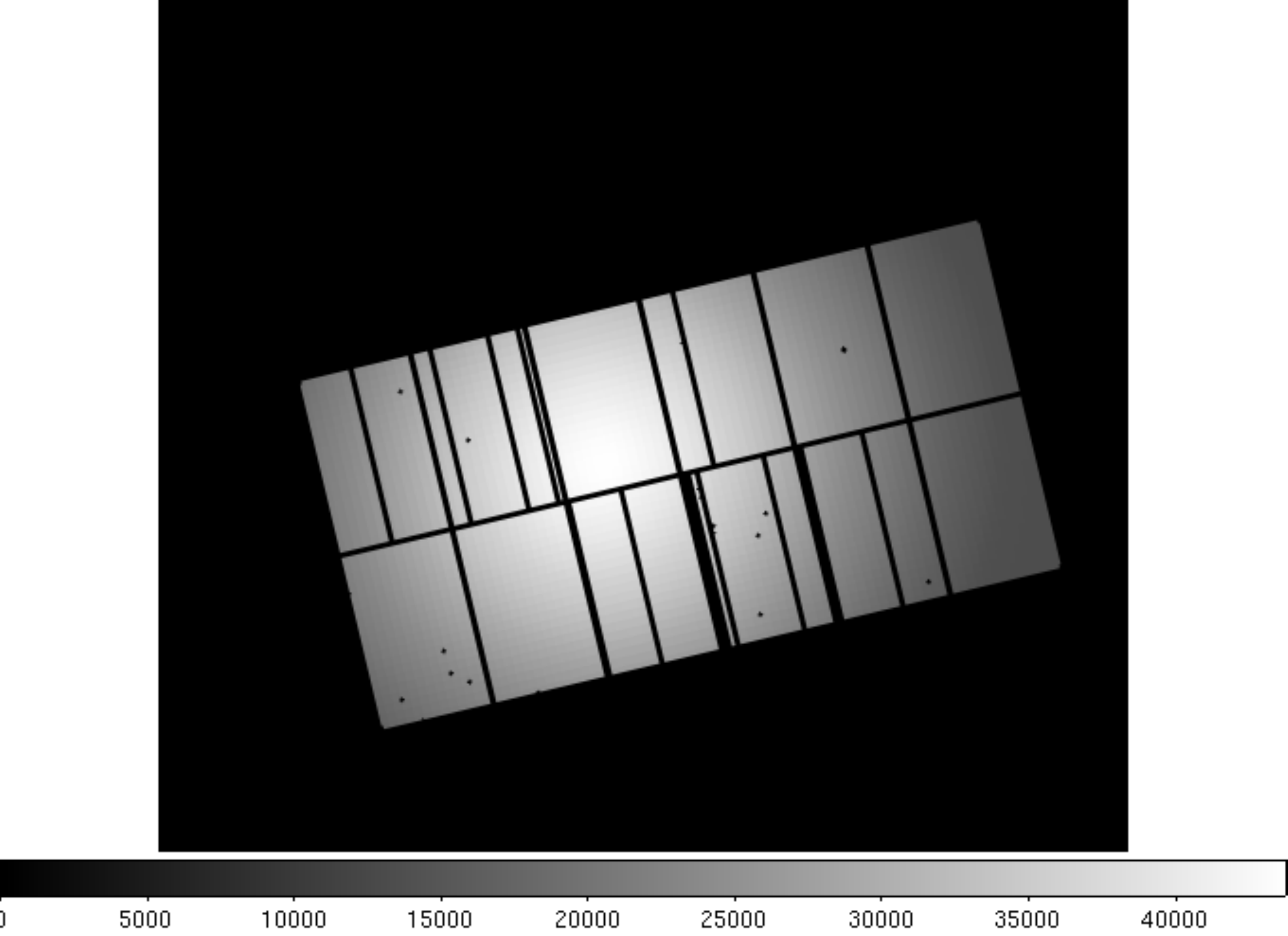}
\caption{Exposure maps relating to differing EPIC observing modes. In
order from top left: MOS1 full window mode, MOS2 full window mode, MOS
fast uncompressed, MOS partial window W2 or W4 mode, MOS partial
window W3 or W5 mode, MOS1 full window mode with CCD6 switched off, pn
full window mode, pn large window mode.}
\label{fig:indiv_exps}
\end{figure*}

\begin{figure*}
\centering
\includegraphics[angle=90, width=0.47\textwidth]{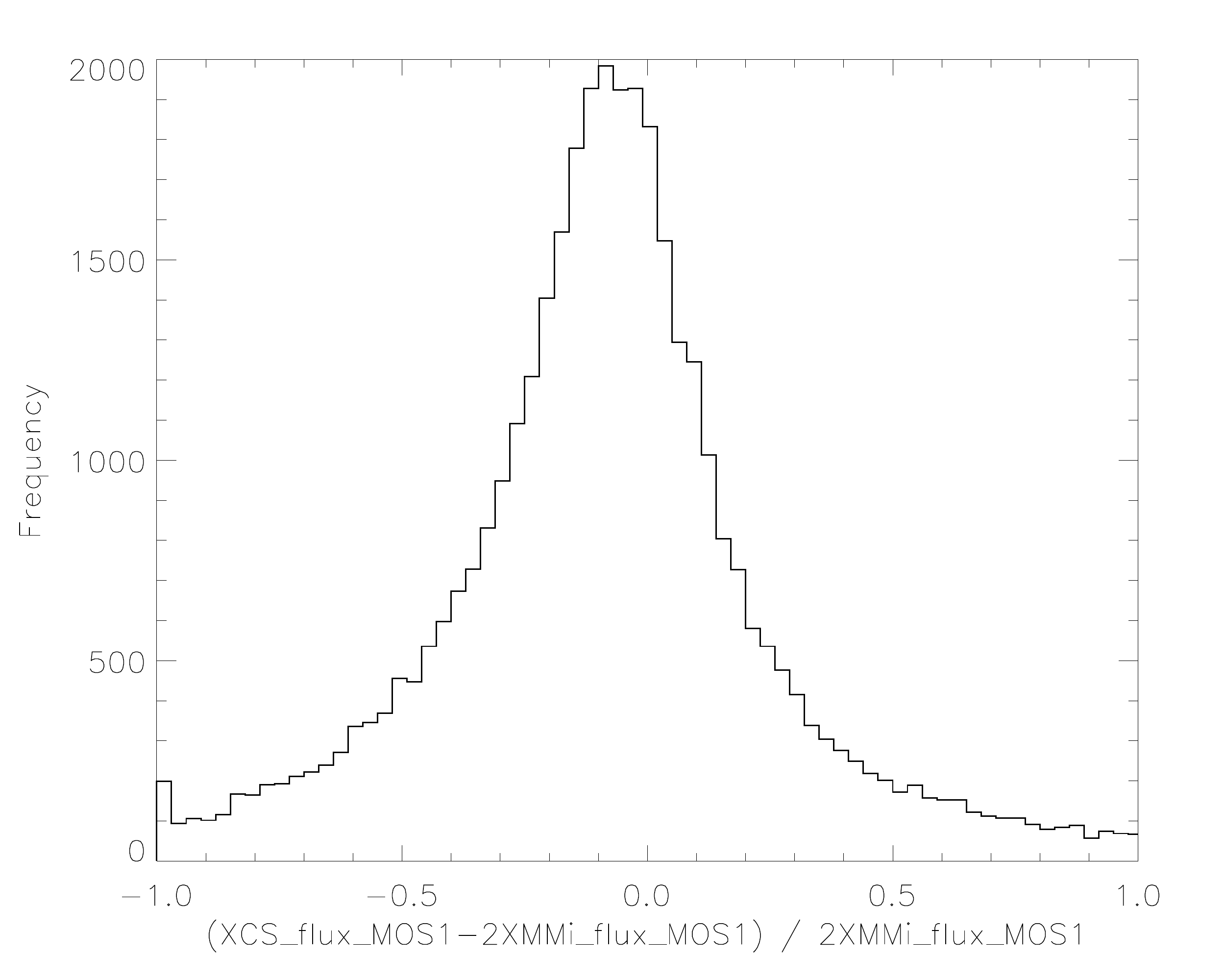}
\includegraphics[angle=90, width=0.47\textwidth]{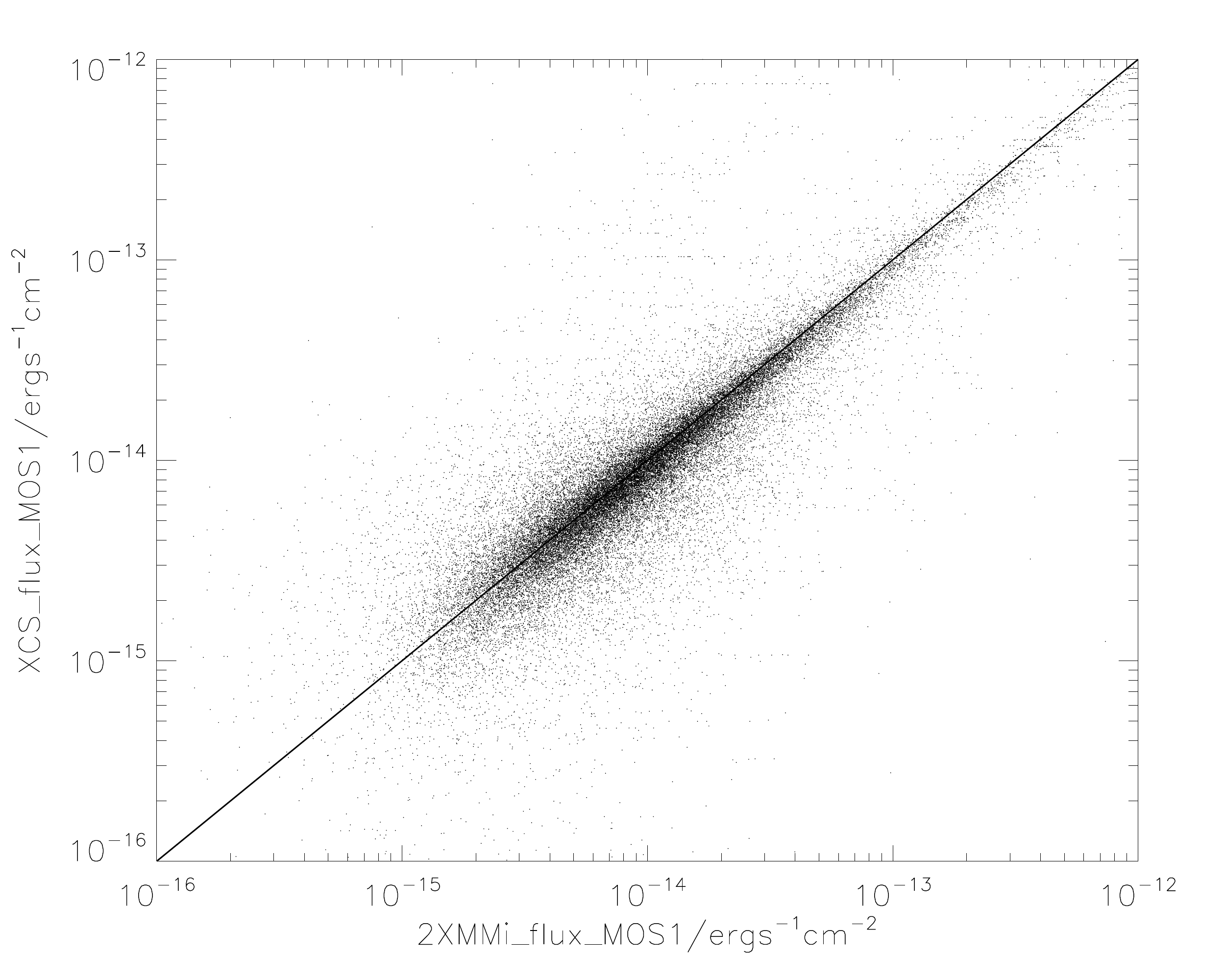}
\includegraphics[angle=90,width=0.47\textwidth]{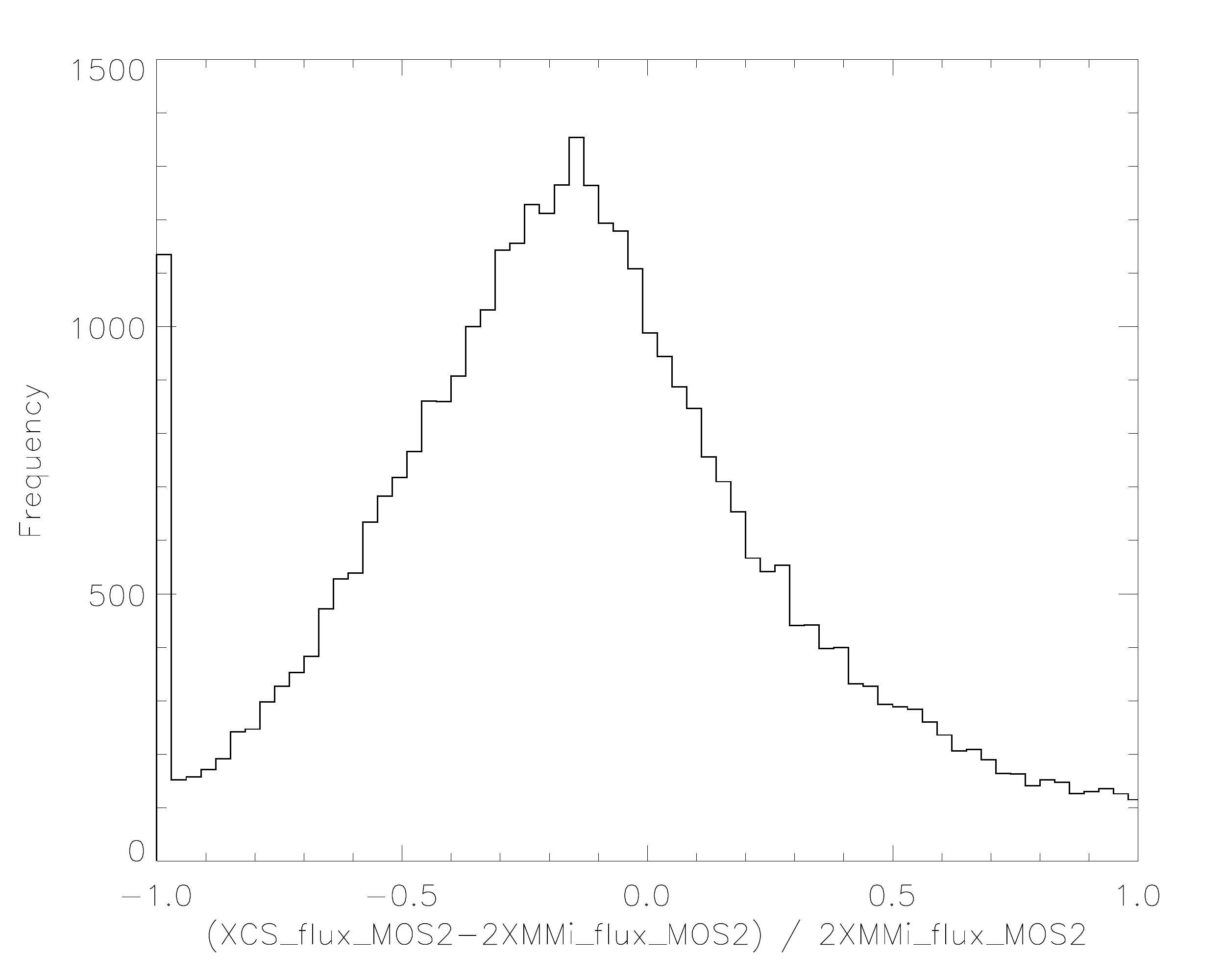}
\includegraphics[angle=90, width=0.47\textwidth]{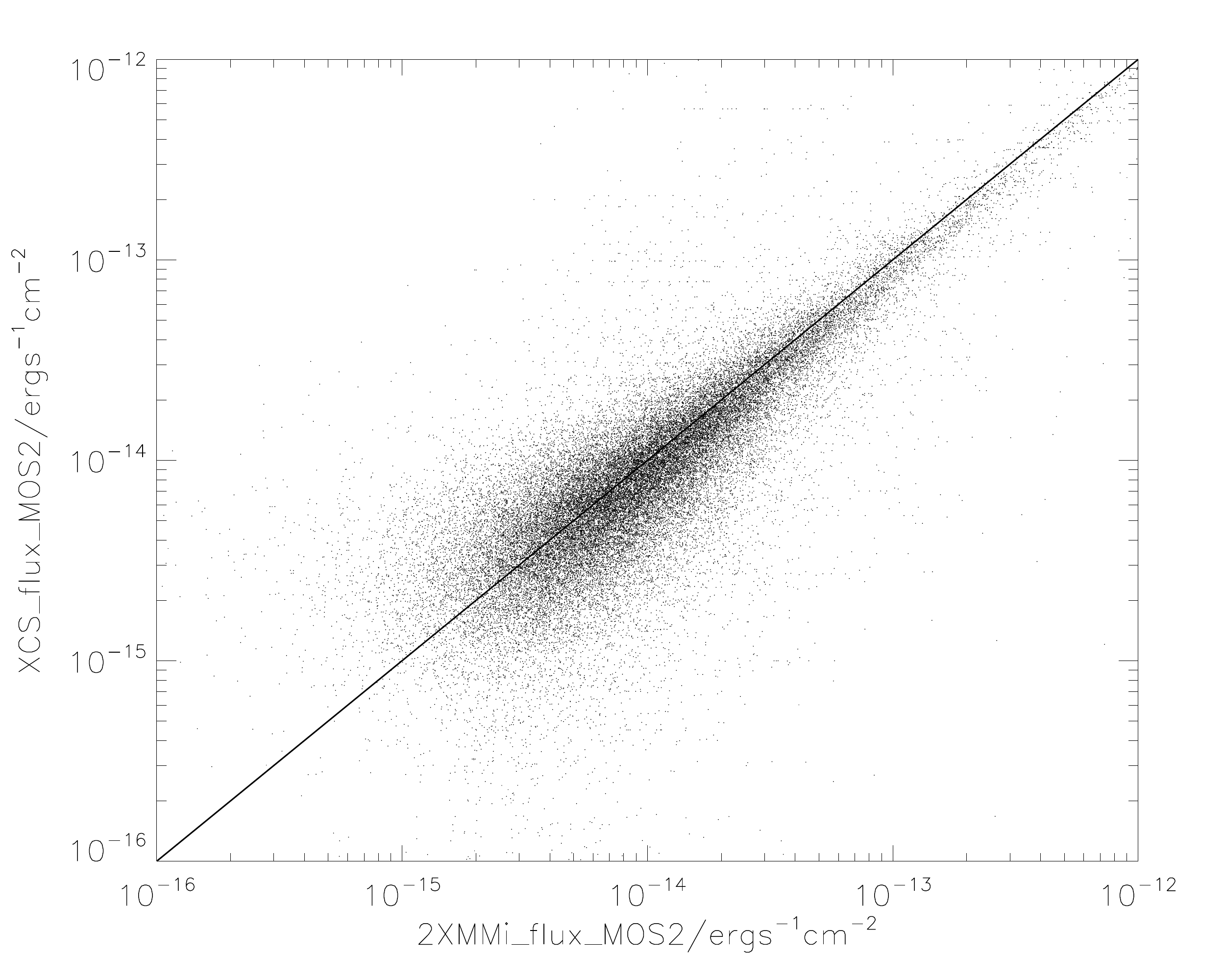}
\includegraphics[angle=90,width=0.47\textwidth]{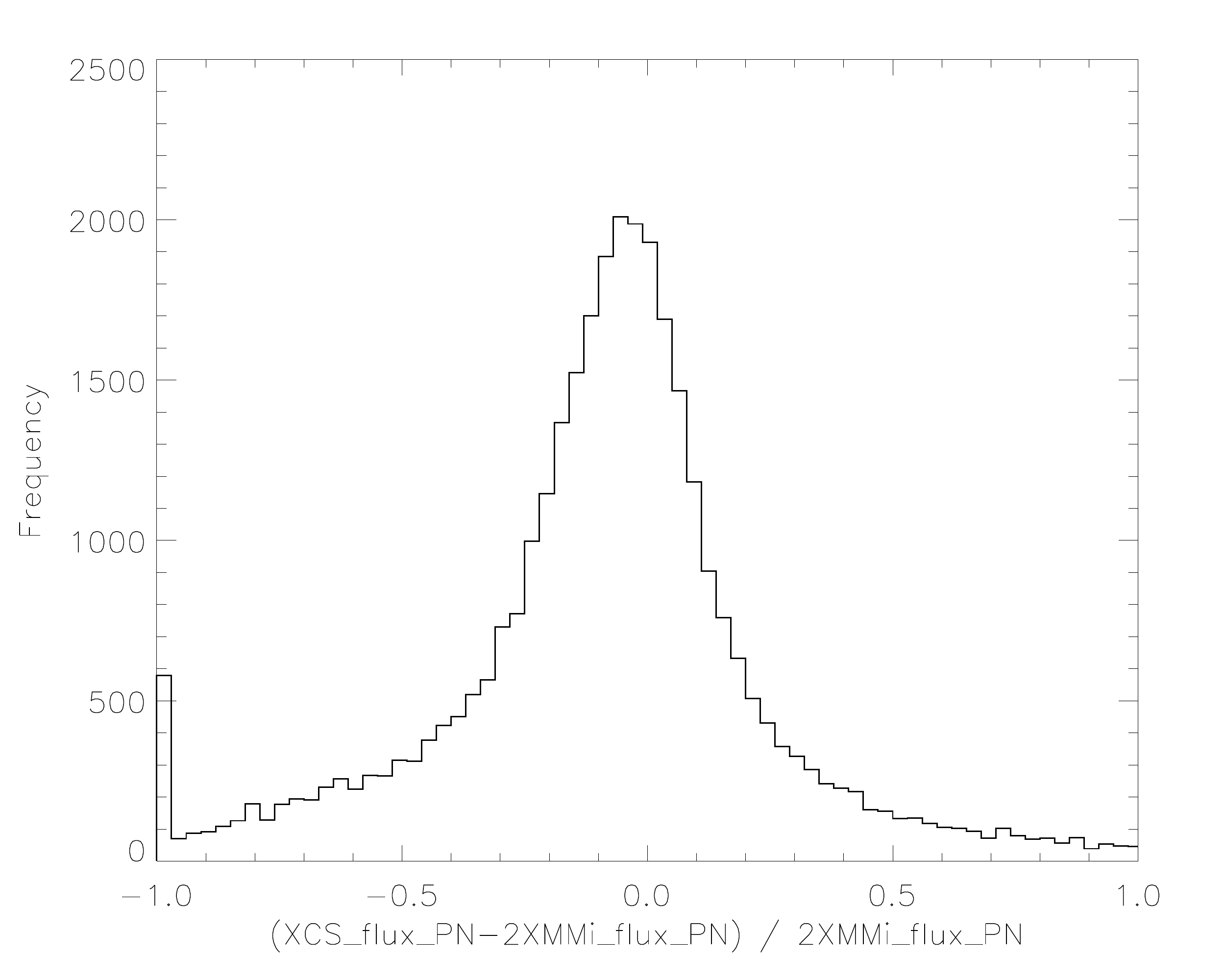}
\includegraphics[angle=90,width=0.47\textwidth]{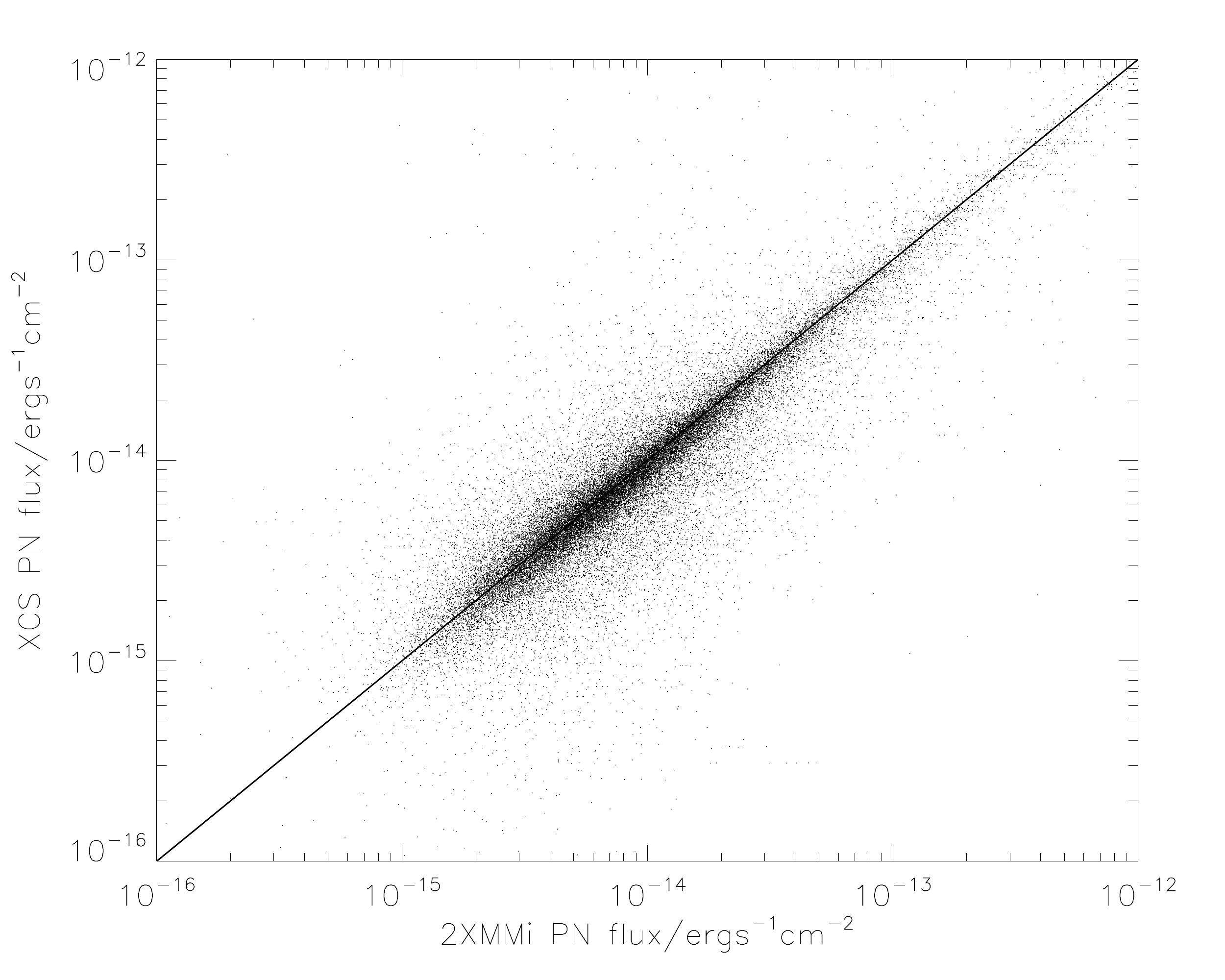}
\caption{A comparison of the individual camera fluxes of XCS sources
  with their matches in the 2XMM catalogue. Top: EPIC-mos1; Middle:
  EPC-mos2; Bottom: EPIC-pn}
\label{fig:2xmm-xcs}
\end{figure*}

\begin{figure*}
\centering
    \includegraphics[width=1.0\textwidth]{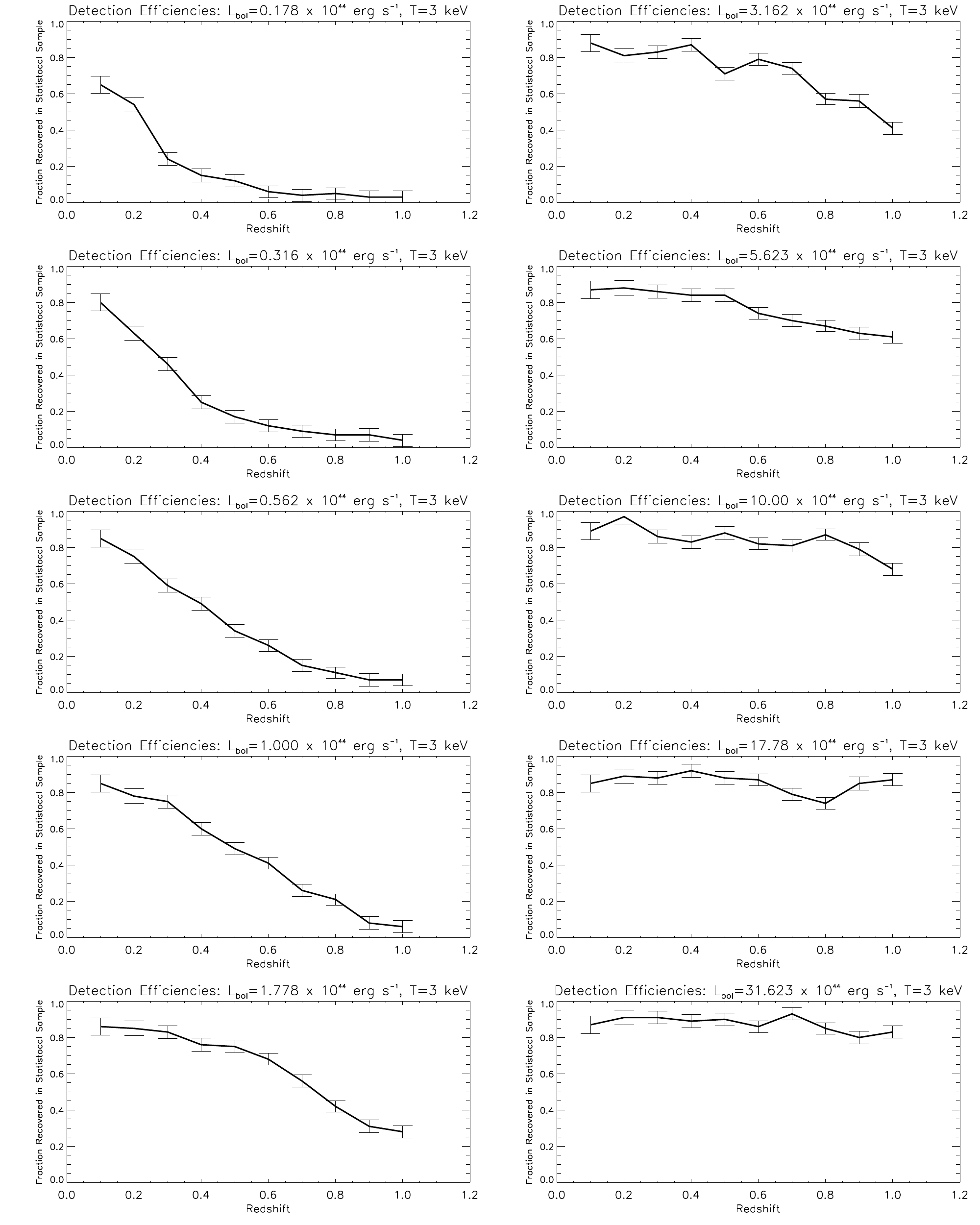}
\caption{Predicted recovery efficiency as a function of redshift for 3
  keV clusters with a range of X-ray luminosities (bolometric). The
  synthetic clusters used for this test had circularly symmetric
  $\beta$-profiles ($\beta=2/3$) with core radii of 160 kpc. The
  typical luminosity of a 3 keV cluster based on the local $L_X$--$T_X$
  relation is 1 to 2$\times10^{44}$ erg~s$^{-1}$.}
\label{fig:betafun13}
\end{figure*}

\begin{figure*}
\centering
    \includegraphics[width=1.0\textwidth]{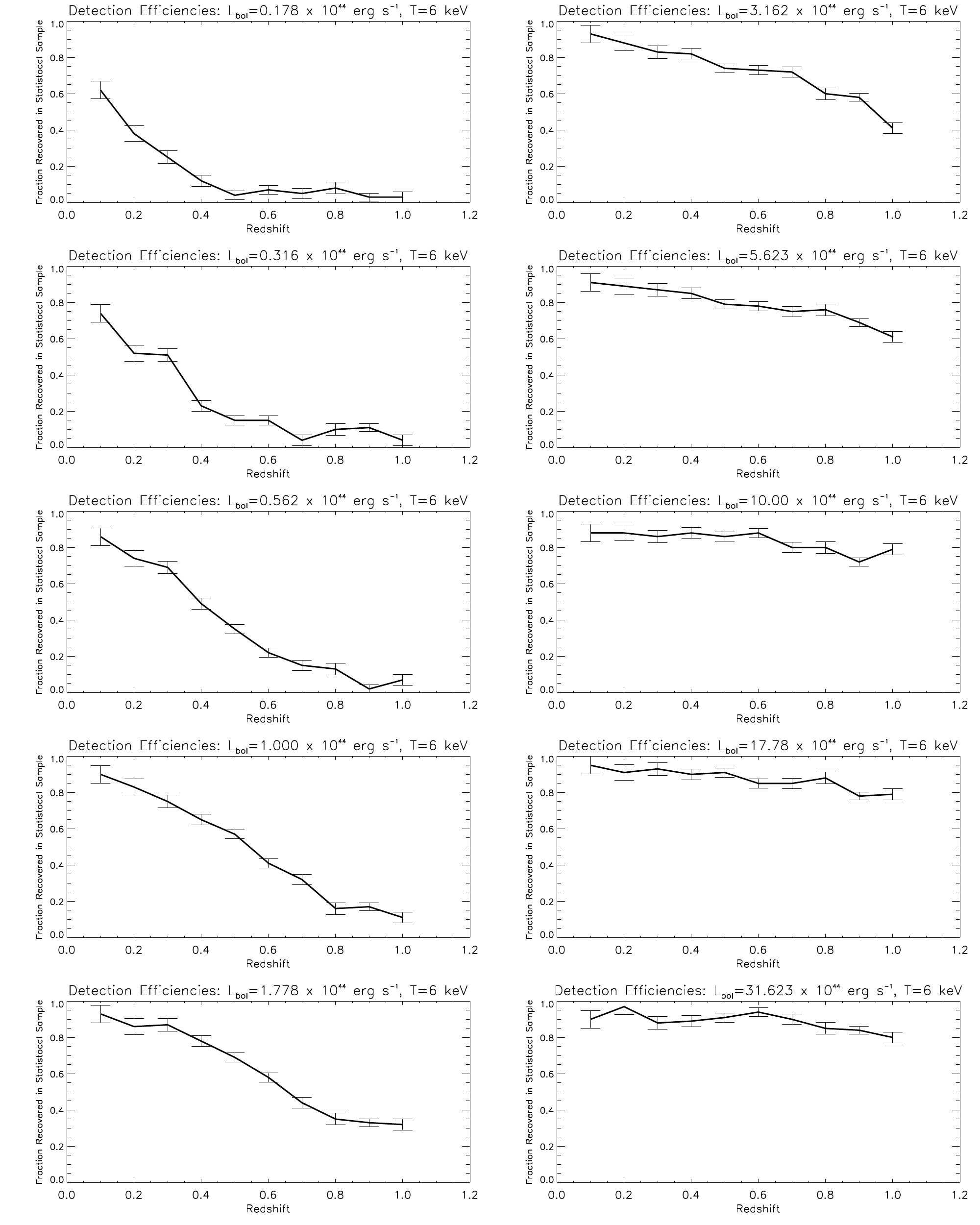}
\caption{Predicted recovery efficiency as a function of redshift for 6
  keV clusters with a range of X-ray luminosities (bolometric). The
  synthetic clusters used for this test had circularly symmetry
  $\beta$-profiles ($\beta=2/3$) with core radii of 160 kpc. The
  typical luminosity of a 6 keV cluster based on the local
  $L_X$--$T_X$ relation is 8 to 15$\times10^{44}$ erg~s$^{-1}$. }
\label{fig:betafun16}
\end{figure*}

\begin{figure*}
\vspace{-0.5cm}
\includegraphics[width=16cm]{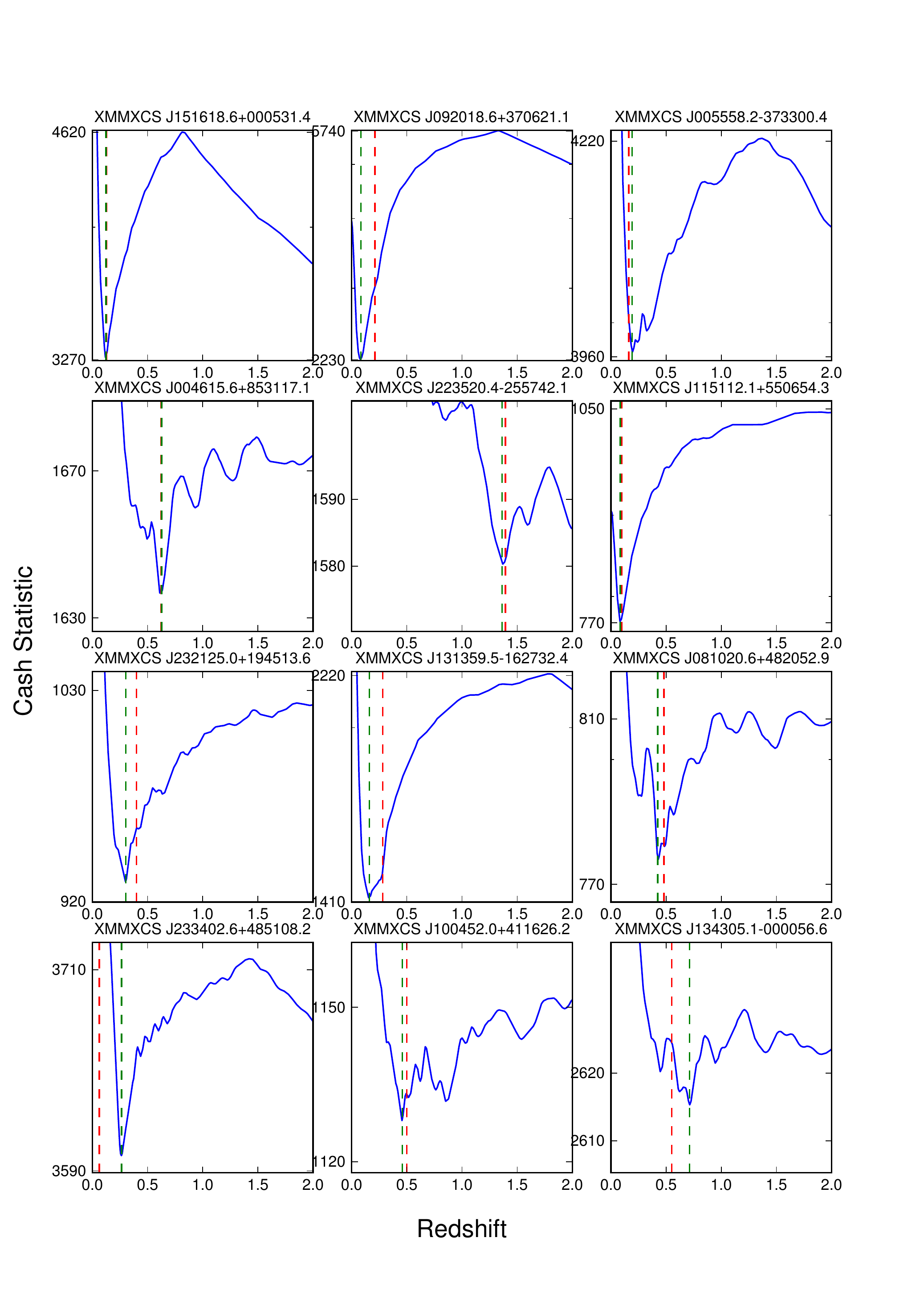}
\caption{Cash statistic output from the X-ray redshift fitting code,
  plotted against redshift. We show 12 XCS clusters that have both
  well measured ($<2.5$\% statistical uncertainty) X-ray
  redshifts and spectroscopically-determined optical redshifts. The optical and X-ray
  redshifts are indicated with red and green dotted lines
  respectively.\label{example_xrz}}
\end{figure*}

%% file: appendix3.tex
\section{XAPA Flowcharts}
\label{sec:xapa_ap}

\begin{figure*}
\includegraphics[width=9.0cm]{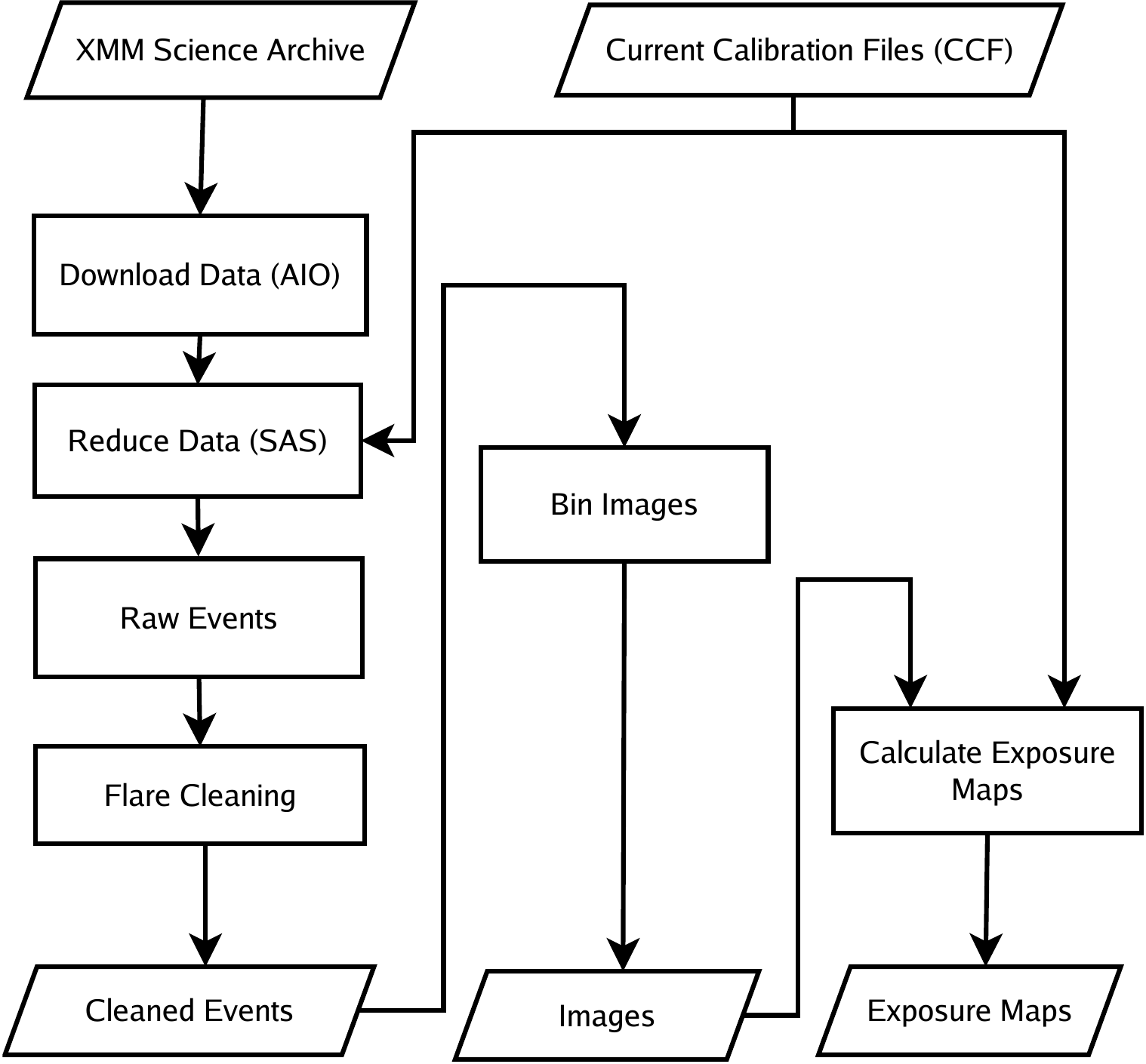}
\caption{Flowchart of the process of XMM data reduction to produce
cleaned event files, images and exposure maps. This
overviews the process by which the XMM data is acquired,
reduced, and cleaned, and the products generated.
\label{reduction_flow}}
\end{figure*}

\begin{figure*}
\includegraphics[width=9.0cm]{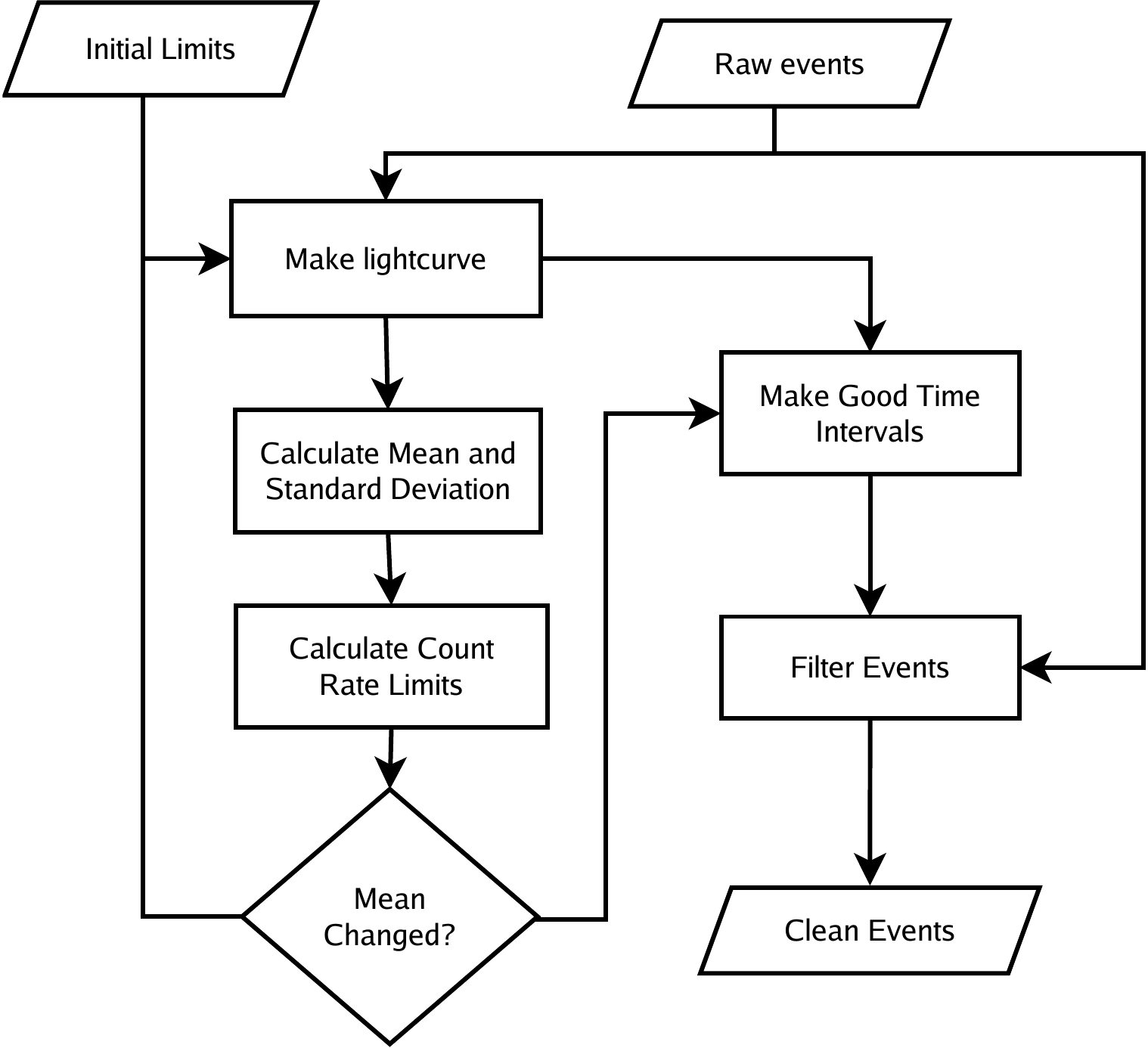}
\caption{Flowchart of the process for removing periods of high background
due to variations in the particle flux to which the instruments are
exposed.  This illustrates the process by which lightcurves of the raw
event files have an iterative 3-$\sigma$ clipping applied to them until
the mean count-rate stops changing. Cleaned event files are then
produced for the good time intervals identified.
\label{cleaning_flow}}
\end{figure*}

\begin{figure*}
\includegraphics[width=11cm]{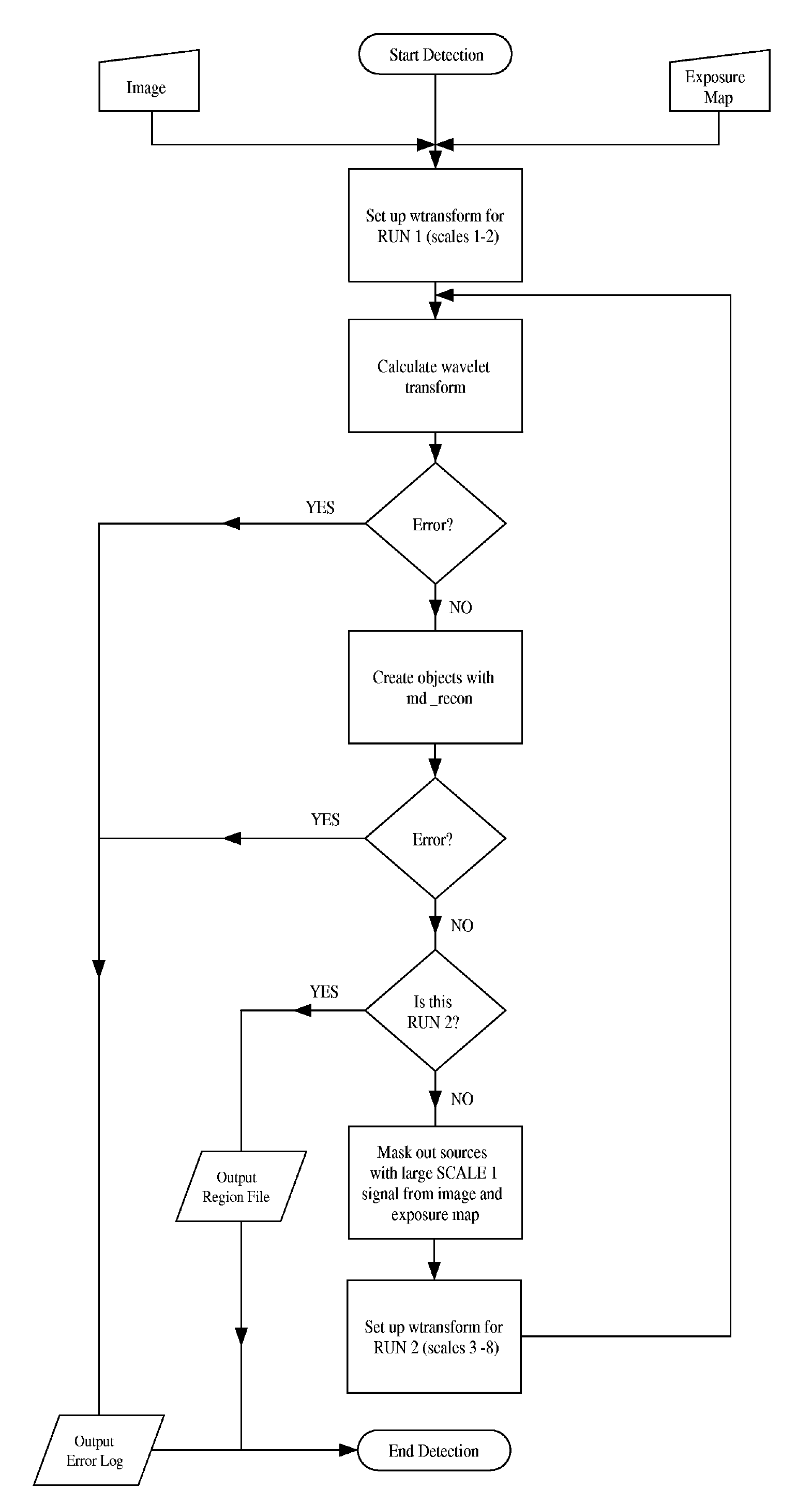}
\caption{Flowchart for the {\tt md\_detect} routine, showing
the two-stage (wavelet transform and reconstruction),
two-pass (to remove pollution of the wavelet signal
by bright, compact sources) process. \label{fig:md_detect_flowchart}}
\end{figure*}

\begin{figure*}
\includegraphics[width=11cm]{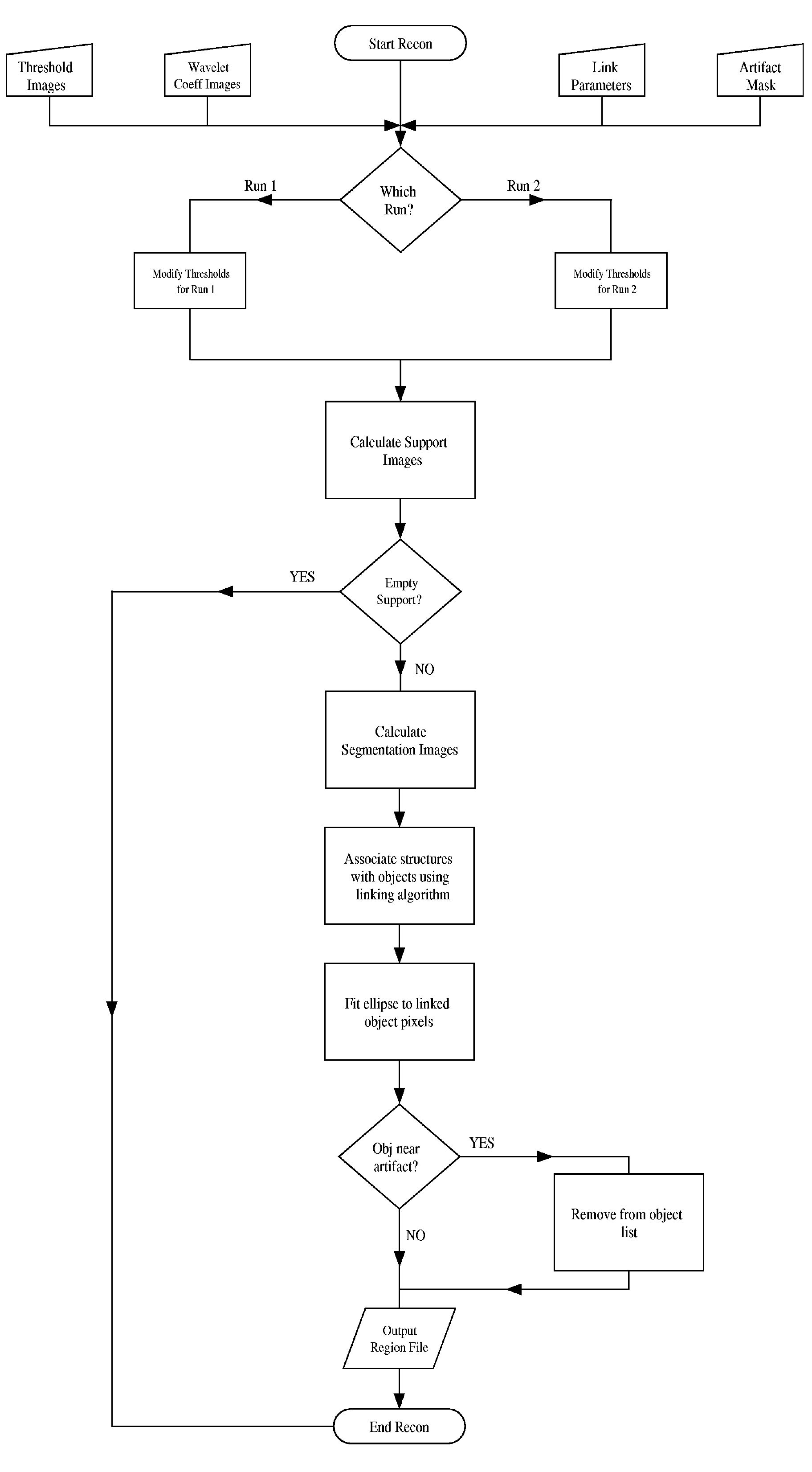}
\caption{Flowchart for the {\tt md\_recon} routine, showing
the stages of the process to reconstruct a source
list from the outputs of {\tt wtransform} on different
scales. \label{fig:md_recon_flowchart}}
\end{figure*}

\begin{figure*}
\includegraphics[width=13cm]{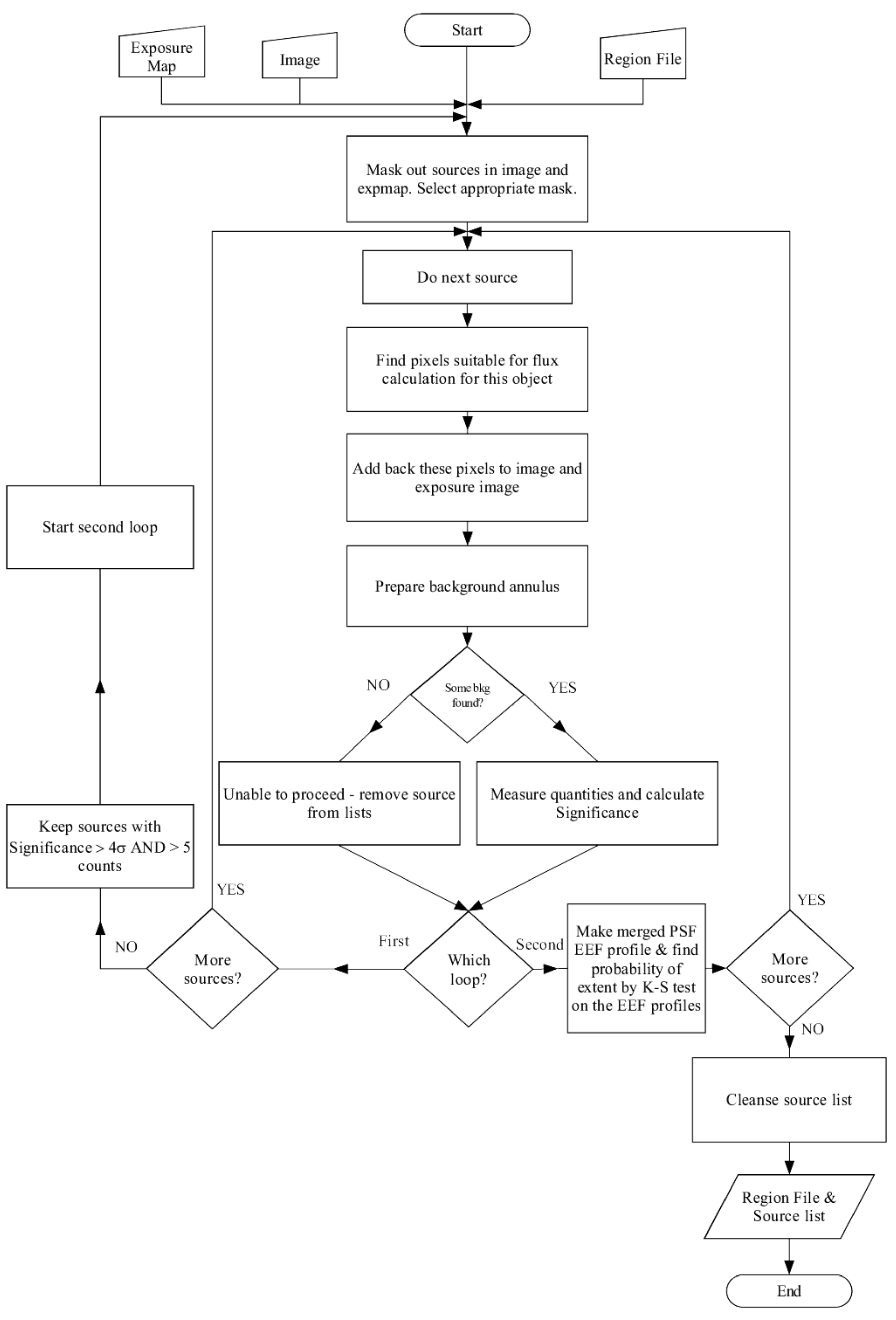}
\caption{Flowchart for the {\tt find\_srcprop} routine, showing the
  stages of the process to derive the properties of each source. The
  routine is run in two stages; the first filters out sources with
  $<4$-$\sigma$ significance, and the second determines the extent
  probability. \label{fig:find_srcprop_flowchart}}
\end{figure*}

\begin{figure*}
\includegraphics[width=11cm]{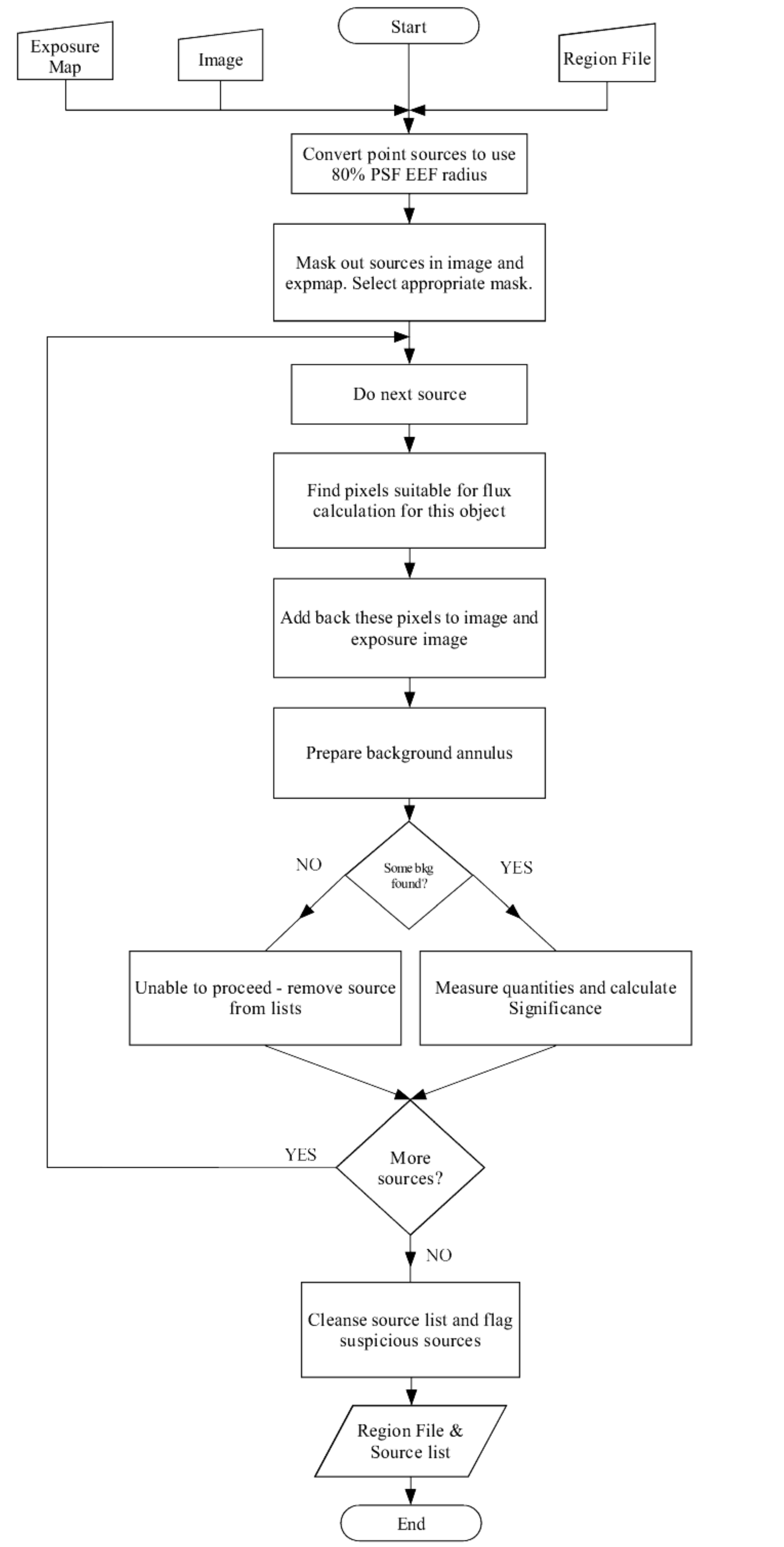}
\caption{Flowchart for the {\tt find\_srcprop\_final} routine. This routine measures properties for every source that is output from  {\tt find\_srcprop}. 
\label{fig:find_srcprop_final_flowchart}}
\end{figure*}

\begin{figure*}
\includegraphics[width=9.5cm]{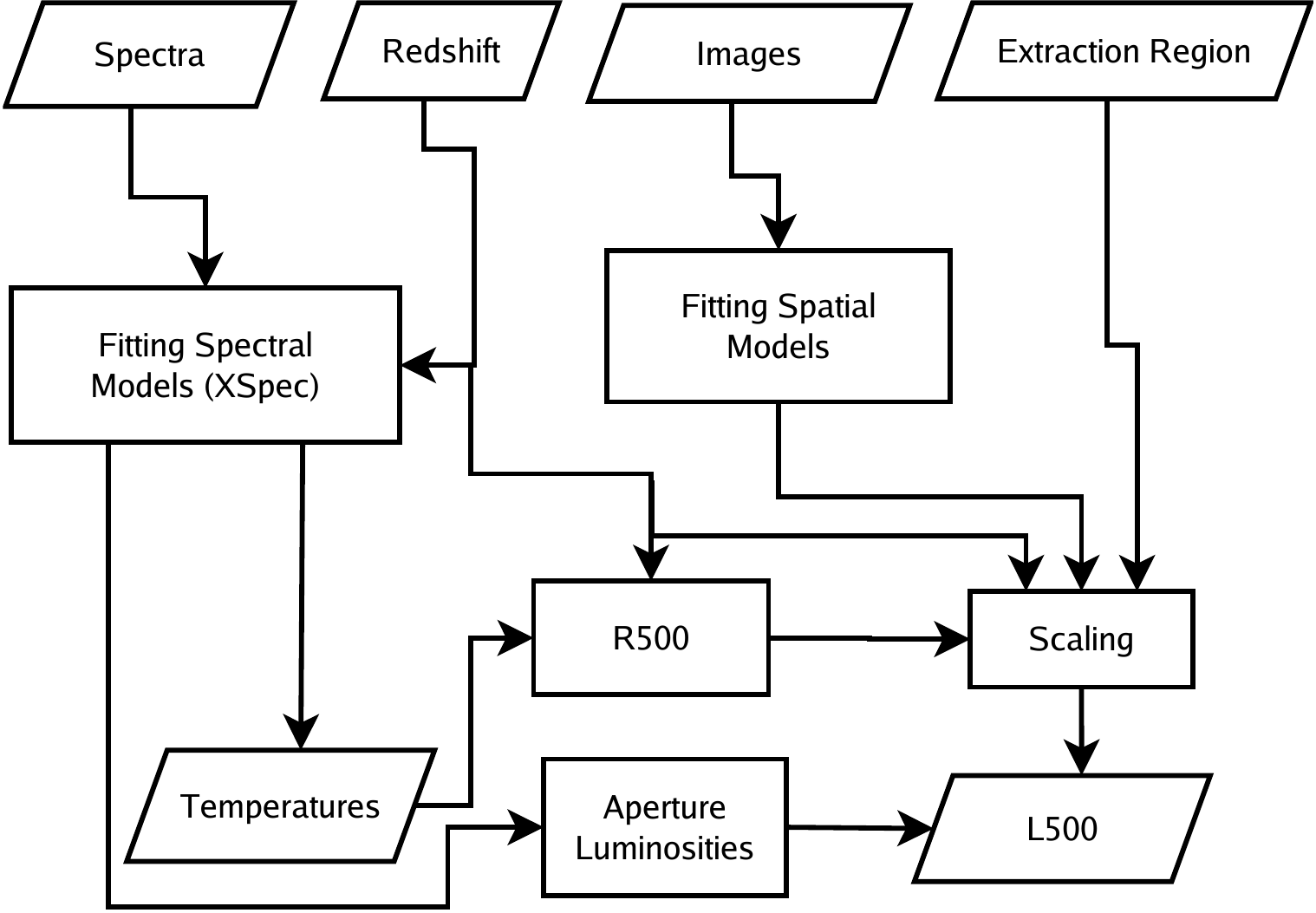}
\caption{Flowchart of process for deriving luminosities and
  temperature from cluster spectra and images. This illustrates the
  process by which models are fitted to the X-ray spectra and images
  to produce temperatures and aperture luminosities, which are
  corrected using the fitted surface-brightness model to produce
  luminosities within $R_{500}$.
\label{analysis_flow}}
\end{figure*}

\begin{figure*}
\includegraphics[width=9.5cm]{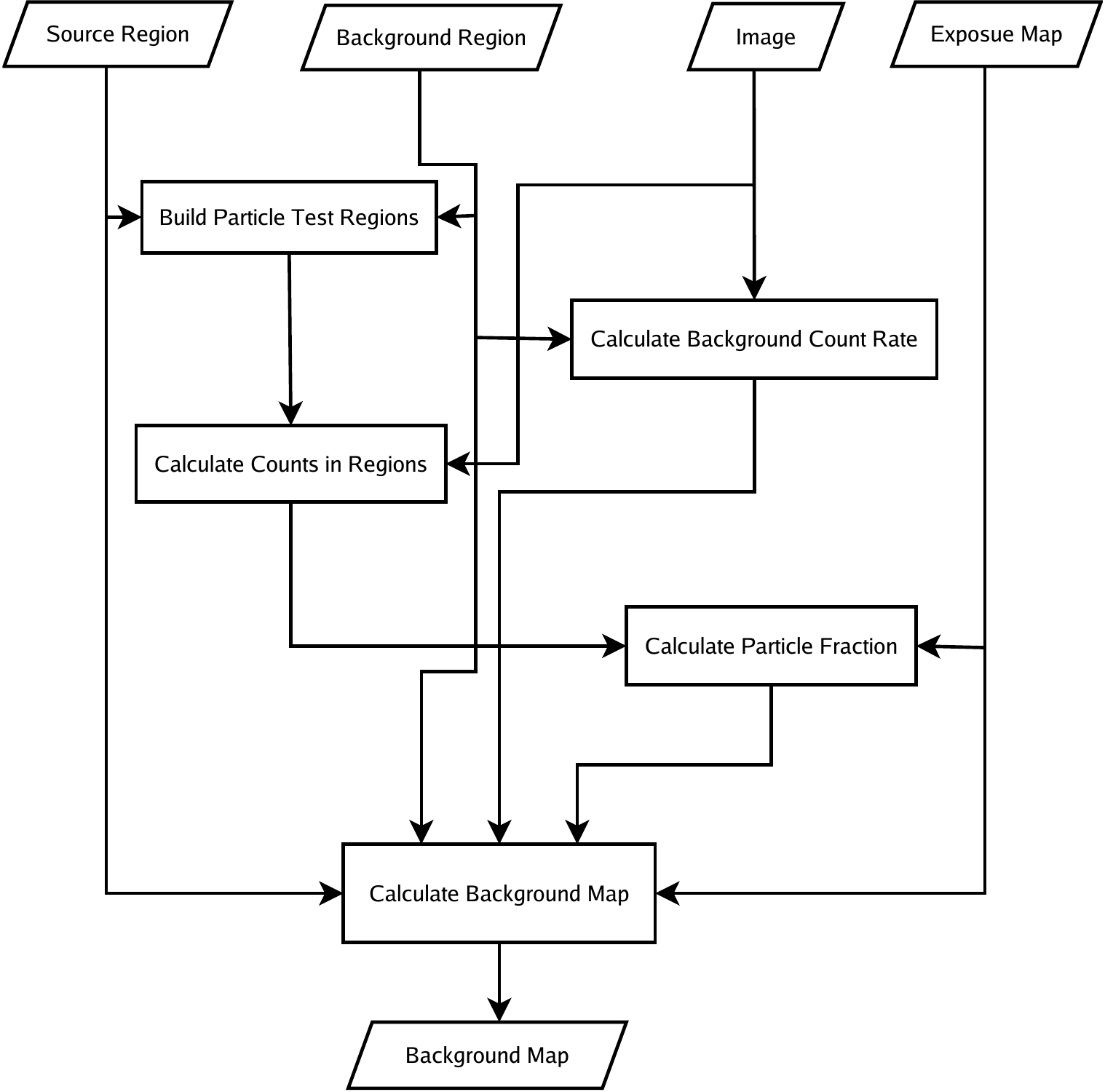}
\caption{Flowchart of process for generating background maps for use
in the XCS surface brightness fitting. This illustrates the process by
which the background measured in an annulus around the source is
extrapolated to all positions in the image, taking into account the
exposure map and the fraction of particles in the background that are
not vignetted by the telescope.
\label{background_flow}}
\end{figure*}

%% file: appendix1.tex
\section{Clusters Used for $T_{\rm X}$ and $L_{\rm X}$ Pipeline Validation}

\begin{table*}
\noindent
\begin{tabular}{llcccccccc}
\hline
Cluster Name& $z$& $n_{\rm H}$ & $T_{\rm X}^{\rm 200}$& $T_{\rm X}^{\rm 300}$ & $T_{\rm X}^{\rm 400}$& $T_{\rm X}^{\rm 500}$ & $T_{\rm X}^{\rm 1000}$ & $T_{\rm X}^{\rm 2000}$\\
&&($10^{20}$ cm$^{-2}$)&(keV)&(keV)&(keV)&(keV)&(keV)&(keV)\\
\hline
XMMXCS J001737.4$-$005235.4 & 0.20 & 2.72 & 6.50$_{-3.23}^{+3.55}$ & 4.82$_{-1.46}^{+2.94}$ & 5.36$_{-1.50}^{+3.71}$ & 4.94$_{-1.39}^{+1.95}$ & 5.24$_{-1.28}^{+0.85}$ & 4.57$_{-0.88}^{+0.30}$\\
XMMXCS J092018.9$+$370617.7 & 0.19 & 1.57 & 2.13$_{-0.57}^{+0.70}$ & 2.51$_{-0.77}^{+0.37}$ & 2.07$_{-0.28}^{+0.67}$ & 2.44$_{-0.34}^{+0.69}$ & 2.28$_{-0.24}^{+0.38}$ & 2.48$_{-0.22}^{+0.24}$\\
XMMXCS J130749.6$+$292549.2 & 0.25 & 1.01 & 2.54$_{-0.72}^{+1.27}$ & 2.93$_{-0.73}^{+0.97}$ & 2.92$_{-0.67}^{+0.92}$ & 2.94$_{-0.50}^{+0.86}$ & 3.17$_{-0.49}^{+0.65}$ & 3.05$_{-0.35}^{+0.45}$\\
XMMXCS J141832.3$+$251104.9 & 0.30 & 1.84 & - & - & 6.81$_{-2.21}^{+7.66}$ & 7.14$_{-2.45}^{+4.95}$ & 7.79$_{-2.56}^{+3.60}$ & 6.30$_{-1.18}^{+1.67}$\\
\hline
\end{tabular}
\caption{Clusters used for the comparison of temperature measurements
  with different numbers of soft-band source counts per spectrum.\label{tab:counts}}
\end{table*}

\begin{table*}
\begin{tabular}{llccc}
\hline
Cluster Name& $z$& $n_{\rm H}$ & On-axis $T_{\rm X}$& Off-axis $T_{\rm X}$\\
&&($10^{20}$ cm$^{-2}$)&(keV)&(keV)\\
\hline
XMMXCS J100304.6$+$325337.9 & 0.42 & 1.55 & 4.11$_{-0.52}^{+0.55}$ & 4.40$_{-2.16}^{+4.37}$\\
XMMXCS J151618.6$+$000531.3 & 0.13 & 4.66 & 5.68$_{-0.21}^{+0.22}$ & 5.11$_{-0.46}^{+0.47}$\\
XMMXCS J184718.3$-$631959.3 & 0.02 & 6.87 & 0.78$_{-0.02}^{+0.01}$ & 0.81$_{-0.03}^{+0.02}$\\
XMMXCS J130832.6$+$534213.8 & 0.33 & 1.58 & 3.66$_{-0.56}^{+0.70}$ & 4.45$_{-1.54}^{+4.36}$\\
XMMXCS J072054.3$+$710900.5 & 0.23 & 3.88 & 2.99$_{-0.92}^{+1.48}$ & 2.93$_{-0.90}^{+1.34}$\\
XMMXCS J132508.7$+$655027.9 & 0.18 & 2.00 & 1.01$_{-0.19}^{+0.20}$ & 0.71$_{-0.19}^{+0.44}$\\
XMMXCS J022403.8$-$041333.4 & 1.05 & 2.51 & 3.73$_{-0.67}^{+0.89}$ & 4.07$_{-1.71}^{+5.92}$\\
XMMXCS J223520.4$-$255742.1 & 1.39 & 1.47 & 9.45$_{-2.44}^{+3.19}$ & 11.29$_{-5.82}^{+14.03}$\\
\hline
\end{tabular}
\caption{Clusters used for the on/off-axis comparison of temperature measurements.\label{tab:axis}}
\end{table*}

\begin{table*}
\begin{tabular}{lcccc}
\hline
Cluster Name &     $z$& $n_{\rm H}$                     & $T_{\rm X}^{\rm lit}$& $T_{\rm X}^{\rm XCS}$\\
                 &            & ($10^{20}$ cm$^{-2}$) &(keV)                           &(keV)                      \\
\hline
 XMMU J131359.7$-$162735 &  0.28 & 4.92 & $3.57_{-0.12}^{+0.12}$ & $3.45_{-0.19}^{+0.19}$\\
 2XMM J100451.6$+$411627 &  0.82 & 0.89 & $4.2_{-0.4}^{+0.4}$  & $5.36_{-0.78}^{+0.96}$\\
XLSS J022045.4$-$032558 & 0.33 & 2.49 & $1.7_{-0.2}^{+0.3}$  & $2.27_{-0.51}^{+0.92}$\\
 XLSS J022145.2$-$034617 &  0.43 & 2.52 & $4.8_{-0.5}^{+0.6}$  & $5.78_{-1.20}^{+1.83}$\\
 XLSS J022404.1$-$041330 &  1.05 & 2.51 & $4.1_{-0.7}^{+0.9}$ & $3.74_{-0.38}^{+0.51}$\\
 XLSS J022433.8$-$041405 &  0.26 & 2.46 & $1.3_{-0.1}^{+0.2}$ & $1.16_{-0.24}^{+0.19}$\\
XLSS J022457.1$-$034856 &  0.61 & 2.49 & $3.2_{-0.3}^{+0.4}$ & $3.87_{-0.85}^{+1.23}$\\
 XLSS J022524.7$-$044039 & 0.26 & 2.49 & $2.0_{-0.2}^{+0.2}$  & $2.27_{-0.53}^{+0.99}$\\
 XLSS J022530.6$-$041420 & 0.14 & 2.35 & $1.34_{-0.06}^{+0.14}$ & $1.22_{-0.13}^{+0.12}$\\
 XLSS J022540.6$-$031121 &  0.14 & 2.66 & $3.5_{-0.5}^{+0.6}$ & $4.16_{-0.51}^{+0.52}$\\
XLSS J022616.3$-$023957 & 0.06 & 2.67 & $0.63_{-0.03}^{+0.03}$ & $0.62_{-0.04}^{+0.04}$\\
 XLSS J022722.4$-$032144 &  0.33 & 2.61 & $2.4_{-0.4}^{+0.5}$  & $2.82_{-0.64}^{+1.02}$\\
 XLSS J022739.9$-$045127 &  0.29 & 2.59 & $1.7_{-0.1}^{+0.1}$ & $1.56_{-0.18}^{+0.33}$\\
 XLSS J022803.4$-$045103 &  0.29 & 2.67 & $2.8_{-0.5}^{+0.6}$ & $2.38_{-0.93}^{+2.13}$\\
\hline
\end{tabular}
\caption{Clusters used for the literature comparison of temperature measurements. Redshifts and $T_{\rm X}$ values for the XLSS clusters 
come from \citet{pacaud07a}, those for the XMMU and 2XMM clusters from \citet{gastaldello07a} and \citet{hoeft08a} respectively.  
\label{tab:lit}}
\end{table*}

\begin{table*}
\begin{tabular}{llccc}
\hline
Cluster Name& $z$& $n_{\rm H}$ & $\beta^{\rm lit}$& $\beta^{\rm XCS}$\\
&&($10^{20}$ cm$^{-2}$)&&\\
\hline
XLSS J022726.0$-$043216 & 0.31 & 2.67 & $0.58_{-0.14}^{+1.25}$ & $0.61_{-0.21}^{+0.11}$\\
XLSS J022356.5$-$030558 & 0.30 & 2.63 & $0.43_{-0.05}^{+0.08}$ & $0.43_{-0.09}^{+0.13}$\\
XLSS J022616.3$-$023957	& 0.06 & 2.67 & $0.69_{-0.05}^{+0.07}$ & $0.79_{-0.04}^{+0.05}$\\
XLSS J022045.4$-$032558 & 0.33 & 2.49 & $0.54_{-0.03}^{+0.03}$ & $0.55_{-0.04}^{+0.06}$\\
\hline
\end{tabular}
\caption{Clusters used for the literature $\beta$ comparison with the
work of \citet{alshino10a}.\label{tab:beta}}
\end{table*}

\begin{table*}
\begin{tabular}{llccc}
\hline
Cluster Name& $z$& $n_{\rm H}$ &  $L_{\rm X, 500}^{\rm lit}$& XCS $L_{\rm X, 500}^{\rm fit}$\\
&&($10^{20}$ cm$^{-2}$)&(10$^{43}$ erg s$^{-1}$)&(10$^{43}$ erg s$^{-1}$)\\
\hline
XLSS J022540.6$-$031121 & 0.14 & 2.66 & $0.93_{-0.06}^{+0.05}$ & $1.1_{-1.1}^{+1.2}$\\
XLSS J022616.3$-$023957 & 0.06 & 2.67 & $0.025_{-0.002}^{+0.002}$ & $0.019_{-0.017}^{+0.050}$\\
XLSS J022404.1$-$041330 & 1.05 & 2.52 & $4.83_{-0.34}^{+0.37}$ & $4.9_{-4.5}^{+5.2}$\\
XLSS J022206.7$-$030314	& 0.49 & 2.52 & $2.89_{-0.15}^{+0.15}$ & $2.3_{-2.1}^{+2.5}$\\
XLSS J022457.1$-$034856 & 0.61 & 2.49 & $3.32_{-0.15}^{+0.15}$ & $3.0_{-2.8}^{+3.2}$\\
XLSS J022045.4$-$032558 & 0.33 & 2.49 & $0.38_{-0.03}^{+0.03}$ & $0.35_{-0.30}^{+0.42}$\\
\hline
\end{tabular}
\caption{Clusters used for the literature luminosity comparison with the
work of \citet{pacaud07a}.\label{tab:lum}}

\label{lastpage}

\end{table*}

%% file: paper.bbl
\begin{thebibliography}{}

\bibitem[\protect\citeauthoryear{{Abazajian} et~al.,}{{Abazajian}
  et~al.}{2009}]{abazajian09a}
{Abazajian} K.~N.,  et~al., 2009, ApJS, 182, 543

\bibitem[\protect\citeauthoryear{{Abbey}, {Carpenter}, {Read}, {Wells}, {Xmm
  Science Centre} \& {Swift Mission Operations Center}}{{Abbey}
  et~al.}{2006}]{abbey06a}
{Abbey} T.,  {Carpenter} J.,  {Read} A.,  {Wells} A.,  {Xmm Science Centre}
  {Swift Mission Operations Center} 2006, in {A.~Wilson} ed., The X-ray
  Universe 2005 Vol.~604 of ESA Special Publication, {Micrometeroid Damage to
  CCDs in XMM-Newton and Swift and its Significance for Future X-ray Missions}.
p.~943

\bibitem[\protect\citeauthoryear{Abell}{Abell}{1958}]{abell58a}
Abell A.~O.,  1958, ApJS, 3, 211

\bibitem[\protect\citeauthoryear{{Adami} et~al.,}{{Adami}
  et~al.}{2011}]{Adami11}
{Adami} C.,  et~al., 2011, A\&A, 526, A18+

\bibitem[\protect\citeauthoryear{{Adami}, {Ulmer}, {Romer}, {Nichol}, {Holden}
  \& {Pildis}}{{Adami} et~al.}{2000}]{adami00a}
{Adami} C.,  {Ulmer} M.~P.,  {Romer} A.~K.,  {Nichol} R.~C.,  {Holden} B.~P.,
   {Pildis} R.~A.,  2000, ApJS, 131, 391

\bibitem[\protect\citeauthoryear{{Ade} et~al.,}{{Ade}
  et~al.}{2011}]{planckcollab11a}
{Ade} P.~A.~R.,  et~al., 2011, arXiv:1101.2024

\bibitem[\protect\citeauthoryear{{Allen}, {Evrard} \& {Mantz}}{{Allen}
  et~al.}{2011}]{allen11a}
{Allen} S.~W.,  {Evrard} A.~E.,    {Mantz} A.~B.,  2011, arXiv:1103.4829

\bibitem[\protect\citeauthoryear{Alshino, Ponman, Pacaud \& Pierre}{Alshino
  et~al.}{2010}]{alshino10a}
Alshino A.,  Ponman T.,  Pacaud F.,    Pierre M.,  2010, preprint,
  arXiv:1005.4958

\bibitem[\protect\citeauthoryear{Altieri, Chen, Gabriel, Gondoin, Lammers, Lumb
  \& Kirsch.}{Altieri et~al.}{2004}]{altieri04a}
Altieri B.,  Chen B.,  Gabriel C.,  Gondoin P.,  Lammers U.,  Lumb D.,
  Kirsch. 2004, Technical Report XMM-PS-GM-20 Issue 3.1, XMM-Newton Calibration
  Access and Data Handbook.
ESA

\bibitem[\protect\citeauthoryear{Anders \& Grevesse}{Anders \&
  Grevesse}{1989}]{anders89a}
Anders E.,  Grevesse N.,  1989, Geochimica et Cosmochimica Acta, 53, 197

\bibitem[\protect\citeauthoryear{{Arabadjis} \& {Bregman}}{{Arabadjis} \&
  {Bregman}}{1999}]{arabadjis99a}
{Arabadjis} J.~S.,  {Bregman} J.~N.,  1999, ApJ, 510, 806

\bibitem[\protect\citeauthoryear{{Arnaud}}{{Arnaud}}{1996}]{arnaud96a}
{Arnaud} K.~A.,  1996, in {G.~H.~Jacoby \& J.~Barnes} ed., Astronomical Data
  Analysis Software and Systems V Vol.~101 of Astronomical Society of the
  Pacific Conference Series, {XSPEC: The First Ten Years}.
p.~17

\bibitem[\protect\citeauthoryear{Arnaud \& Evrard}{Arnaud \&
  Evrard}{1999}]{arnaud99a}
Arnaud M.,  Evrard A.~E.,  1999, MNRAS, 305, 631

\bibitem[\protect\citeauthoryear{{Arnaud}, {Pointecouteau} \& {Pratt}}{{Arnaud}
  et~al.}{2005}]{arnaud05a}
{Arnaud} M.,  {Pointecouteau} E.,    {Pratt} G.~W.,  2005, A\&A, 441, 893

\bibitem[\protect\citeauthoryear{{Arnaud}, {Pratt}, {Piffaretti},
  {B{\"o}hringer}, {Croston} \& {Pointecouteau}}{{Arnaud}
  et~al.}{2010}]{arnaud10a}
{Arnaud} M.,  {Pratt} G.~W.,  {Piffaretti} R.,  {B{\"o}hringer} H.,  {Croston}
  J.~H.,    {Pointecouteau} E.,  2010, A\&A, 517, A92

\bibitem[\protect\citeauthoryear{{Arviset}, {Guainazzi}, {Hernandez}, {Dowson},
  {Osuna} \& {Venet}}{{Arviset} et~al.}{2002}]{arviset02a}
{Arviset} C.,  {Guainazzi} M.,  {Hernandez} J.,  {Dowson} J.,  {Osuna} P.,
  {Venet} A.,  2002, ArXiv Astrophysics e-prints

\bibitem[\protect\citeauthoryear{{Arviset}, {Osuna} \& {Salgado}}{{Arviset}
  et~al.}{2004}]{arviset04a}
{Arviset} C.,  {Osuna} P.,    {Salgado} J.,  2004, in {F.~Ochsenbein,
  M.~G.~Allen, \& D.~Egret} ed., Astronomical Data Analysis Software and
  Systems (ADASS) XIII Vol.~314 of Astronomical Society of the Pacific
  Conference Series, {ESA ISO and XMM-Newton Archives Inter-Operability and VO
  services}.
p.~574

\bibitem[\protect\citeauthoryear{{Barkhouse} et~al.,}{{Barkhouse}
  et~al.}{2006}]{barkhouse06a}
{Barkhouse} W.~A.,  et~al., 2006, ApJ, 645, 955

\bibitem[\protect\citeauthoryear{{Bertin} \& {Arnouts}}{{Bertin} \&
  {Arnouts}}{1996}]{bertin96a}
{Bertin} E.,  {Arnouts} S.,  1996, A\&AS, 117, 393

\bibitem[\protect\citeauthoryear{Bird, Mushotzky \& Metzler}{Bird
  et~al.}{1995}]{bird95a}
Bird C.,  Mushotzky R.~F.,    Metzler C.~A.,  1995, ApJ, 453, 40

\bibitem[\protect\citeauthoryear{{Blackburn}}{{Blackburn}}{1995}]{blackburn95a}
{Blackburn} J.~K.,  1995, in {R.~A.~Shaw, H.~E.~Payne, \& J.~J.~E.~Hayes} ed.,
  Astronomical Data Analysis Software and Systems IV Vol.~77 of Astronomical
  Society of the Pacific Conference Series, {FTOOLS: A FITS Data Processing and
  Analysis Software Package}.
p.~367

\bibitem[\protect\citeauthoryear{{Blanton}, {Gregg}, {Helfand}, {Becker} \&
  {White}}{{Blanton} et~al.}{2003}]{blanton03a}
{Blanton} E.~L.,  {Gregg} M.~D.,  {Helfand} D.~J.,  {Becker} R.~H.,    {White}
  R.~L.,  2003, AJ, 125, 1635

\bibitem[\protect\citeauthoryear{{B{\"o}hringer} et~al.,}{{B{\"o}hringer}
  et~al.}{2000}]{bohringer00a}
{B{\"o}hringer} H.,  et~al., 2000, ApJS, 129, 435

\bibitem[\protect\citeauthoryear{{B{\"o}hringer} et~al.,}{{B{\"o}hringer}
  et~al.}{2004}]{bohringer04a}
{B{\"o}hringer} H.,  et~al., 2004, A\&A, 425, 367

\bibitem[\protect\citeauthoryear{Boyle \& Smith}{Boyle \&
  Smith}{1970}]{boyle70a}
Boyle W.~S.,  Smith G.~E.,  1970, Bell Sys. Tech. J., 49, 587

\bibitem[\protect\citeauthoryear{{Bremer} et~al.,}{{Bremer}
  et~al.}{2006}]{bremer06a}
{Bremer} M.~N.,  et~al., 2006, MNRAS, 371, 1427

\bibitem[\protect\citeauthoryear{{Burenin}, {Vikhlinin}, {Hornstrup},
  {Ebeling}, {Quintana} \& {Mescheryakov}}{{Burenin} et~al.}{2007}]{burenin07a}
{Burenin} R.~A.,  {Vikhlinin} A.,  {Hornstrup} A.,  {Ebeling} H.,  {Quintana}
  H.,    {Mescheryakov} A.,  2007, ApJS, 172, 561

\bibitem[\protect\citeauthoryear{{Burke}, {Collins}, {Sharples}, {Romer} \&
  {Nichol}}{{Burke} et~al.}{2003}]{burke03a}
{Burke} D.~J.,  {Collins} C.~A.,  {Sharples} R.~M.,  {Romer} A.~K.,    {Nichol}
  R.~C.,  2003, MNRAS, 341, 1093

\bibitem[\protect\citeauthoryear{{Cash}}{{Cash}}{1979}]{cash79a}
{Cash} W.,  1979, ApJ, 228, 939

\bibitem[\protect\citeauthoryear{Cavaliere \& Fusco-Femiano}{Cavaliere \&
  Fusco-Femiano}{1976}]{cavaliere76a}
Cavaliere A.,  Fusco-Femiano R.,  1976, A\&A, 49, 137

\bibitem[\protect\citeauthoryear{{Clavel}}{{Clavel}}{1998}]{clavel98a}
{Clavel} J.,  1998, in Science with XMM, {XMM science archive}

\bibitem[\protect\citeauthoryear{{Collins} et~al.,}{{Collins}
  et~al.}{2009}]{collins09a}
{Collins} C.~A.,  et~al., 2009, nat, 458, 603

\bibitem[\protect\citeauthoryear{{Croston} et~al.,}{{Croston}
  et~al.}{2008}]{croston08a}
{Croston} J.~H.,  et~al., 2008, A\&A, 487, 431

\bibitem[\protect\citeauthoryear{{Cruddace} et~al.,}{{Cruddace}
  et~al.}{2002}]{cruddace02a}
{Cruddace} R.,  et~al., 2002, ApJS, 140, 239

\bibitem[\protect\citeauthoryear{{Cunha}, {Huterer} \& {Frieman}}{{Cunha}
  et~al.}{2009}]{cunha09a}
{Cunha} C.,  {Huterer} D.,    {Frieman} J.~A.,  2009, Phys.~Rev.~D, 80, 063532

\bibitem[\protect\citeauthoryear{Davidson}{Davidson}{2006}]{davidson06a}
Davidson M.,  2006, PhD thesis, IfA, University of Edinburgh.

\bibitem[\protect\citeauthoryear{{Deponte Evans}, {Cresitello-Dittmar}, {Doe},
  {Evans}, {Germain}, {Glotfelty} \& {Overly}}{{Deponte Evans}
  et~al.}{2008}]{deponteevans08a}
{Deponte Evans} J.,  {Cresitello-Dittmar} M.,  {Doe} S.,  {Evans} I.~N.,
  {Germain} G.,  {Glotfelty} K.,    {Overly} J.,  2008, in {R.~W.~Argyle,
  P.~S.~Bunclark, \& J.~R.~Lewis} ed., Astronomical Data Analysis Software and
  Systems XVII Vol.~394 of Astronomical Society of the Pacific Conference
  Series, {CIAO 4 Infrastructure --- Moving in a Modular Direction}.
p.~627

\bibitem[\protect\citeauthoryear{Dickey \& Lockman}{Dickey \&
  Lockman}{1990}]{dickey90a}
Dickey J.~M.,  Lockman F.~J.,  1990, ARA\&A, 28, 215

\bibitem[\protect\citeauthoryear{{Doe}, {Noble} \& {Smith}}{{Doe}
  et~al.}{2001}]{doe01a}
{Doe} S.,  {Noble} M.,    {Smith} R.,  2001, in {F.~R.~Harnden Jr.,
  F.~A.~Primini, \& H.~E.~Payne} ed., Astronomical Data Analysis Software and
  Systems X Vol.~238 of Astronomical Society of the Pacific Conference Series,
  {Interactive Analysis and Scripting in CIAO 2.0}.
p.~310

\bibitem[\protect\citeauthoryear{{Dunkley} et~al.,}{{Dunkley}
  et~al.}{2010}]{dunkley10a}
{Dunkley} J.,  et~al., 2010, arXiv:1009.0866

\bibitem[\protect\citeauthoryear{{Dupke} \& {Bregman}}{{Dupke} \&
  {Bregman}}{2001a}]{dupke01b}
{Dupke} R.~A.,  {Bregman} J.~N.,  2001a, ApJ, 562, 266

\bibitem[\protect\citeauthoryear{{Dupke} \& {Bregman}}{{Dupke} \&
  {Bregman}}{2001b}]{dupke01a}
{Dupke} R.~A.,  {Bregman} J.~N.,  2001b, ApJ, 547, 705

\bibitem[\protect\citeauthoryear{Ebeling, Edge, Allen, Crawford, Fabian \&
  Huchra}{Ebeling et~al.}{2000}]{ebeling00a}
Ebeling H.,  Edge A.,  Allen S.,  Crawford C.,  Fabian A.,    Huchra J.,  2000,
  MNRAS, 318, 333

\bibitem[\protect\citeauthoryear{Ebeling, Edge, B\dd{o}hringer, Allen,
  Crawford, Fabian, Voges \& Huchra}{Ebeling et~al.}{1998}]{ebeling98a}
Ebeling H.,  Edge A.~C.,  B\dd{o}hringer H.,  Allen S.~W.,  Crawford C.~S.,
  Fabian A.~C.,  Voges W.,    Huchra J.~P.,  1998, MNRAS, 301, 881

\bibitem[\protect\citeauthoryear{{Ebeling}, {Edge} \& {Henry}}{{Ebeling}
  et~al.}{2001}]{ebeling01a}
{Ebeling} H.,  {Edge} A.~C.,    {Henry} J.~P.,  2001, ApJ, 553, 668

\bibitem[\protect\citeauthoryear{{Ebeling}, {Mullis} \& {Tully}}{{Ebeling}
  et~al.}{2002}]{ebeling02a}
{Ebeling} H.,  {Mullis} C.~R.,    {Tully} R.~B.,  2002, ApJ, 580, 774

\bibitem[\protect\citeauthoryear{{Fassbender}, {B{\"o}hringer}, {Lamer},
  {Mullis}, {Rosati}, {Schwope}, {Kohnert} \& {Santos}}{{Fassbender}
  et~al.}{2008}]{fassbender08a}
{Fassbender} R.,  {B{\"o}hringer} H.,  {Lamer} G.,  {Mullis} C.~R.,  {Rosati}
  P.,  {Schwope} A.,  {Kohnert} J.,    {Santos} J.~S.,  2008, A\&A, 481, L73

\bibitem[\protect\citeauthoryear{{Fassbender} et~al.,}{{Fassbender}
  et~al.}{2010}]{fassbender10a}
{Fassbender} R.,  et~al., 2010, arXiv:1009.0264

\bibitem[\protect\citeauthoryear{{Finoguenov} et~al.,}{{Finoguenov}
  et~al.}{2007}]{finoguenov07a}
{Finoguenov} A.,  et~al., 2007, ApJS, 172, 182

\bibitem[\protect\citeauthoryear{{Finoguenov} et~al.,}{{Finoguenov}
  et~al.}{2010}]{finoguenov10a}
{Finoguenov} A.,  et~al., 2010, MNRAS, 403, 2063

\bibitem[\protect\citeauthoryear{{Foley} et~al.,}{{Foley}
  et~al.}{2011}]{Foley11}
{Foley} R.~J.,  et~al., 2011, ApJ, 731, 86

\bibitem[\protect\citeauthoryear{{Freeman}, {Kashyap}, {Rosner} \&
  {Lamb}}{{Freeman} et~al.}{2002}]{freeman02a}
{Freeman} P.~E.,  {Kashyap} V.,  {Rosner} R.,    {Lamb} D.~Q.,  2002, ApJS,
  138, 185

\bibitem[\protect\citeauthoryear{{Frenk}, {White}, {Efstathiou} \&
  {Davis}}{{Frenk} et~al.}{1990}]{frenk90a}
{Frenk} C.~S.,  {White} S.~D.~M.,  {Efstathiou} G.,    {Davis} M.,  1990, ApJ,
  351, 10

\bibitem[\protect\citeauthoryear{{Fruscione} et~al.,}{{Fruscione}
  et~al.}{2006}]{fruscione06a}
{Fruscione} A.,  et~al., 2006, in SPIE Conference Series Vol.~6270 of SPIE
  Conference Series, {CIAO: Chandra's data analysis system}

\bibitem[\protect\citeauthoryear{{Gabriel} et~al.,}{{Gabriel}
  et~al.}{2004}]{gabriel04a}
{Gabriel} C.,  et~al., 2004, in {F.~Ochsenbein, M.~G.~Allen, \& D.~Egret} ed.,
  Astronomical Data Analysis Software and Systems (ADASS) XIII Vol.~314 of
  Astronomical Society of the Pacific Conference Series, {The XMM-Newton SAS -
  Distributed Development and Maintenance of a Large Science Analysis System: A
  Critical Analysis}.
p.~759

\bibitem[\protect\citeauthoryear{{Gastaldello}, {Buote}, {Humphrey},
  {Zappacosta}, {Seigar}, {Barth}, {Brighenti} \& {Mathews}}{{Gastaldello}
  et~al.}{2007}]{gastaldello07a}
{Gastaldello} F.,  {Buote} D.~A.,  {Humphrey} P.~J.,  {Zappacosta} L.,
  {Seigar} M.~S.,  {Barth} A.~J.,  {Brighenti} F.,    {Mathews} W.~G.,  2007,
  ApJ, 662, 923

\bibitem[\protect\citeauthoryear{Ghizzardi}{Ghizzardi}{2001}]{ghizzardi01a}
Ghizzardi S.,  2001, Technical Report EPIC-MCT-TN-011, In Flight Calibration of
  the PSF for the MOS1 and MOS2 Cameras.
ESA

\bibitem[\protect\citeauthoryear{Ghizzardi}{Ghizzardi}{2002}]{ghizzardi02a}
Ghizzardi S.,  2002, Technical Report EPIC-MCT-TN-012, In Flight Calibration of
  the PSF for the PN Camera.
ESA

\bibitem[\protect\citeauthoryear{{Gioia}, {Henry}, {Mullis}, {B{\"o}hringer},
  {Briel}, {Voges} \& {Huchra}}{{Gioia} et~al.}{2003}]{gioia03a}
{Gioia} I.~M.,  {Henry} J.~P.,  {Mullis} C.~R.,  {B{\"o}hringer} H.,  {Briel}
  U.~G.,  {Voges} W.,    {Huchra} J.~P.,  2003, ApJS, 149, 29

\bibitem[\protect\citeauthoryear{{Gioia}, {Maccacaro}, {Schild}, {Wolter},
  {Stocke}, {Morris} \& {Henry}}{{Gioia} et~al.}{1990}]{gioia90a}
{Gioia} I.~M.,  {Maccacaro} T.,  {Schild} R.~E.,  {Wolter} A.,  {Stocke} J.~T.,
   {Morris} S.~L.,    {Henry} J.~P.,  1990, ApJS, 72, 567

\bibitem[\protect\citeauthoryear{{Gladders} \& {Yee}}{{Gladders} \&
  {Yee}}{2000}]{gladders00a}
{Gladders} M.~D.,  {Yee} H.~K.~C.,  2000, AJ, 120, 2148

\bibitem[\protect\citeauthoryear{{Gobat} et~al.,}{{Gobat}
  et~al.}{2011}]{gobat11}
{Gobat} R.,  et~al., 2011, A\&A, 526, A133+

\bibitem[\protect\citeauthoryear{{Gondoin}, {Aschenbach}, {Erd}, {Lumb},
  {Majerowicz}, {Neumann} \& {Sauvageot}}{{Gondoin} et~al.}{2000}]{gondoin00a}
{Gondoin} P.,  {Aschenbach} B.,  {Erd} C.,  {Lumb} D.~H.,  {Majerowicz} S.,
  {Neumann} D.,    {Sauvageot} J.~L.,  2000, in {K.~A.~Flanagan \&
  O.~H.~Siegmund} ed., Society of Photo-Optical Instrumentation Engineers
  (SPIE) Conference Series Vol.~4140 of Presented at the Society of
  Photo-Optical Instrumentation Engineers (SPIE) Conference, {In-orbit
  calibration of the XMM-Newton telescopes}.
pp 1--12

\bibitem[\protect\citeauthoryear{{Gondoin}, {Aschenbach}, {Beijersbergen},
  {Egger}, {Jansen}, {Stockman} \& {Tock}}{{Gondoin} et~al.}{1998}]{gondoin98a}
{Gondoin} P.,  {Aschenbach} B.~R.,  {Beijersbergen} M.~W.,  {Egger} R.,
  {Jansen} F.~A.,  {Stockman} Y.,    {Tock} J.,  1998, in {R.~B.~Hoover \&
  A.~B.~Walker} ed., Society of Photo-Optical Instrumentation Engineers (SPIE)
  Conference Series Vol.~3444 of Presented at the Society of Photo-Optical
  Instrumentation Engineers (SPIE) Conference, {Calibration of the first XMM
  flight mirror module: I. Image quality}.
pp 278--289

\bibitem[\protect\citeauthoryear{{Gondoin}, {Beijersbergen}, {Willingale}, {de
  Chambure}, {van Katwijk}, {Lumb} \& {Peacock}}{{Gondoin}
  et~al.}{1996}]{gondoin96a}
{Gondoin} P.,  {Beijersbergen} M.,  {Willingale} R.,  {de Chambure} D.,  {van
  Katwijk} K.,  {Lumb} D.~H.,    {Peacock} A.~J.,  1996, in {O.~H.~Siegmund \&
  M.~A.~Gummin} ed., Society of Photo-Optical Instrumentation Engineers (SPIE)
  Conference Series Vol.~2808 of Presented at the Society of Photo-Optical
  Instrumentation Engineers (SPIE) Conference, {Simulation of the XMM mirror
  performance based on metrology data}.
pp 376--388

\bibitem[\protect\citeauthoryear{{Gonzalez}, {Zaritsky} \&
  {Zabludoff}}{{Gonzalez} et~al.}{2007}]{gonzalez07a}
{Gonzalez} A.~H.,  {Zaritsky} D.,    {Zabludoff} A.~I.,  2007, ApJ, 666, 147

\bibitem[\protect\citeauthoryear{{Haberl}}{{Haberl}}{2007}]{haberl07a}
{Haberl} F.,  2007, Ap\&SS, 308, 181

\bibitem[\protect\citeauthoryear{{Hashimoto}, {Barcons}, {B{\"o}hringer},
  {Fabian}, {Hasinger}, {Mainieri} \& {Brunner}}{{Hashimoto}
  et~al.}{2004}]{hashimoto04a}
{Hashimoto} Y.,  {Barcons} X.,  {B{\"o}hringer} H.,  {Fabian} A.~C.,
  {Hasinger} G.,  {Mainieri} V.,    {Brunner} H.,  2004, A\&A, 417, 819

\bibitem[\protect\citeauthoryear{{Henry}}{{Henry}}{2004}]{henry04a}
{Henry} J.~P.,  2004, ApJ, 609, 603

\bibitem[\protect\citeauthoryear{{Henry} et~al.,}{{Henry}
  et~al.}{2010}]{henry10}
{Henry} J.~P.,  et~al., 2010, ApJ, 725, 615

\bibitem[\protect\citeauthoryear{{Henry}, {Mullis}, {Voges}, {B{\"o}hringer},
  {Briel}, {Gioia} \& {Huchra}}{{Henry} et~al.}{2006}]{henry06a}
{Henry} J.~P.,  {Mullis} C.~R.,  {Voges} W.,  {B{\"o}hringer} H.,  {Briel}
  U.~G.,  {Gioia} I.~M.,    {Huchra} J.~P.,  2006, ApJS, 162, 304

\bibitem[\protect\citeauthoryear{{Hilton} et~al.,}{{Hilton}
  et~al.}{2007}]{hilton07a}
{Hilton} M.,  et~al., 2007, ApJ, 670, 1000

\bibitem[\protect\citeauthoryear{{Hilton} et~al.,}{{Hilton}
  et~al.}{2009}]{hilton09a}
{Hilton} M.,  et~al., 2009, ApJ, 697, 436

\bibitem[\protect\citeauthoryear{{Hilton} et~al.,}{{Hilton}
  et~al.}{2010}]{hilton10a}
{Hilton} M.,  et~al., 2010, ApJ, 718, 133

\bibitem[\protect\citeauthoryear{{Hoeft}, {Lamer}, {Kohnert} \&
  {Schwope}}{{Hoeft} et~al.}{2008}]{hoeft08a}
{Hoeft} M.,  {Lamer} G.,  {Kohnert} J.,    {Schwope} A.,  2008, arXiv:0809.2265

\bibitem[\protect\citeauthoryear{{Horner}}{{Horner}}{2001}]{horner01a}
{Horner} D.~J.,  2001, PhD thesis, University of Maryland College Park

\bibitem[\protect\citeauthoryear{{Horner}, {Perlman}, {Ebeling}, {Jones},
  {Scharf}, {Wegner}, {Malkan} \& {Maughan}}{{Horner} et~al.}{2008}]{horner08a}
{Horner} D.~J.,  {Perlman} E.~S.,  {Ebeling} H.,  {Jones} L.~R.,  {Scharf}
  C.~A.,  {Wegner} G.,  {Malkan} M.,    {Maughan} B.,  2008, ApJS, 176, 374

\bibitem[\protect\citeauthoryear{Hosmer}{Hosmer}{2010}]{hosmer10a}
Hosmer M.,  2010, PhD thesis, Astronomy Centre, University of Sussex

\bibitem[\protect\citeauthoryear{{Hoyle}, {Jimenez} \& {Verde}}{{Hoyle}
  et~al.}{2010}]{hoyle10a}
{Hoyle} B.,  {Jimenez} R.,    {Verde} L.,  2010, arXiv:1009.3884

\bibitem[\protect\citeauthoryear{{James} \& {Roos}}{{James} \&
  {Roos}}{1975}]{james75a}
{James} F.,  {Roos} M.,  1975, Computer Physics Communications, 10, 343

\bibitem[\protect\citeauthoryear{{Jansen} \& {Laine}}{{Jansen} \&
  {Laine}}{1997}]{jansen97a}
{Jansen} F.~A.,  {Laine} R.,  1997, in Bulletin of the American Astronomical
  Society Vol.~29 of Bulletin of the American Astronomical Society, {The XMM
  Observatory, a technical and scientific overview}.
p.~1365

\bibitem[\protect\citeauthoryear{Jones \& Forman}{Jones \&
  Forman}{1984}]{jones84a}
Jones C.,  Forman W.,  1984, ApJ, 276, 38

\bibitem[\protect\citeauthoryear{Kay, da Silva, Aghanim, Blanchard, Liddle A.
  R.and~Puget, Sadat \& Thomas}{Kay et~al.}{2007}]{kay07a}
Kay S.~T.,  da Silva A.~C.,  Aghanim N.,  Blanchard A.,  Liddle A. R.and~Puget
  J.-L.,  Sadat R.,    Thomas P.~A.,  2007, MNRAS, 377, 317

\bibitem[\protect\citeauthoryear{{Kessler} et~al.,}{{Kessler}
  et~al.}{2009}]{kessler09a}
{Kessler} R.,  et~al., 2009, ApJS, 185, 32

\bibitem[\protect\citeauthoryear{{Koester} et~al.,}{{Koester}
  et~al.}{2007}]{koester07a}
{Koester} B.~P.,  et~al., 2007, ApJ, 660, 239

\bibitem[\protect\citeauthoryear{{Kolokotronis}, {Georgakakis}, {Basilakos},
  {Kitsionas}, {Plionis}, {Georgantopoulos} \& {Gaga}}{{Kolokotronis}
  et~al.}{2006}]{kolokotronis06a}
{Kolokotronis} V.,  {Georgakakis} A.,  {Basilakos} S.,  {Kitsionas} S.,
  {Plionis} M.,  {Georgantopoulos} I.,    {Gaga} T.,  2006, MNRAS, 366, 163

\bibitem[\protect\citeauthoryear{{Lamer}, {Hoeft}, {Kohnert}, {Schwope} \&
  {Storm}}{{Lamer} et~al.}{2008a}]{lamer08a}
{Lamer} G.,  {Hoeft} M.,  {Kohnert} J.,  {Schwope} A.,    {Storm} J.,  2008a,
  A\&A, 487, L33

\bibitem[\protect\citeauthoryear{{Lamer}, {Hoeft}, {Kohnert}, {Schwope} \&
  {Storm}}{{Lamer} et~al.}{2008b}]{lamar08a}
{Lamer} G.,  {Hoeft} M.,  {Kohnert} J.,  {Schwope} A.,    {Storm} J.,  2008b,
  A\&A, 487, L33

\bibitem[\protect\citeauthoryear{{Larson} et~al.,}{{Larson}
  et~al.}{2010}]{larson10a}
{Larson} D.,  et~al., 2010, arXiv:1001.4635

\bibitem[\protect\citeauthoryear{Lloyd-Davies, Ponman \& Cannon}{Lloyd-Davies
  et~al.}{2000}]{lloyd-davies00a}
Lloyd-Davies E.~J.,  Ponman T.~J.,    Cannon D.~B.,  2000, MNRAS, 315, 689

\bibitem[\protect\citeauthoryear{{Lumb}, {Warwick}, {Page} \& {De Luca}}{{Lumb}
  et~al.}{2002}]{lumb02a}
{Lumb} D.~H.,  {Warwick} R.~S.,  {Page} M.,    {De Luca} A.,  2002, A\&A, 389,
  93

\bibitem[\protect\citeauthoryear{{Majumdar} \& {Mohr}}{{Majumdar} \&
  {Mohr}}{2004}]{majumdar04a}
{Majumdar} S.,  {Mohr} J.~J.,  2004, ApJ, 613, 41

\bibitem[\protect\citeauthoryear{{Mantz}, {Allen}, {Ebeling}, {Rapetti} \&
  {Drlica-Wagner}}{{Mantz} et~al.}{2010}]{mantz10a}
{Mantz} A.,  {Allen} S.~W.,  {Ebeling} H.,  {Rapetti} D.,    {Drlica-Wagner}
  A.,  2010, MNRAS, 406, 1773

\bibitem[\protect\citeauthoryear{{Mantz}, {Allen} \& {Rapetti}}{{Mantz}
  et~al.}{2010}]{mantz10b}
{Mantz} A.,  {Allen} S.~W.,    {Rapetti} D.,  2010, MNRAS, 406, 1805

\bibitem[\protect\citeauthoryear{{Marriage} et~al.,}{{Marriage}
  et~al.}{2010}]{marriage10a}
{Marriage} T.~A.,  et~al., 2010, arXiv:1010.1065

\bibitem[\protect\citeauthoryear{Maughan}{Maughan}{2007}]{maughan07a}
Maughan B.~J.,  2007, ApJ, 668, 772

\bibitem[\protect\citeauthoryear{{Maughan}, {Jones}, {Forman} \& {Van
  Speybroeck}}{{Maughan} et~al.}{2008}]{maughan08a}
{Maughan} B.~J.,  {Jones} C.,  {Forman} W.,    {Van Speybroeck} L.,  2008,
  ApJS, 174, 117

\bibitem[\protect\citeauthoryear{Mehrtens et~al.,}{Mehrtens
  et~al.}{2011}]{mehrtens10a}
Mehrtens N.,  et~al., 2011, in prep.

\bibitem[\protect\citeauthoryear{{Melin}, {Bartlett} \& {Delabrouille}}{{Melin}
  et~al.}{2005}]{melin05a}
{Melin} J.,  {Bartlett} J.~G.,    {Delabrouille} J.,  2005, A\&A, 429, 417

\bibitem[\protect\citeauthoryear{{Menanteau} et~al.,}{{Menanteau}
  et~al.}{2010}]{menanteau10a}
{Menanteau} F.,  et~al., 2010, arXiv:1006.5126

\bibitem[\protect\citeauthoryear{Mewe, Lemen \& van~den Oord}{Mewe
  et~al.}{1986}]{mewe86a}
Mewe R.,  Lemen J.~R.,    van~den Oord G.,  1986, A\&A, 65, 511

\bibitem[\protect\citeauthoryear{{Miller} et~al.,}{{Miller}
  et~al.}{2005}]{miller05a}
{Miller} C.~J.,  et~al., 2005, AJ, 130, 968

\bibitem[\protect\citeauthoryear{{Morrison} \& {McCammon}}{{Morrison} \&
  {McCammon}}{1983}]{morrison83a}
{Morrison} R.,  {McCammon} D.,  1983, ApJ, 270, 119

\bibitem[\protect\citeauthoryear{{Mullis} et~al.,}{{Mullis}
  et~al.}{2003}]{mullis03a}
{Mullis} C.~R.,  et~al., 2003, ApJ, 594, 154

\bibitem[\protect\citeauthoryear{{Mullis}, {Rosati}, {Lamer}, {B{\"o}hringer},
  {Schwope}, {Schuecker} \& {Fassbender}}{{Mullis} et~al.}{2005}]{mullis05a}
{Mullis} C.~R.,  {Rosati} P.,  {Lamer} G.,  {B{\"o}hringer} H.,  {Schwope} A.,
  {Schuecker} P.,    {Fassbender} R.,  2005, ApJ, 623, L85

\bibitem[\protect\citeauthoryear{{Mushotzky}, {Done} \& {Pounds}}{{Mushotzky}
  et~al.}{1993}]{mushotzky93a}
{Mushotzky} R.~F.,  {Done} C.,    {Pounds} K.~A.,  1993, ARA\&A, 31, 717

\bibitem[\protect\citeauthoryear{{Olsen} et~al.,}{{Olsen}
  et~al.}{2008}]{olsen08a}
{Olsen} L.~F.,  et~al., 2008, A\&A, 478, 93

\bibitem[\protect\citeauthoryear{Onuora, Kay \& Thomas}{Onuora
  et~al.}{2003}]{onuora03a}
Onuora L.~I.,  Kay S.~T.,    Thomas P.~A.,  2003, MNRAS, 341, 1246

\bibitem[\protect\citeauthoryear{{Oukbir} \& {Blanchard}}{{Oukbir} \&
  {Blanchard}}{1992}]{oukbir92a}
{Oukbir} J.,  {Blanchard} A.,  1992, A\&A, 262, L21

\bibitem[\protect\citeauthoryear{Pacaud, Pierre, Adami, Altieri, Andreon,
  Chiappetti, Detal et~al.,}{Pacaud et~al.}{2007}]{pacaud07a}
Pacaud F.,  Pierre M.,  Adami C.,  Altieri B.,  Andreon S.,  Chiappetti L.,
  Detal A.,    et~al., 2007, MNRAS, 382, 1289

\bibitem[\protect\citeauthoryear{{Papovich} et~al.,}{{Papovich}
  et~al.}{2010}]{papovich10a}
{Papovich} C.,  et~al., 2010, ApJ, 716, 1503

\bibitem[\protect\citeauthoryear{{Perlman}, {Horner}, {Jones}, {Scharf},
  {Ebeling}, {Wegner} \& {Malkan}}{{Perlman} et~al.}{2002}]{perlman02a}
{Perlman} E.~S.,  {Horner} D.~J.,  {Jones} L.~R.,  {Scharf} C.~A.,  {Ebeling}
  H.,  {Wegner} G.,    {Malkan} M.,  2002, ApJS, 140, 265

\bibitem[\protect\citeauthoryear{{Piccinotti}, {Mushotzky}, {Boldt}, {Holt},
  {Marshall}, {Serlemitsos} \& {Shafer}}{{Piccinotti}
  et~al.}{1982}]{piccinotti82a}
{Piccinotti} G.,  {Mushotzky} R.~F.,  {Boldt} E.~A.,  {Holt} S.~S.,  {Marshall}
  F.~E.,  {Serlemitsos} P.~J.,    {Shafer} R.~A.,  1982, ApJ, 253, 485

\bibitem[\protect\citeauthoryear{{Pierre} et~al.,}{{Pierre}
  et~al.}{2006}]{pierre06a}
{Pierre} M.,  et~al., 2006, MNRAS, 372, 591

\bibitem[\protect\citeauthoryear{{Ponman}, {Cannon} \& {Navarro}}{{Ponman}
  et~al.}{1999}]{ponman99a}
{Ponman} T.~J.,  {Cannon} D.~B.,    {Navarro} J.~F.,  1999, Nature, 397, 135

\bibitem[\protect\citeauthoryear{{Pradas} \& {Kerp}}{{Pradas} \&
  {Kerp}}{2005}]{pradas05a}
{Pradas} J.,  {Kerp} J.,  2005, aap, 443, 721

\bibitem[\protect\citeauthoryear{{Pratt}, {Croston}, {Arnaud} \&
  {B{\"o}hringer}}{{Pratt} et~al.}{2009}]{pratt09a}
{Pratt} G.~W.,  {Croston} J.~H.,  {Arnaud} M.,    {B{\"o}hringer} H.,  2009,
  A\&A, 498, 361

\bibitem[\protect\citeauthoryear{{Predehl} et~al.,}{{Predehl}
  et~al.}{2006}]{predehl06a}
{Predehl} P.,  et~al., 2006, in Society of Photo-Optical Instrumentation
  Engineers (SPIE) Conference Series Vol.~6266 of Society of Photo-Optical
  Instrumentation Engineers (SPIE) Conference Series, {eROSITA}

\bibitem[\protect\citeauthoryear{{Rapetti}, {Allen}, {Mantz} \&
  {Ebeling}}{{Rapetti} et~al.}{2010}]{rapetti10a}
{Rapetti} D.,  {Allen} S.~W.,  {Mantz} A.,    {Ebeling} H.,  2010, MNRAS, 406,
  1796

\bibitem[\protect\citeauthoryear{Read}{Read}{2004}]{read04a}
Read A.~M.,  2004, Technical Report XMM-CCF-REL-167, PSF of the X-ray
  telescopes.
ESA

\bibitem[\protect\citeauthoryear{{Read} \& {Ponman}}{{Read} \&
  {Ponman}}{2003}]{read03a}
{Read} A.~M.,  {Ponman} T.~J.,  2003, A\&A, 409, 395

\bibitem[\protect\citeauthoryear{Read, Saxton, Guainazzi, Rosen \&
  Stuhlinger}{Read et~al.}{2010}]{read10a}
Read A.~M.,  Saxton R.,  Guainazzi M.,  Rosen S.,    Stuhlinger M.,  2010,
  Technical Report XMM-CCF-REL-263, 2-D PSF parameterisation.
ESA

\bibitem[\protect\citeauthoryear{Reiprich \& B{\"{o}}hringer}{Reiprich \&
  B{\"{o}}hringer}{2002}]{reiprich02a}
Reiprich T.~H.,  B{\"{o}}hringer H.,  2002, ApJ, 567, 716

\bibitem[\protect\citeauthoryear{{Romer} et~al.,}{{Romer}
  et~al.}{2000}]{romer00a}
{Romer} A.~K.,  et~al., 2000, ApJS, 126, 209

\bibitem[\protect\citeauthoryear{{Romer}, {Viana}, {Liddle} \& {Mann}}{{Romer}
  et~al.}{2001}]{romer01a}
{Romer} A.~K.,  {Viana} P.~T.~P.,  {Liddle} A.~R.,    {Mann} R.~G.,  2001, ApJ,
  547, 594

\bibitem[\protect\citeauthoryear{{Rosati}, {della Ceca}, {Norman} \&
  {Giacconi}}{{Rosati} et~al.}{1998}]{rosati98a}
{Rosati} P.,  {della Ceca} R.,  {Norman} C.,    {Giacconi} R.,  1998, ApJ, 492,
  L21

\bibitem[\protect\citeauthoryear{{Rosati} et~al.,}{{Rosati}
  et~al.}{2009}]{rosati09a}
{Rosati} P.,  et~al., 2009, A\&A, 508, 583

\bibitem[\protect\citeauthoryear{{Rozo} et~al.,}{{Rozo}
  et~al.}{2010}]{rozo10a}
{Rozo} E.,  et~al., 2010, ApJ, 708, 645

\bibitem[\protect\citeauthoryear{Sahl\'{e}n et~al.,}{Sahl\'{e}n
  et~al.}{2009}]{sahlen09a}
Sahl\'{e}n M.,  et~al., 2009, MNRAS, 397, 577

\bibitem[\protect\citeauthoryear{{Santos} et~al.,}{{Santos}
  et~al.}{2009}]{santos09a}
{Santos} J.~S.,  et~al., 2009, A\&A, 501, 49

\bibitem[\protect\citeauthoryear{{Schneider} et~al.,}{{Schneider}
  et~al.}{2007}]{schneider07a}
{Schneider} D.~P.,  et~al., 2007, aj, 134, 102

\bibitem[\protect\citeauthoryear{{Schwope} et~al.,}{{Schwope}
  et~al.}{2010}]{schwope10a}
{Schwope} A.~D.,  et~al., 2010, A\&A, 513, L10

\bibitem[\protect\citeauthoryear{{Schwope}, {Lamer}, {Burke}, {Elvis},
  {Watson}, {Schulze}, {Szokoly} \& {Urrutia}}{{Schwope}
  et~al.}{2004}]{schwope04a}
{Schwope} A.~D.,  {Lamer} G.,  {Burke} D.,  {Elvis} M.,  {Watson} M.~G.,
  {Schulze} M.~P.,  {Szokoly} G.,    {Urrutia} T.,  2004, Advances in Space
  Research, 34, 2604

\bibitem[\protect\citeauthoryear{{Sehgal} et~al.,}{{Sehgal}
  et~al.}{2011}]{sehgal11a}
{Sehgal} N.,  et~al., 2011, ApJ, 732, 44

\bibitem[\protect\citeauthoryear{{Slezak}, {Bijaoui} \& {Mars}}{{Slezak}
  et~al.}{1990}]{slezak90a}
{Slezak} E.,  {Bijaoui} A.,    {Mars} G.,  1990, A\&A, 227, 301

\bibitem[\protect\citeauthoryear{Stanford et~al.,}{Stanford
  et~al.}{2006}]{stanford06a}
Stanford S.~A.,  et~al., 2006, ApJ, 646, L13

\bibitem[\protect\citeauthoryear{{Staniszewski} et~al.,}{{Staniszewski}
  et~al.}{2009}]{staniszewski09a}
{Staniszewski} Z.,  et~al., 2009, ApJ, 701, 32

\bibitem[\protect\citeauthoryear{{Stark}, {Gammie}, {Wilson}, {Bally}, {Linke},
  {Heiles} \& {Hurwitz}}{{Stark} et~al.}{1992}]{stark92a}
{Stark} A.~A.,  {Gammie} C.~F.,  {Wilson} R.~W.,  {Bally} J.,  {Linke} R.~A.,
  {Heiles} C.,    {Hurwitz} M.,  1992, ApJS, 79, 77

\bibitem[\protect\citeauthoryear{{Stockman}, {Mazy}, {Tock}, {de Chambure} \&
  {Gondoin}}{{Stockman} et~al.}{1997}]{stockman97a}
{Stockman} Y.,  {Mazy} E.,  {Tock} J.~P.,  {de Chambure} D.,    {Gondoin} P.,
  1997, in {T.-D.~Guyenne} ed., Environmental Testing for Space Programms
  Vol.~408 of ESA Special Publication, {X-Ray and EUV Characterisation of the
  First XMM Flight Mirror Module}.
p.~169

\bibitem[\protect\citeauthoryear{{Stott} et~al.,}{{Stott}
  et~al.}{2010}]{stott10a}
{Stott} J.~P.,  et~al., 2010, ApJ, 718, 23

\bibitem[\protect\citeauthoryear{Suhada et~al.,}{Suhada
  et~al.}{2010}]{suhada10a}
Suhada R.,  et~al., 2010, A\&A, 514, L3

\bibitem[\protect\citeauthoryear{Suhada et~al.,}{Suhada
  et~al.}{2011}]{Suhada11}
Suhada R.,  et~al., 2011, arXiv:1104.4888

\bibitem[\protect\citeauthoryear{Sunyaev \& Zeldovich}{Sunyaev \&
  Zeldovich}{1972}]{sunyaev72a}
Sunyaev R.~A.,  Zeldovich Y.~B.,  1972, Comments Astrophys. Space Phys., 4, 173

\bibitem[\protect\citeauthoryear{{Tanaka}, {Finoguenov} \& {Ueda}}{{Tanaka}
  et~al.}{2010}]{tanaka10a}
{Tanaka} M.,  {Finoguenov} A.,    {Ueda} Y.,  2010, ApJ, 716, L152

\bibitem[\protect\citeauthoryear{{Vanderlinde} et~al.,}{{Vanderlinde}
  et~al.}{2010}]{vanderlinde10a}
{Vanderlinde} K.,  et~al., 2010, arXiv:1003.0003

\bibitem[\protect\citeauthoryear{{V{\'e}ron-Cetty} \&
  {V{\'e}ron}}{{V{\'e}ron-Cetty} \& {V{\'e}ron}}{2006}]{veron06a}
{V{\'e}ron-Cetty} M.,  {V{\'e}ron} P.,  2006, A\&A, 455, 773

\bibitem[\protect\citeauthoryear{{Vikhlinin} et~al.,}{{Vikhlinin}
  et~al.}{2009a}]{vikhlinin09a}
{Vikhlinin} A.,  et~al., 2009a, ApJ, 692, 1033

\bibitem[\protect\citeauthoryear{{Vikhlinin} et~al.,}{{Vikhlinin}
  et~al.}{2009b}]{vikhlinin09b}
{Vikhlinin} A.,  et~al., 2009b, ApJ, 692, 1060

\bibitem[\protect\citeauthoryear{{Villa} et~al.,}{{Villa}
  et~al.}{1996}]{villa96a}
{Villa} G.~E.,  et~al., 1996, in {O.~H.~Siegmund \& M.~A.~Gummin} ed., Society
  of Photo-Optical Instrumentation Engineers (SPIE) Conference Series
  Vol.~2808, {EPIC system onboard the ESA XMM}.
pp 402--413

\bibitem[\protect\citeauthoryear{Voit}{Voit}{2005}]{voit05a}
Voit G.~M.,  2005, Rev. Mod. Phys., 77, 207

\bibitem[\protect\citeauthoryear{{Watson} et~al.,}{{Watson}
  et~al.}{2009}]{watson09a}
{Watson} M.~G.,  et~al., 2009, A\&A, 493, 339

\bibitem[\protect\citeauthoryear{{Weisskopf}}{{Weisskopf}}{1999}]{weisskopf99a}
{Weisskopf} M.~C.,  1999, arXiv:astro-ph/9912097

\bibitem[\protect\citeauthoryear{{Werner} et~al.,}{{Werner}
  et~al.}{2007}]{werner07a}
{Werner} N.,  et~al., 2007, A\&A, 474, 707

\bibitem[\protect\citeauthoryear{{Williamson} et~al.,}{{Williamson}
  et~al.}{2011}]{williamson11a}
{Williamson} R.,  et~al., 2011, arXiv:1101.1290

\bibitem[\protect\citeauthoryear{Wilms, Allen \& McCray}{Wilms
  et~al.}{2000}]{wilms00a}
Wilms J.,  Allen A.,    McCray R.,  2000, ApJ, 542, 914

\bibitem[\protect\citeauthoryear{{Wilson} et~al.,}{{Wilson}
  et~al.}{2009}]{wilson09a}
{Wilson} G.,  et~al., 2009, ApJ, 698, 1943

\bibitem[\protect\citeauthoryear{{Wittman}, {Margoniner}, {Tyson}, {Cohen},
  {Becker} \& {Dell'Antonio}}{{Wittman} et~al.}{2003}]{wittman03a}
{Wittman} D.,  {Margoniner} V.~E.,  {Tyson} J.~A.,  {Cohen} J.~G.,  {Becker}
  A.~C.,    {Dell'Antonio} I.~P.,  2003, ApJ, 597, 218

\bibitem[\protect\citeauthoryear{Wolter}{Wolter}{1952a}]{wolter52b}
Wolter H.,  1952a, Ann. Physik, 10, 286

\bibitem[\protect\citeauthoryear{Wolter}{Wolter}{1952b}]{wolter52a}
Wolter H.,  1952b, Ann. Physik, 10, 94

\bibitem[\protect\citeauthoryear{{Wu}, {Rozo} \& {Wechsler}}{{Wu}
  et~al.}{2010}]{wu10a}
{Wu} H.,  {Rozo} E.,    {Wechsler} R.~H.,  2010, ApJ, 713, 1207

\bibitem[\protect\citeauthoryear{{Yu}, {Tozzi}, {Borgani}, {Rosati} \&
  {Zhu}}{{Yu} et~al.}{2011}]{yu11a}
{Yu} H.,  {Tozzi} P.,  {Borgani} S.,  {Rosati} P.,    {Zhu} Z.-H.,  2011, A\&A,
  529, A65+

\bibitem[\protect\citeauthoryear{Zwicky, Herzog \& Wild}{Zwicky
  et~al.}{1968}]{zwicky68a}
Zwicky F.,  Herzog E.,    Wild P.,  1968, Catalogue of galaxies and of clusters
  of galaxies.
California Institute of Technology, Pasadena

\end{thebibliography}
